\documentclass[fleqn,10pt,authoryear]{elsarticle}
\usepackage{amsmath}
\usepackage{amssymb}
\usepackage[margin=0.8in]{geometry}
\usepackage{graphics}
\usepackage{graphicx}
\usepackage{epsfig}
\usepackage{overpic}
\usepackage{subfigure}
\usepackage{epstopdf}
\usepackage{float}
\usepackage{setspace}
\usepackage{color}
\usepackage{xcolor}
\usepackage{listings}
\usepackage{lipsum}
\usepackage{algorithm}
\usepackage{algpseudocode}

\DeclareMathOperator{\Tr}{Tr}
\DeclareMathOperator{\Div}{Div}

\DeclareMathOperator*{\sgn}{sgn}

\usepackage[font=small,labelfont=bf,labelsep=period,width=.9\textwidth]{caption}

\usepackage[colorinlistoftodos,prependcaption,textsize=tiny]{todonotes}

\graphicspath{{./figures/}}

\pdfoutput=1

\usepackage{hyperref}
\hypersetup{
    unicode=false,          
    pdftoolbar=true,        
    pdfmenubar=true,        
    pdffitwindow=false,     
    pdfstartview={FitH},    
    pdftitle={My title},    
    pdfauthor={Author},     
    pdfsubject={Subject},   
    pdfcreator={Creator},   
    pdfproducer={Producer}, 
    pdfkeywords={keyword1, key2, key3}, 
    pdfnewwindow=true,      
    colorlinks=false,       
    linkcolor=red,          
    citecolor=green,        
    filecolor=magenta,      
    urlcolor=cyan           
}

\newcommand{\strain}{\varepsilon}
\newcommand{\dmg}{d}

\begin{document}

\begin{frontmatter}
\title{A phase-field formulation for dynamic cohesive fracture}


\journal{Computer Methods in Applied Mechanics and Engineering}

\author[d1]{Rudy J.M. Geelen}
\author[d2]{Yingjie Liu}
\author[d1]{Tianchen Hu}
\author[s]{Michael R. Tupek}
\author[d1,d2]{John E. Dolbow}
\ead{jdolbow@duke.edu}

\address[d1]{Department of Mechanical Engineering and Materials Science, Duke University, Durham, NC 27708, USA}
\address[d2]{Department of Civil and Environmental Engineering, Duke University, Durham, NC 27708, USA}
\address[s]{Sandia National Laboratories, Albuquerque, New Mexico, USA}

\begin{abstract}
We extend a phase-field/gradient damage formulation for cohesive fracture to the dynamic case.  The model is characterized by a regularized fracture energy that is linear in the damage field, as well as non-polynomial degradation functions.  Two categories of degradation functions are examined, and a process to derive a given degradation function based on a local stress-strain response in the cohesive zone is presented. The resulting model is characterized by a linear elastic regime prior to the onset of damage, and controlled strain-softening thereafter. The governing equations are derived according to macro- and microforce balance theories, naturally accounting for the irreversible nature of the fracture process by introducing suitable constraints for the kinetics of the underlying microstructural changes. The model is complemented by an efficient staggered solution scheme based on an augmented Lagrangian method. Numerical examples demonstrate that the proposed model is a robust and effective method for simulating cohesive crack propagation, with particular emphasis on dynamic fracture.  

\end{abstract}


\end{frontmatter}



\section{Introduction}

In recent years diffuse crack approaches have become increasingly popular for predicting crack propagation in engineering materials. This is largely due to their potential for capturing the evolution of complex crack patterns without the need for algorithmic crack-front tracking methods. There has been a proliferation of modeling techniques developed around a variational formulation of, most commonly, a Griffith-type description of fracture. In this manuscript, we take an alternative approach and develop a phase-field model of dynamic fracture that approximates a cohesive type of response. Critically, our model extends the work of \cite{LORENTZ201120, Lorentz2012} to the dynamic regime and to a broad set of local cohesive behaviors. In contrast to regularized formulations based on Griffith-type descriptions of fracture, the effective macroscopic fracture parameters can be held fixed as the regularization length scale is decreased. We employ Lagrange multipliers to restrict the evolution of the damage to satisfy physically-motivated constraints. The robustness of the resulting formulation is demonstrated for multi-dimensional quasi-static and dynamic problems in fracture and fragmentation.  

Phase-field models for fracture rely on the regularization of a sharp crack surface by means of an elliptical functional. In \cite{ambrosio1990approximation}, a phase-field approximation to discontinuous fields was proposed, inspired by the work on image segmentation by \cite{CPA:CPA3160420503}. In \cite{FRANCFORT19981319} and \cite{BOURDIN2000797, bourdin2008variational} a phase-field approximation was proposed to a variational formulation of brittle fracture based on the same potential. In these variational frameworks, which have largely become the standard in the phase-field for fracture community, sharp crack topologies are regularized by their diffusive counterparts and governed by a scalar continuous field variable. More recently, alternative derivations have been proposed in \cite{Miehe20102765,NME:NME2861} and \cite{DASILVA20132178}, which have broadened the scope of phase-field modeling for fracture. With respect to dynamic fracture, phase-field models have been considered in \cite{Bourdin2011}, \cite{BORDEN201277}, \cite{NME:NME4387}, \cite{Schluter2014} and \cite{NME:NME5262}, just to name a few. The vast majority of these works have been based on a regularization of a Griffith type description of fracture, in which the stress field at the crack tip is singular.  

\begin{figure}[!tbp]
\centering \scriptsize
\begin{overpic}[width=.46\linewidth]{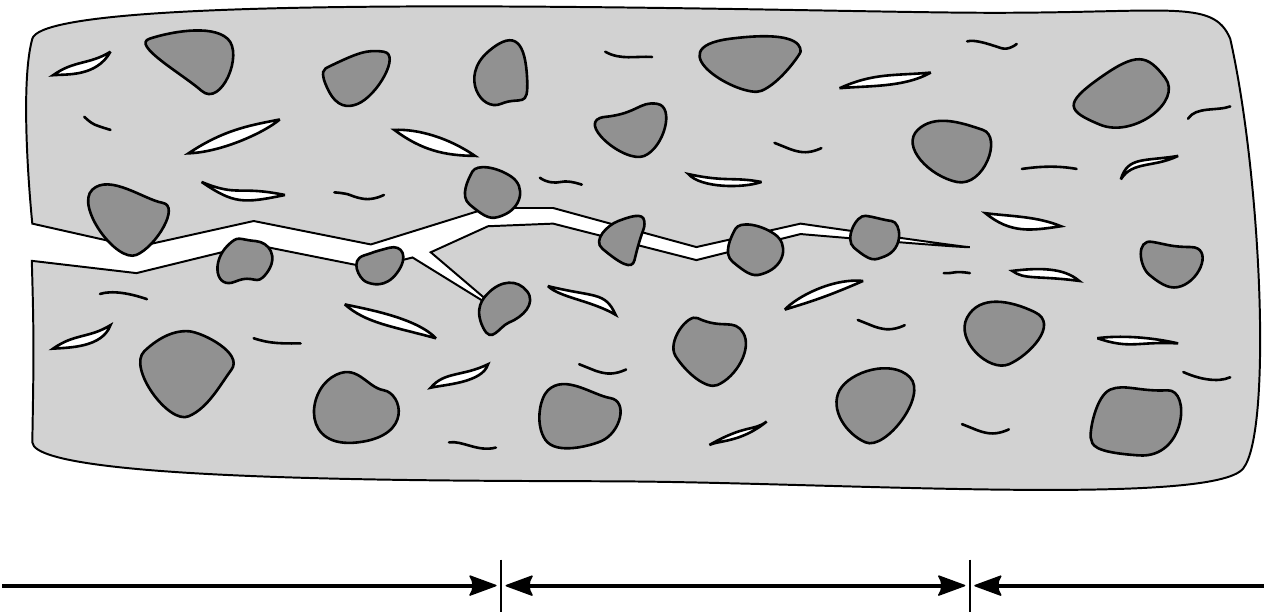}
\put(9 ,4){Traction-free}
\put(41.75,4){Fracture process zone}
\put(84,4){Intact}
\end{overpic}
\caption{Schematic representation of a fracture process zone.}
\label{fig:fpz}
\end{figure}

In fact, it bears emphasis that most materials are not perfectly brittle in the Griffith sense, but display some ductility after reaching their ultimate strength. Often, there exists a \textit{process zone} in the vicinity of the crack tip. This region is typically characterized by small-scale yielding, micro-crack initiation, growth and coalescence, as graphically depicted in Figure \ref{fig:fpz}. In those cases in which the fracture process zone is sufficiently small compared to the representative size of the structure, the use of a Griffith description of fracture is justified. In the more general case, cohesive forces in the fracture process zone must be taken into account, and the utility of a Griffith model diminishes.  


In addition to this restriction, there are clear disadvantages to phase-field models based on a Griffith description of fracture. It is well known that the regularization length is intrinsically tied to the material properties, e.g.\ see \cite{BORDEN201277}, which poses a significant limitation for many important crack propagation problems. In addition, such formulations typically exhibit nonlinear behavior immediately at the onset of loading. In the quest to develop a remedy to these issues, researchers have adopted a number of different perspectives. In \cite{Pham2013} a modification to the phase-field potential was proposed which results in a purely elastic response up to the onset of damage. In \cite{MIEHE2015486, MIEHE2015449} the same modification was considered and complemented by a generalization of the crack driving forces, allowing for a large spectrum of multi-field problems to be studied in the phase-field framework. In particular, it is characterized by a \textit{threshold} energy in the elastic regime, after which stress-based failure criteria are defined to govern crack growth in the strain-softening regime. The same motivation lead \cite{Borden2016130} to employ a cubic degradation function instead of the commonly adopted quadratic functional. This function essentially provides a stress-strain response prior to crack growth that more closely approximates linear elasticity. Despite the quasi-linear behavior, however, there is no means of calibrating the effective cohesive response to different materials. Therefore we view this approach as somewhat limited in use in the context of cohesive fracture problems. In \cite{TANNE201880} it was shown that both crack nucleation and propagation can be predicted for a large variety of materials by carefully choosing the length scale in a standard phase-field formulation. However, for materials with large process zone sizes relative to the geometry, this strategy results in regularized damage bands that are extremely wide.

An extension of the variational formulation of brittle fracture to cohesive fracture has been considered in \cite{bourdin2008variational}, but the development of a phase-field approximation is widely regarded as non-trivial. Several recent efforts have been made to establish such a formulation. In \cite{Lorentz20111927} a gradient damage model was proposed that decouples the regularization length from the macroscopic fracture parameters. The same formulation was demonstrated to converge to a cohesive zone model in the limit of vanishing regularization length  in \cite{LORENTZ201120}. However, its extension to the multidimensional case remains, from a conceptual standpoint, a non-trivial challenge. In \cite{NME:NME4553}, an attempt was made to construct a phase-field model for cohesive fracture by casting the cohesive zone approach, pioneered by \cite{dugdale1960yielding} and \cite{barenblatt1962mathematical}, in an energetic framework. The approach relies on the construction of an auxiliary field to capture the displacement jump across the regularized fracture surface.   

With particular regard to the gradient damage model from \cite{Lorentz20111927}, we caution against establishing too strong a connection between regularized descriptions of cohesive fracture and the cohesive zone concept. In a conventional cohesive zone model, tractions are transmitted across a two-dimensional surface, which is embedded in a three-dimensional continuum. The relevant kinematic quantities are the crack opening displacements in the normal and shear directions. However, a kinematic quantity that represents the \textit{stretching} of the fracture plane, i.e. the strain component parallel to the crack, is lacking in classical cohesive zone models. Generally speaking, these in-plane components cannot be ignored. For this reason, the cohesive zone model should be regarded as a strictly \textit{uniaxial} model, as argued in \cite{BAZANT2002165}. In fact, a number of approaches have recently been proposed to overcome this limitation of cohesive zone models, such as the finite band method from \cite{HUESPE20092349} and the cohesive band model from \cite{Remmers2013}, just to mention a few. It is well known that conventional zero thickness cohesive zone models are very limited in use, despite their conceptual simplicity. The aforementioned works make an attempt to include stress triaxiality effects in the constitutive response, which play an important role in the context of the ductile failure of materials. While the approach proposed in this paper embraces a \textit{diffuse} description of fracture, the formulation is seen to be more in line with these contributions than with conventional cohesive zone formulations.

In this contribution, we explore the fundamental properties of dynamic cohesive fracture in a phase-field setting. The particular choice of the crack surface density functional will be motivated and various degradation functions are derived and examined. 
We examine the issue of convergence of these types of models as the regularization length scale is reduced, while maintaining sufficient numerical resolution. While some evidence of mesh insensitivity has previously been demonstrated in \cite{Lorentz2012}, a formal numerical validation and convergence study has yet to be provided.
Finally, we examine various options for enforcing the irreversibility of the damage field as fracture progresses.  Several benchmark problems in quasi-static and dynamic fracture are considered to demonstrate the robustness and efficacy of the models and methods.  

The paper is structured as follows. In Section \ref{sec:formulation} a phase-field/gradient damage formulation is presented that is particularly well-suited for cohesive fracture mechanics. Special attention is given to the elastic and dissipation functionals, after which the governing equations are derived using macro- and microforce balance theories. The section concludes with a discussion on the crack driving forces and the stiffness degradation function. Section \ref{sec:analysis} provides some analysis for the model described in Section \ref{sec:formulation}. Some of the fundamental properties of the method are illustrated by means of a one-dimensional analysis. We then present a process for deriving degradation functions that give rise to particular stress-strain behaviors in the process zone, with an emphasis on linear decay.  
 Finally, we discuss some aspects concerning the regularization of crack topologies and the issue of $\Gamma$-convergence. Section \ref{sec:finite_element_implementation} discusses the finite element implementation of the proposed formulation. The augmented Lagrangian method for enforcing irreversibility is reviewed and the coupled problem is formulated by means of the Galerkin method. The numerical formulation is completed by the definition of a robust solution scheme based on operator splits. Section \ref{sec:numerical_examples} provides a series of numerical experiments under quasi-static and dynamic loading conditions. Finally, a summary and concluding remarks are presented in Section \ref{sec:conclusions}.
 
\section{Cohesive fracture in a phase-field setting}
\label{sec:formulation}

We now define a phase-field/gradient damage formulation for dynamic cohesive fracture in elastic solids under small-strains and iso-thermal conditions. The theory is set up around materials for which a Griffith's description of fracture is no longer justified due to the size of the fracture process zone. The governing equations are obtained following macro- and microforce balance theories, which have been commonly applied in the development of phase-field theories. The theory developed here incorporates irreversibility by introducing suitable constraints on the fracture evolution.

\subsection{General considerations}

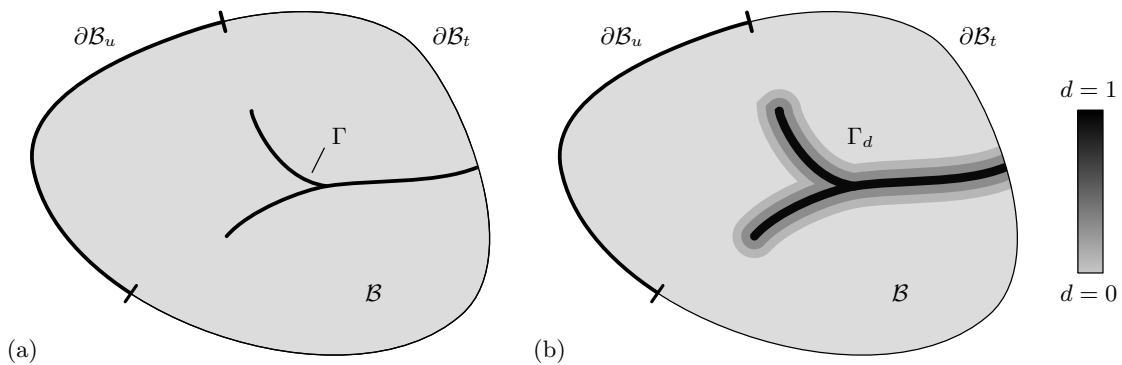
\begin{figure}[!bp]
\centering \small
\begin{overpic}[width=.85\linewidth]{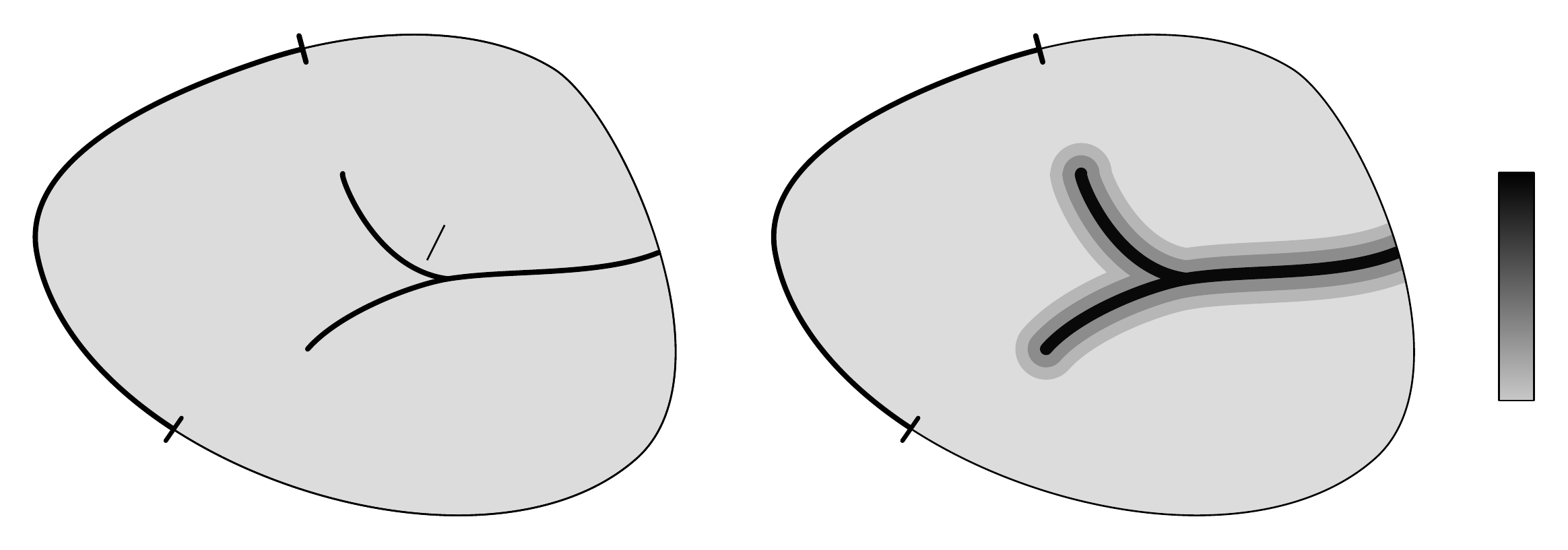}
\put(29,21){$\Gamma$}
\put(75,21){$\Gamma_d$}
\put(6 ,30){$\partial \mathcal{B}_u$}
\put(53,30){$\partial \mathcal{B}_u$}
\put(38,30){$\partial \mathcal{B}_t$}
\put(85,30){$\partial \mathcal{B}_t$}
\put(32,7){$\mathcal{B}$}
\put(79,7){$\mathcal{B}$}
\put(94,7 ){$d=0$}
\put(94,25.25){$d=1$}
\put(0 ,2){(a)}
\put(47,2){(b)}
\end{overpic}
\caption{ (a) Sketch of a body, $\mathcal{B}$, with an internal discontinuity $\Gamma$. (b) A regularized representation of the internal discontinuity.}
\label{fig:body}
\end{figure}

Consider a body $\mathcal{B} \subset \mathbb{R}^n$ (with $n \in  \left\lbrace 1,2,3 \right\rbrace$) with external boundary $\partial \mathcal{B}$ with $\partial\mathcal{B}_t \cap \partial \mathcal{B}_u = \emptyset$, $\partial\mathcal{B}_t \cup \partial \mathcal{B}_u = \partial \mathcal{B}$ and an internal discontinuity boundary $\Gamma$ as shown in Figure \ref{fig:body}a.  
The state of the system is described by two independent variables, the vector displacement field $\mathbf{u}$ and a scalar damage field $d$.  As in standard phase-field descriptions, the damage plays the role of approximating a given crack $\Gamma$ by its regularized counterpart $\Gamma_d$, as shown in Figure \ref{fig:body}b. Following continuum damage mechanics conventions, it takes values in $[0,1]$, with $d=0$ away from the crack surface and $d=1$ inside the crack.  Small deformations and deformation gradients are assumed. The infinitesimal strain tensor $\boldsymbol{\varepsilon}$ is defined as \begin{equation}
\boldsymbol{\varepsilon} = \frac{1}{2} \left( \nabla \mathbf{u} + (\nabla \mathbf{u})^T \right).
\end{equation}
We assume that the damage field acts only to degrade the tensile resistance of the body and that crack propagation is prohibited under compression.  Following  \cite{Miehe20102765}, this is effected by employing a spectral decomposition of $\boldsymbol{\varepsilon}$ into positive and negative components, via 
\begin{equation}
\boldsymbol{\varepsilon}_{\pm} := \sum_{a=1}^{n_{sd}} \left\langle \varepsilon_a \right\rangle_{\pm} \mathbf{M}_a, \qquad \mathbf{M}_a = \mathbf{n}_a \otimes \mathbf{n}_a,
\end{equation}
where $\lbrace \varepsilon_a \rbrace_{a=1...n_{sd}}$ are the principal strains and $\lbrace \mathbf{n}_a \rbrace_{a=1...n_{sd}}$ the principal strain directions.  The positive and negative operators $\langle \,\cdot\,\rangle_+$ and $\langle \,\cdot\,\rangle_-$ are defined in accord with
\begin{equation}
\langle x \rangle_+ = 
\begin{cases}
~x & \text{if}~x\geq 0,  \\
~0 & \text{otherwise,}
\end{cases}
\quad \text{and} \quad
\langle x \rangle_- = 
\begin{cases}
~x & \text{if}~x\leq 0, \\
~0 & \text{otherwise.}
\end{cases}
\end{equation}
The decomposition of the strain makes it possible to decompose the strain energy density $\psi_0$ into tensile and compressive contributions. We restrict attention to isotropic linear elasticity, and define these strain energy densities by \begin{equation}
\label{eq:strain_nrg}
\psi_0^+(\boldsymbol{\varepsilon}) = \dfrac{1}{2} \lambda_s  \left\langle\Tr{\boldsymbol{\varepsilon}} \right\rangle_+^2
+ \mu_s \boldsymbol{\varepsilon}_+:\boldsymbol{\varepsilon}_+ \quad \text{and} \quad
\psi_0^-(\boldsymbol{\varepsilon}) = \dfrac{1}{2} \lambda_s  \left\langle\Tr{\boldsymbol{\varepsilon}} \right\rangle_-^2
+ \mu_s \,\boldsymbol{\varepsilon}_-: \boldsymbol{\varepsilon}_-,
\end{equation}
where $\lambda_s>0$ and $\mu_s>0$ are the Lam\'{e} coefficients.

We assume that the total potential energy $\Psi$ of a body $\mathcal{B}$ consists of bulk and fracture contributions. In accordance with an energetic description of fractured bodies, its total free energy can be written as
\begin{equation}
\Psi(\boldsymbol{\varepsilon},\Gamma) = \int_{\mathcal{B}} \psi_0(\boldsymbol{\varepsilon})~d\mathcal{B} + \int_{\Gamma} \mathcal{G}_c ~d\Gamma,
\label{total_energy1}
\end{equation}
where $\mathcal{G}_c$ was introduced as the critical fracture energy per unit area and $\Gamma$ equals the unknown fracture surface. As in \cite{bourdin2008variational}, we approximate this functional by
\begin{equation}
\widetilde{\Psi}(\boldsymbol{\varepsilon},d,\nabla d) = \int_{\mathcal{B}} \psi_{\text{bulk}}(\boldsymbol{\varepsilon},d)~d\mathcal{B} + \int_{\mathcal{B}} \mathcal{G}_c \gamma_L ~d\mathcal{B},
\label{total_energy2}
\end{equation} 
where $\psi_{\text{bulk}}$ is the \textit{degrading} elastic bulk energy and $\gamma_L$ the crack surface density functional. To account for a tension-compression asymmetry, the elastic bulk energy is frequently decomposed as
\begin{equation}
\psi_{\text{bulk}}(\boldsymbol{\varepsilon},d) = g(d) \psi^+_0 (\boldsymbol{\varepsilon}) + \psi^-_0(\boldsymbol{\varepsilon}),
\label{eq:elas}
\end{equation}
where $g(d)$ is a \textit{stiffness degradation function}. The degradation function is assumed to be a monotonically decreasing function of damage, with $g(0)=1$ (initial damage) and $g(1)=0$ (complete loss of stiffness). Explicit forms of this function are provided in Section~\ref{sec:gfunc}.  

With respect to the energetic contribution from the crack surface, we propose the use of the following crack surface density functional:
\begin{equation}
\gamma_L(d,\nabla d) := \dfrac{3}{4L} \left[ d + \dfrac{L^2}{4} \nabla d \cdot \nabla d \right],
\label{eq:crack_surface_density}
\end{equation}
in which the regularization length scale $L$ is introduced. We note that, in contrast to the commonly employed phase-field approximiation from \cite{bourdin2008variational}, the damage field enters the energy through a linear term. Similar formulations have been discussed, amongst others, in \cite{FREMOND19961083}, \cite{doi:10.1177/1056789510386852}, \cite{MIEHE2011898} and \cite{Lorentz20111927,Lorentz2012}, in the context of gradient damage mechanics. It will be demonstrated that such an approximation is central in the development of a phase-field description of cohesive fracture.

\subsection{Evolution equations}
\label{subsec:force_balances}

We suppose the microstructural changes, governed by the physics of the underlying problem, act to enforce a crack irreversibility condition that can be expressed as the inequality 
\begin{equation}
\dot{d} \ge 0,
\label{eq:var_inequality}
\end{equation}
where the superposed dot denotes a differentiation with respect to time.  

We follow the approach outlined in \cite{DASILVA20132178} based on the existence of a set of internal constraints. This leads to the definition of a macroscopic momentum balance
\begin{equation}
\Div \left( \dfrac{\partial\widetilde{\psi} (\boldsymbol{\varepsilon},d,\nabla d)}{\partial \boldsymbol{\varepsilon}} \right) = \rho \ddot{\mathbf{u}},
\label{macroforce}
\end{equation}
and a microscopic force balance
\begin{equation}
\Div \left( \dfrac{\partial \widetilde{\psi}(\boldsymbol{\varepsilon},d,\nabla d)}{\partial \nabla d} \right) - \dfrac{\partial \widetilde{\psi} (\boldsymbol{\varepsilon},d,\nabla d)}{\partial d} = 
\begin{cases}
\phantom{-}\beta \dot{d} & \text{if } \dot{d} > 0, \\
-\pi_r & \text{if } \dot{d} = 0,
\end{cases}
\label{microforce}
\end{equation}
where $\widetilde{\psi}$ denotes a combined strain energy density and fracture energy density, given by
\begin{equation}
\label{eq:nrg_density}
\widetilde{\psi}(\boldsymbol{\varepsilon}, d,\nabla d) = \psi_{\text{bulk}}(\boldsymbol{\varepsilon},d) + \mathcal{G}_c \gamma_L(d,\nabla d),
\end{equation}
and $\beta \geq 0$ is a kinetic modulus.
The right-hand side of \eqref{microforce} is essential as it embodies the irreversibility of damage evolution. It takes the form $\dot{d} \geq 0$ and is complemented by a reactive microforce $\pi_r$ to enforce that constraint. In particular, $\pi_r$ vanishes for $\dot{d} > 0$ and is given by the left-hand side of \eqref{microforce} for $\dot{d}=0$. Given a positive kinetic modulus $\beta$, the damage only increases if the left-hand side of \eqref{microforce} is positive. 

We now turn our focus to the kinetic modulus $\beta$. In \cite{Miehe20102765} a similar term was employed, which was coined the \textit{viscous regularization}, with the purpose of stabilizing the numerical treatment of the algebraic equations. While it is clearly demonstrated how such a regularization has an impact, it remains unclear, generally speaking, how to determine such a parameter in practice.  Here, we simply consider the rate-independent case where $\beta = 0$. Taking the above considerations and definitions into account, and using \eqref{eq:elas}, \eqref{eq:crack_surface_density}, and \eqref{eq:nrg_density},  we arrive at the  following coupled system of evolution equations:\begin{equation}
(\mathcal{S})~
\begin{cases}
\quad
\begin{aligned}
\Div \boldsymbol{\sigma}  &= \rho \ddot{\mathbf{u}} & \text{on}~ \mathcal{B}  \times ]0,T[,\\
\dfrac{3 \mathcal{G}_c}{4L} \left( \dfrac{L^2}{2}\Delta d - 1 \right) - g'(d)\psi_0^+(\boldsymbol{\varepsilon}) &= 0 & \text{on}~ \mathcal{B}  \times ]0,T[, \\
\dot{d} &\geq 0 & \text{on}~ \mathcal{B} \times ]0,T[,
\end{aligned}
\end{cases}
\label{eq:strong_form1}  
\end{equation}
where the Cauchy stress tensor is given by
\begin{equation}
\boldsymbol{\sigma} = g(d) \dfrac{\partial\psi_0^+}{\partial \boldsymbol{\varepsilon}} + \dfrac{\partial \psi_0^-}{\partial \boldsymbol{\varepsilon}}.
\label{cauchy_stress}
\end{equation}
The strong form of the governing equations is complemented by the following boundary conditions
\begin{equation}
(\mathcal{S}:BC)\phantom{I}
\begin{cases}
\quad
\begin{aligned}
\mathbf{u} &= \bar{\mathbf{u}}  & \text{on}~ \partial\mathcal{B}{u}  \times ]0,T[,\\
\boldsymbol{\sigma} \cdot \mathbf{n} &= \bar{\mathbf{t}}  & \text{on}~  \partial \mathcal{B}_{t}  \times ]0,T[,\\
\nabla d \cdot \mathbf{n} &= 0       & \text{on}~  \partial \mathcal{B} \times ]0,T[,
\end{aligned}
\end{cases}
\end{equation}
where $\bar{\mathbf{u}}$ and $\bar{\mathbf{t}}$ are the prescribed boundary displacements and surface tractions, respectively.  Additionally, we supplement \eqref{eq:strong_form1} with initial conditions
\begin{equation}
(\mathcal{S}:IC)\phantom{B}
\begin{cases}
\quad
\begin{aligned}
\mathbf{u}(\mathbf{x},0) &= \mathbf{u}_0(\mathbf{x}) &\quad \mathbf{x} \in \mathcal{B}, \\
\dot{\mathbf{u}}(\mathbf{x},0) &= \dot{\mathbf{u}}_0(\mathbf{x}) &\quad \mathbf{x} \in \mathcal{B}, \\
d(\mathbf{x},0) &= d_0(\mathbf{x}) &\quad \mathbf{x} \in \mathcal{B},
\end{aligned}
\end{cases}
\label{eq:IC}
\end{equation}
for both the displacement field and the damage field. As a final remark in this subsection, we note that the formulation for quasi-static fracture can be retrieved by simply omitting the inertial term in \eqref{eq:strong_form1}$_1$.

\subsection{Damage initiation and degradation functions}
\label{sec:gfunc}

The evolution equation for the damage, \eqref{eq:strong_form1}$_2$ allows for the construction of a threshold level of the tensile strain energy  before the onset of damage. As a means to trivially satisfy this equation for tensile strain energies below the threshold, we employ a crack driving force given by 
\begin{equation}
\widetilde{D}\left(\boldsymbol{\varepsilon}\right) = \max \left( \psi_0^+(\boldsymbol{\varepsilon}), \psi_c \right),
\label{eq:driving_state_functional}
\end{equation}
where we have introduced $\psi_c$ as the \textit{critical fracture energy} per unit volume of material
\begin{equation}
\psi_c = \dfrac{\sigma_c}{2E},
\end{equation}
where $\sigma_c$ is the critical tensile strength and $E$ the Young's modulus.

Alternative motivations for such functions can be found in the work of \cite{MIEHE2015486, MIEHE2015449}, for example, where analogous expressions have been used to generalize the onset of damage to a broader class of thresholds.  For the moment, we simply note that this function is used in place of the tensile strain energy in \eqref{eq:strong_form1}$_2$, 
arriving at the following system of coupled equations:
\begin{equation}
(\mathcal{S})~
\begin{cases}
\quad
\begin{aligned}
\Div \boldsymbol{\sigma}  &= \rho \ddot{\mathbf{u}} & \text{on}~ \mathcal{B}  \times ]0,T[,\\
\dfrac{3 \mathcal{G}_c}{4L} \left( \dfrac{L^2}{2}\Delta d - 1 \right) - g'(d)\widetilde{D}(\boldsymbol{\varepsilon}) &= 0 & \text{on}~ \mathcal{B}  \times ]0,T[, \\
\dot{d} &\geq 0 & \text{on}~ \mathcal{B} \times ]0,T[.
\end{aligned}
\end{cases}
\label{eq:strong_form2}  
\end{equation}
The utility of this crack driving force as a threshold for damage initiation is discussed further in Section \ref{subsec:analytical}.


\begin{figure}[!bp]
\centering \small
\begin{overpic}[width=0.31\linewidth]{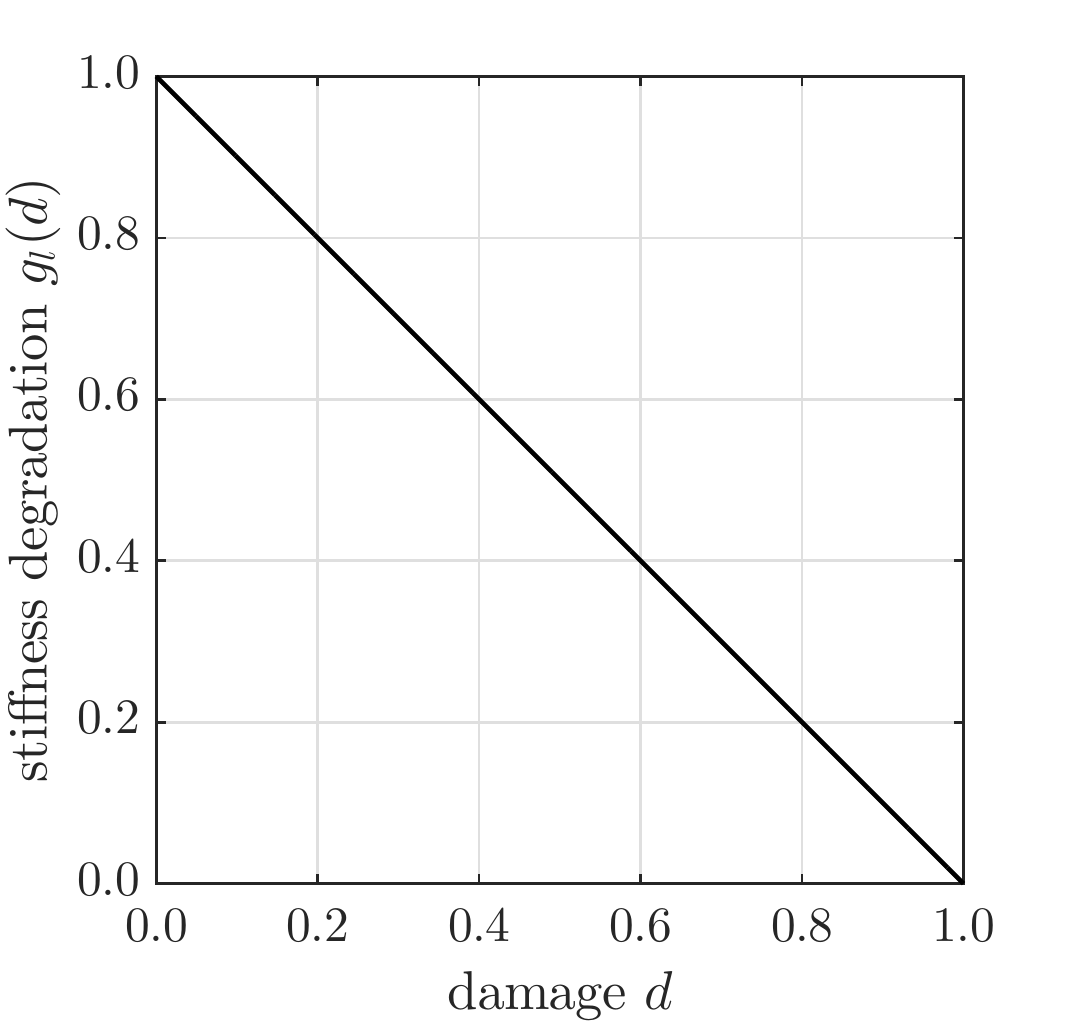}
\put(0,2){(a)}
\end{overpic} 
\begin{overpic}[width=0.31\linewidth]{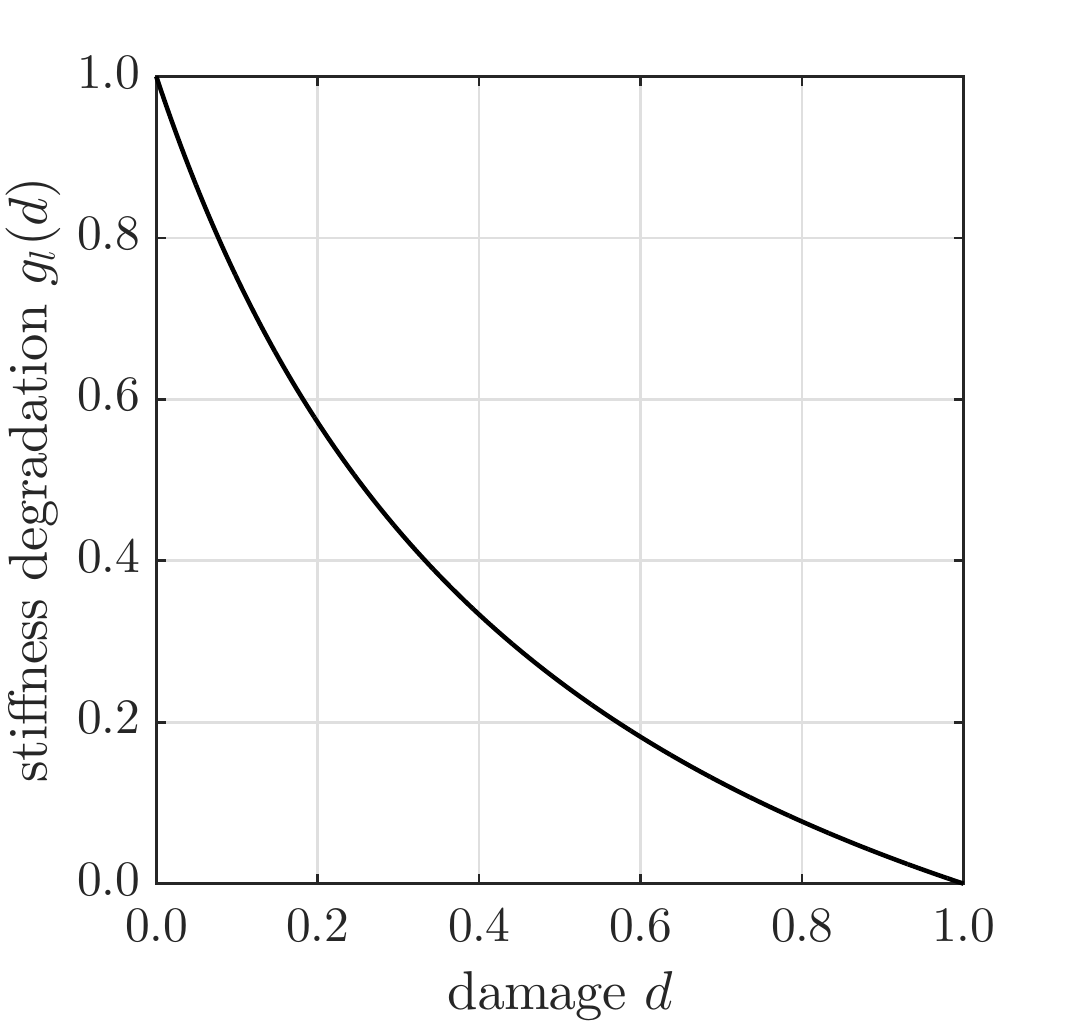}
\put(0,2){(b)}
\end{overpic}
\begin{overpic}[width=0.31\linewidth]{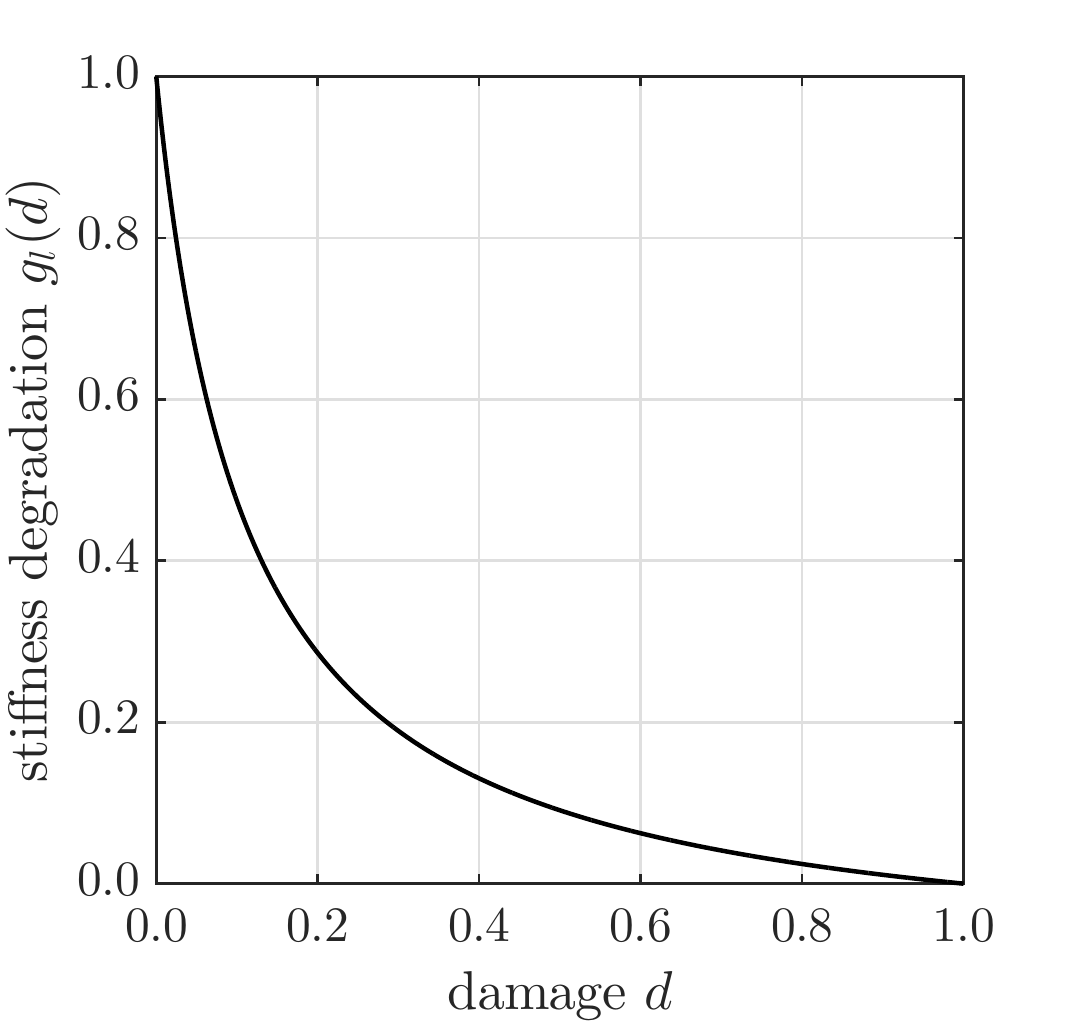}
\put(0,2){(c)}
\end{overpic}
\phantom{
\begin{overpic}[width=0.3\linewidth]{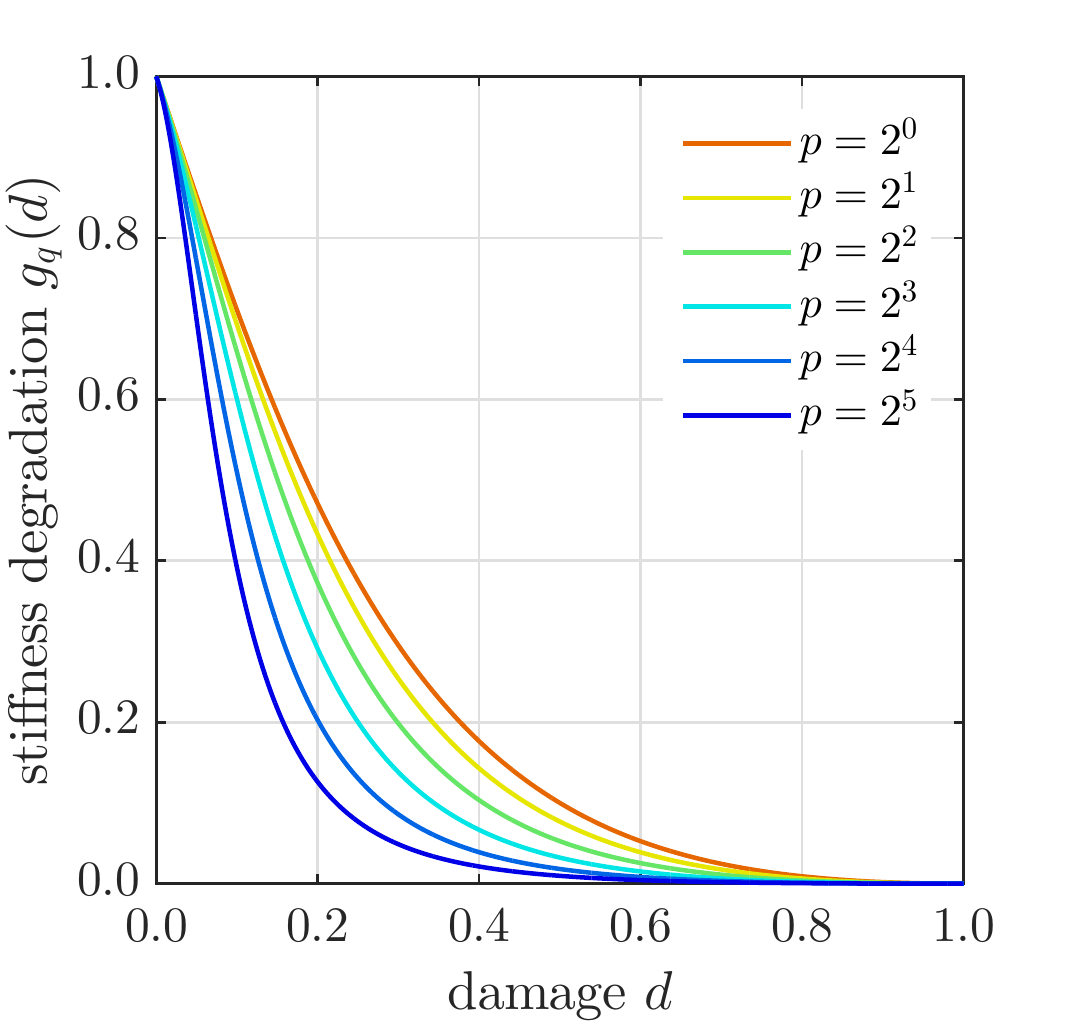}
\end{overpic}}
\begin{overpic}[width=0.31\linewidth]{gq3-eps-converted-to.pdf}
\put(0,2){(d)}
\put(39,37){\vector(-2,-1){18}}
\put(40.5,37.5){$p$}
\end{overpic}
\begin{overpic}[width=0.31\linewidth]{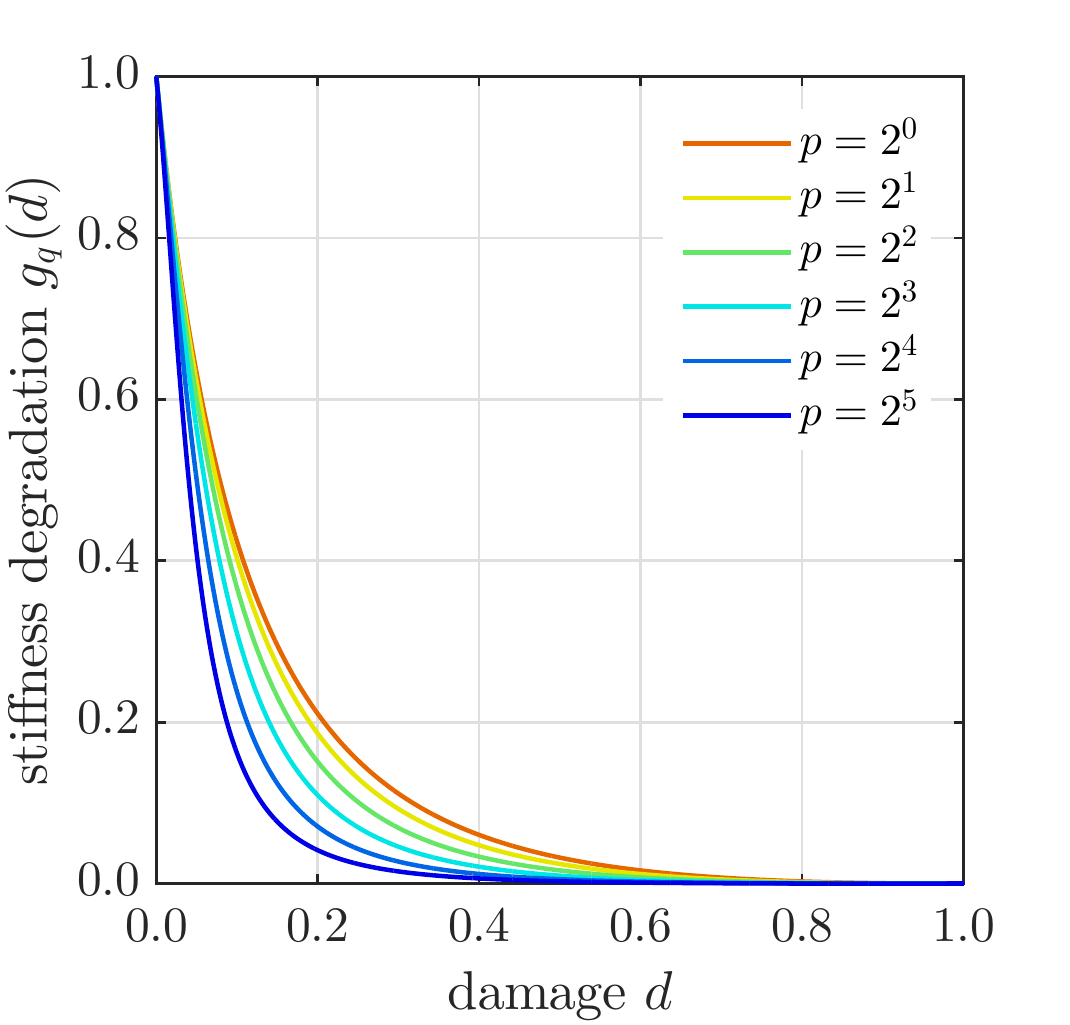}
\put(0,2){(e)}
\end{overpic}
\caption{Influence of the constant $m$ on the quasi-linear degradation function $g_l(d)$ (top) and the quasi-quadratic degradation function $g_q(d)$ (bottom). For illustration purposes $m=1$ (a); $m=3$ (b,d) and $m=10$ (c,e) are considered.}
\label{fig:gds}
\end{figure}

The formulation is completed by the definition of a stiffness degradation function. In this paper, we are concerned with two particular functions, which hereafter we refer to as the quasi-linear and quasi-quadratic degradation functions. 

The quasi-linear and quasi-quadratic degradation functions are both rational functions of the damage field, with the numerator being either a linear function of damage or a quadratic function of damage, respectively. Each of the functions has an associated upper bound on the regularization length scale that can be used in conjunction with it. The quasi-linear degradation function is given by
\begin{equation}
\label{eq:gl}
g_l(d) = \dfrac{1-d}{1-d+md}, \quad \text{with} \quad L < \dfrac{3}{2}\dfrac{E\mathcal{G}_c}{\sigma_c^2},
\end{equation}
where $m \ge 1$ is a constant. This function and the associated bounds on $m$ and the regularization length $L$ are derived in Section~\ref{sec:ql-deriv}.

The quasi-quadratic degradation function follows the work of \cite{LORENTZ201120} and \cite{Lorentz2017}, and is given by
\begin{equation}
g_q(d) = \dfrac{(1-d)^2}{(1-d)^2 + m d(1+pd)}, \quad \text{with} \quad p \geq 1 \quad \text{and} \quad L \leq \dfrac{3}{2\left(p+2\right)} \dfrac{E\mathcal{G}_c}{\sigma_c^2},
\label{eq:gq}
\end{equation}
where $p$ is a shape parameter. The dependence of both degradation functions on the constant $m$ and shape parameter $p$ is graphically depicted in Figure \ref{fig:gds}, for illustration purposes. We note that both degradation functions satisfy the condition that 
\begin{equation}
g^{\prime}(0) = -m.
\end{equation}
In the present work, we require that \begin{equation}
m = \dfrac{3\mathcal{G}_c}{4L\psi_c}.
\label{eq:m}
\end{equation}
The justification for this choice is provided in Section~\ref{subsec:analytical}. This leaves only the shape parameter $p$ to be defined in \eqref{eq:gq}.  In the remainder of this paper, unless otherwise noted, we set $p=1$.  

In general, different choices of the shape parameter $p$ will give rise to different fracture responses, and influence the load-deflection response of the structure as well as the fracture evolution.  A thorough examination of this effect is beyond the scope of this manuscript, and we refer the reader to the discussions in \cite{Lorentz2017} as to how multiple shape parameters can be employed to effectively calibrate the model to individual materials. 

The constraints on the regularization length $L$ and shape parameter $p$ that go along with \eqref{eq:gl} and \eqref{eq:gq} ensure an increasing damage band width, and a decreasing stress. The latter essentially stems from the objectivity condition in cohesive zone modeling, as discussed in \cite{BAZANT2002165}. Finally, we note that the upper bound on the regularization length is closely related to the characteristic length of the fracture process zone, which was approximated as $\ell_{\text{FPZ}}=E\mathcal{G}_c/\sigma_c^2$ in \cite{HILLERBORG1976773}.

\section{Analysis \& discussion}
\label{sec:analysis} 
We now provide some analysis of the model described in Section~\ref{sec:formulation}. We begin by examining a simple one-dimensional problem. We then provide the derivation of the quasi-linear degradation function. Finally, some aspects concerning the regularization of crack topologies and the issue of $\Gamma$-convergence are discussed.


\subsection{Analytical solution of the one-dimensional rate-independent problem}
\label{subsec:analytical}
To illustrate the fundamental properties of our formulation, we begin by studying a particular boundary value problem. Consider a one-dimensional bar subjected to a uniaxial tensile load. A symmetrical solution for the localization band is expected in a domain of interest $\mathcal{B} = [-L;L]$. Without a loss of generality, this point of symmetry is taken at center of the bar, i.e. $x=0$. We assume a non-negative strain-field and ignore inertial effects. Under these assumptions \eqref{eq:strong_form2} simplifies to
\begin{equation}
\begin{cases}
\quad
\begin{aligned}
\dfrac{\text{d}\sigma}{\text{d}x} = 0 & \qquad \text{ on } \mathcal{B} \times ]0,T[,  \\
\dfrac{3\mathcal{G}_c}{4L} \left( 
\dfrac{L^2}{2}\dfrac{\text{d}^2 d}{\text{d}x^2} -1 \right) - g'(d)\widetilde{D}(\strain) = 0 & \qquad \text{ on } \mathcal{B} \times ]0,T[, \\
\end{aligned}
\end{cases}
\label{1dode}
\end{equation}
with $\sigma = g(d)E\varepsilon$, where $E$ denotes Young's modulus.  

Provided that the strain $\strain$ and damage $d$ are monotonically increasing functions of time at every point in the domain, $\widetilde{D}$ can be written as 
\begin{equation}
\label{eq:driveforcecases}
\widetilde{D}(\strain) = \begin{cases}
\begin{aligned}
~\psi_c & \quad\mbox{for}~ \psi_0^+ < \psi_c, \\
~\dfrac{1}{2}E \varepsilon^2 & \quad\mbox{for}~ \psi_0^+ \geq \psi_c. \\ 
\end{aligned}
\end{cases}
\end{equation}
We begin our analysis by considering the situation before the onset of any damage in the bar.   In such a configuration, the damage $d=0$ everywhere, and the stress is spatially constant and below the critical stress, i.e.\ $\sigma \le \sigma_c$.  Under these conditions \eqref{1dode}$_2$ and \eqref{eq:driveforcecases} simplify to  
\begin{equation}
-\dfrac{3\mathcal{G}_c}{4L} - g'(0) \psi_c = 0.
\end{equation}
In order for this equation to be satisfied, the two terms must balance.  This is trivial, provided that 
\begin{equation}
g'(0) \psi_c = -\dfrac{3\mathcal{G}_c}{4L}.
\end{equation}
We note that, given the choice of $m$ in \eqref{eq:m}, this constraint is satisfied for both the quasi-linear, \eqref{eq:gl}, and quasi-quadratic, \eqref{eq:gq}, degradation functions. By contrast, the use of a simple quadratic degradation function of $g(d) = (1-d)^2$ requires the use of a particular value of $\psi_c$ for this constraint to be satisfied.  

\begin{figure}[!tbp]
\centering \scriptsize
\begin{overpic}[width=.95\linewidth]{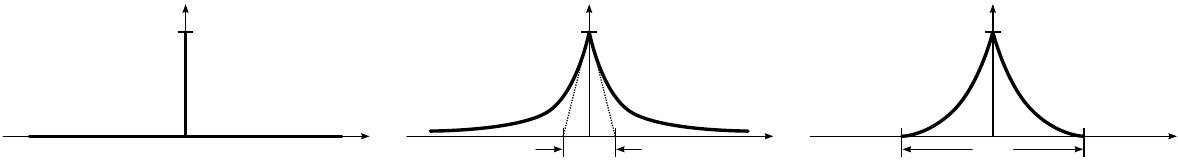}
\put(17,12){$d(x)$}
\put(13.5,10){1}
\put(51.5,12){$d(x) = \exp{ \left(-\dfrac{|x|}{l} \right)}$}
\put(48,10){1}
\put(85.75,12){$d(x) = \left( \dfrac{|x|}{L}-1 \right)^2$}
\put(82.25,10){1}
\put(83.5,0.4){$2L$}
\put(49.25,0.4){$2l$}
\put(0,-1){\small (a)}
\put(34.5,-1){\small (b)}
\put(69,-1){\small (c)}
\end{overpic}
\vspace{1em}
\caption{Sharp and diffuse crack topologies for a crack at $x=0$. (a) Sharp crack. (b) Smeared crack modeled with a `standard' phase-field approach. (c) Smeared crack modeled with the proposed formulation.}
\label{fig:topologies}
\end{figure}
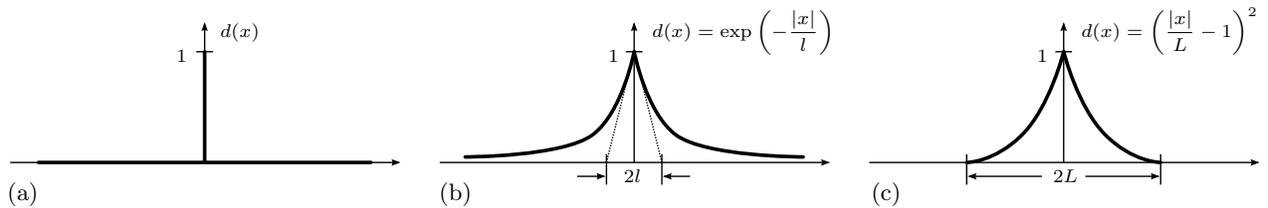

We turn now to considering the onset of damage and the post-critical behavior. If we want damage to begin when $\sigma = \sigma_c$, we can effect this by setting 
\begin{equation}
\psi_c = \psi_0^+(\sigma=\sigma_c) := \dfrac{\sigma_c^2}{2g(0)^2E}  \quad \rightarrow \quad \psi_c = \dfrac{\sigma_c^2}{2E},
\end{equation}
which relates the critical fracture energy per unit volume $\psi_c$ to the tensile strength $\sigma_c$. At the onset of damage and beyond, \eqref{1dode} and \eqref{eq:driveforcecases} simplify to the following nonlinear ordinary differential equation:
\begin{equation}
\dfrac{3\mathcal{G}_c}{4L} \left( 
\dfrac{L^2}{2}\dfrac{\text{d}^2 d}{\text{d}x^2} -1 \right) - g'(d) \dfrac{\sigma^2}{2g^2(d)E} = 0.
\label{eq:1dode_0}
\end{equation}
A solution for the damage field can be found by making use of the fact that the stress across the bar is constant and multiplying \eqref{eq:1dode_0} with $\frac{\text{d}d}{\text{d}x}$, to obtain
\begin{equation}
\dfrac{\text{d}}{\text{d}x} \left[ \dfrac{3\mathcal{G}_c}{4L} \left( \dfrac{L^2}{4} \left( \dfrac{\text{d}d}{\text{d}x} \right)^2 - d \right) + \dfrac{\sigma^2}{2g(d)E} \right] = 0.
\label{eq:1dode_1}
\end{equation}
Because of symmetry about $x=0$, \eqref{eq:1dode_1} is integrated from $x$ to $L$ for positive values of $x$, and from $-L$ to $x$ when $x$ is negative. Since we have specified the crack to be centered at $x = 0$, we require the damage field to have a maximum at $x = 0$. From this requirement, we obtain
\begin{equation}
\dfrac{\text{d}d}{\text{d}x} = \sgn{(x)} \sqrt{\dfrac{4}{L^2} \left( d - \dfrac{2L\sigma^2}{3\mathcal{G}_cg(d)E} - a \right)},
\label{eq:1dode_2}
\end{equation}
where $a$ is given by
\begin{equation}
a = d_{\text{hom}} - \dfrac{2L\sigma^2}{3\mathcal{G}_cg(d_{\text{hom}})E}.
\end{equation}
By definition, substitution of the homogeneous solution $d_{\text{hom}}$ into \eqref{eq:1dode_2} yields a zero damage gradient. For a fully developed crack, i.e. $\sigma=0$ and $d_{\text{hom}}(0) = 0$, \eqref{eq:1dode_2} reduces to
\begin{equation}
\dfrac{\text{d}d}{\text{d}x} = \sgn{(x)} \dfrac{2}{L} \sqrt{d},
\end{equation}
and by applying the boundary conditions $d(0)=1$ and $d(L)=0$, we find 
\begin{equation}
d(x) =
\begin{cases}
\quad
\begin{aligned}
\left( \dfrac{|x|}{L}-1 \right)^2 , &\quad\mbox{for}~ |x| \leq L,\\
0\quad  , &\quad\mbox{for}~ |x| > L,
\end{aligned}
\end{cases}
\end{equation}
as the solution that satisfies the specified boundary conditions. We note that this regularization is different from what is traditionally used in the phase-field modeling of brittle fracture. The different ultimate damage distributions are compared  in Figure \ref{fig:topologies}. We refer the reader to \cite{BORDEN201277} for an analogous investigation in the context of phase-field models for brittle fracture.

Given a closed-form analytical solution for the ultimate damage profile in a one-dimensional setting, we calculate the corresponding dissipated energy in the system
\begin{equation}
E_{\text{diss.}} = \int_{\mathcal{B}} \mathcal{G}_c \gamma_L \,d\mathcal{B} = \int_{-L}^L \left( \dfrac{3\mathcal{G}_c}{4L} \left[ d + \dfrac{L^2}{4} \left( \dfrac{\text{d}d}{\text{d}x} \right)^2 \right]  \right) dx.
\label{eq:diss0}
\end{equation}
For a fully developed crack we can deduce from \eqref{eq:1dode_1} that this differential equation reduces to:
\begin{equation}
d(x) = \dfrac{L^2}{4}\left( \dfrac{\text{d}d}{\text{d}x} \right)^2, 
\end{equation}
with $d(x)$ the ultimate damage distribution. This reduces \eqref{eq:diss0} to
\begin{equation}
E_{\text{diss.}} = \mathcal{G}_c,
\end{equation}
which effectively demonstrates that \eqref{eq:crack_surface_density} indeed constitutes a crack surface density functional, in the phase-field for fracture sense. The same line of reasoning was presented in the gradient damage framework of \cite{Lorentz20111927}.

\subsection{Derivation of quasi-linear degradation function} 
\label{sec:ql-deriv}

Up to this point in this Section, the only conditions that have been placed on the degradation function concern its slope at the onset of damage.   In this subsection, we demonstrate how a desired stress-strain behavior in the damage zone can be effected through a particular choice of the degradation function.  Attention is focused on obtaining a relatively simple, linearly decaying traction-separation law in the damaged region.  Such a response  is common in standard cohesive methods for fracture, see e.g.\  \cite{ortiz1999}.  Here, we show how such a response gives rise to the quasi-linear degradation function \eqref{eq:gl}. 




A general discussion on the requirements of a degradation function can be found in the work of \cite{Pham2013}. Generally speaking,  a valid degradation function is assumed to satisfy the following set of conditions:
\begin{equation}
g(0) = 1; \quad g(1) = 0; \quad \text{and} \quad  g^\prime(\dmg) \leq 0 \quad \text{for} \quad 0 \leq \dmg \leq 1. 
\end{equation}
We consider the local response (i.e.\ within the localization band) in a one-dimensional system, and assume that all fields are spatially uniform.  In this setting, the stress-strain relationship simplifies to 
\begin{equation}
\sigma(\strain,\dmg) = g(\dmg) E \strain.
\label{eq:linear1}
\end{equation}
The critical stress $\sigma_c$ at which damage initiates can be associated with a critical strain $\varepsilon_c$ through $\sigma_c = E\varepsilon_c$.  Assume that the desired stress-strain relationship in the post-critical regime is given by the following linear decay
\begin{equation}
\sigma(\strain) = E \varepsilon_c \left( \dfrac{\varepsilon_f-\varepsilon}{\varepsilon_f-\varepsilon_c} \right) \quad\mbox{for}~\epsilon > \epsilon_c,
\label{eq:linear2}
\end{equation}
where $\varepsilon_f$ denotes the final strain at which the stress vanishes.   This contrasts to most other phase-field approaches in which the stress vanishes only in the limit as $\varepsilon \rightarrow \infty$. 

Combining  \eqref{eq:linear1} and \eqref{eq:linear2}, we obtain
\begin{equation}
\varepsilon_c \left( \dfrac{\varepsilon_f-\varepsilon}{\varepsilon_f-\varepsilon_c} \right) = g(d) \varepsilon \quad \rightarrow \quad \varepsilon = \dfrac{\varepsilon_c\varepsilon_f}{g(d)\left( \varepsilon_f - \varepsilon_c \right) + \varepsilon_c}.
\label{eq:1d_local_strain}
\end{equation}
As the elastic regime is overstepped in this simplified one-dimensional setting, \eqref{1dode}$_2$ reduces to
\begin{equation}
\dfrac{3\mathcal{G}_c}{4L} + g'(d) \dfrac{E\varepsilon^2}{2} = 0.
\label{eq:1d_local_pf}
\end{equation}
Inserting the strain \eqref{eq:1d_local_strain} into this local evolution equation, we obtain a nonlinear ordinary differential equation for the degradation function of the form
\begin{equation}
\alpha_1 g'(d) = \left( g(d) + \alpha_2 \right)^2,
\end{equation}
where $\alpha_{1,2}$ are constants which depend on $\varepsilon_c, \varepsilon_f, E, \mathcal{G}_c$ and $L$. It is easy to show that the quasi-linear degradation function \eqref{eq:gl} is a generic solution to this equation, satisfying $g(0)=1$ and $g(1)=0$.


As the critical strain is naturally accounted for by fixing the critical strength, we are left to determine the final strain $\strain_f$, which is reached when $d=1$. In that case, \eqref{eq:1d_local_pf} becomes
\begin{equation}
\dfrac{3\mathcal{G}_c}{4L} + g'(1) \dfrac{E\varepsilon_f^2}{2} = 0,
\label{eq:1d_strain_f}
\end{equation}
which by means of inserting \eqref{eq:gl},  reduces to
\begin{equation}
\varepsilon_f = m \varepsilon_c.
\end{equation}
For this model to have a clear physical interpretation, it is obviously required that $m \geq$ 1.  This constraint can be rewritten in terms of an upper bound on the regularization length scale $L$, given by 
\begin{equation}
L < \dfrac{3E\mathcal{G}_c}{2\sigma_c^2}.
\end{equation}
In case the damage has reached unity and the stress has dropped to zero, the ``opening displacement" within the damage band can be approximated as
\begin{equation}
u_f \approx \strain_f L = \frac{3 G_c}{2 \sigma_c}.
\end{equation}
This quantity which has dimensions of displacement does not depend on the regularization length scale.  This is consistent with the desire to asymptotically approach a cohesive model with a known maximum opening displacement.  As the damage localization must occur over a smaller length as $L \rightarrow 0$, the average strain over this localization zone must increase to compensate. To confirm that the post-critical stress-strain behavior matches \eqref{eq:linear2}, a series of numerical experiments are conducted in Section~\ref{sec:numerical_examples}. 

While the above derivation was employed to extract a linear decay of the stress-strain behavior in the localization band, more complex constitutive relationships can obviously be obtained by following a similar procedure.  
Finally, we note that the form of the resulting quasi-linear degradation function \eqref{eq:gl} differs markedly from the quasi-quadratic function \eqref{eq:gq} as $d\rightarrow 1$.  In particular, the slope does not vanish as complete damage is approached.  An unfortunate consequence of this feature is that there is no means to constrain the damage from exceeding unity, and we will find the need to prevent the overshoot in our discrete formulation.  



\subsection{On the regularization of crack topologies}
\label{sec:regularized_crack_topologies}

We now investigate the approximations of the fracture energy that are employed by the proposed cohesive model. Our investigation follows the work described in \cite{NME:NME2861} for the numerical investigation of error in regularized crack surface representations. 
In particular, we examine a dimensionless version of the evolution equation \eqref{eq:strong_form2}$_2$, given by  
\begin{equation}
\left( \dfrac{L^2}{2}\Delta d-1 \right)  - g'(d) \widetilde{D}_p = 0 \text{ in } \mathcal{B} \quad \text{with} \quad \nabla d \cdot \mathbf{n} = 0 \text{ on } \partial \mathcal{B},
\label{eq:gen1}
\end{equation}
where $\widetilde{D}_p$ is a prescribed form of the driving force.  

We consider two separate boundary-value problems.  In the first, the prescribed driving force is set to $\widetilde{D}_p = 1$, and 
the Dirichlet constraint
\begin{equation}
d(\mathbf{x},t) = 1 \quad \text{at} \quad \mathbf{x} \in \Gamma,
\label{eq:gen2}
\end{equation}
is enforced on the damage field along the sharp crack surface.  
A standard approach is to generate a finite-element approximation to the solution of \eqref{eq:gen1}-\eqref{eq:gen2}, and then examine the ability of the corresponding (dimensionless) fracture energy to capture the exact energy, as a function of regularization length and mesh spacing. 

In the second boundary value problem, we replace the Dirichlet boundary condition \eqref{eq:gen2} with a prescribed driving 
force $\widetilde{D}_p$ that is sufficiently large in the vicinity of the sharp crack.   In effect, the magnitude of the driving force is increased significantly (approximating a Dirac distribution)  at quadrature points along the sharp crack surface $\Gamma$ to drive the regularization.  
By adopting such an approach, the regularized topologies are expected to be more representative of those obtained in fully coupled displacement-damage calculations.  


\begin{figure}[!bbp]
\centering \small
\begin{overpic}[width=0.85\linewidth]{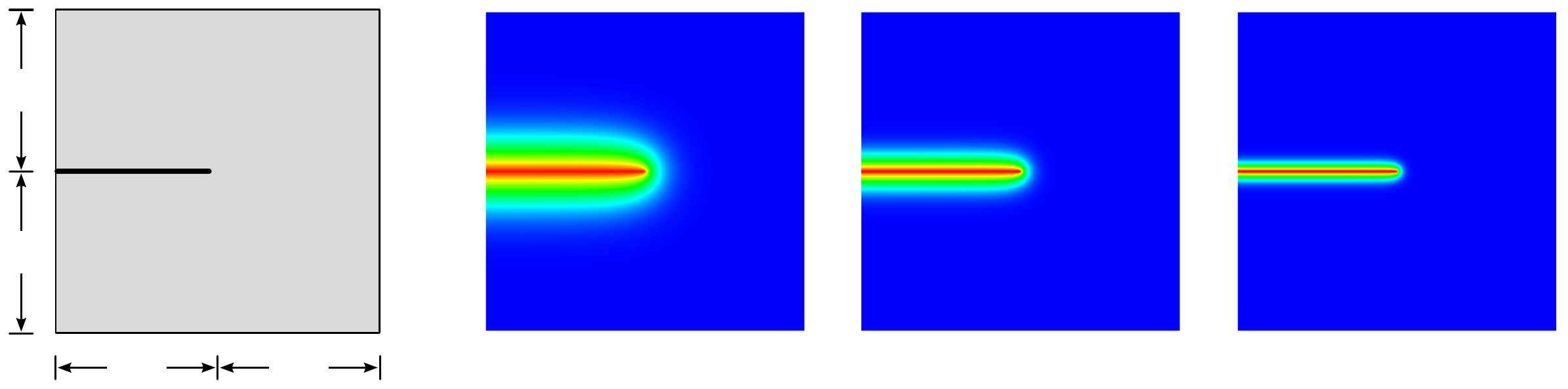}
\put(-1,0.5){(a)}
\put(39.5,0.5){(b)}
\put(63.75,0.5){(c)}
\put(88,0.5){(d)}
\put(17.8,0.75){0.5}
\put(7.4,0.75){0.5}
\put(0,8.15){0.5}
\put(0,18.4){0.5}
\put(18.5,7){$\mathcal{B}$}
\put(5.25,15){$\Gamma = 0.5$}
\end{overpic}
\caption{(a) The model problem from the work of \cite{NME:NME2861}. A damage field is approximated using a uniform 401x401 finite element mesh and a large, prescribed driving force along the sharp crack front.  Results are shown for the regularized crack topologies $\Gamma_l(d)$ and regularization lengths of (a) $L=0.200$, (b) $L=0.100$ and (c) $L=0.050$.}
\label{fig:model_problem}
\end{figure}

\begin{figure}[!bp]
\centering \small
\begin{overpic}[width=0.31\linewidth]{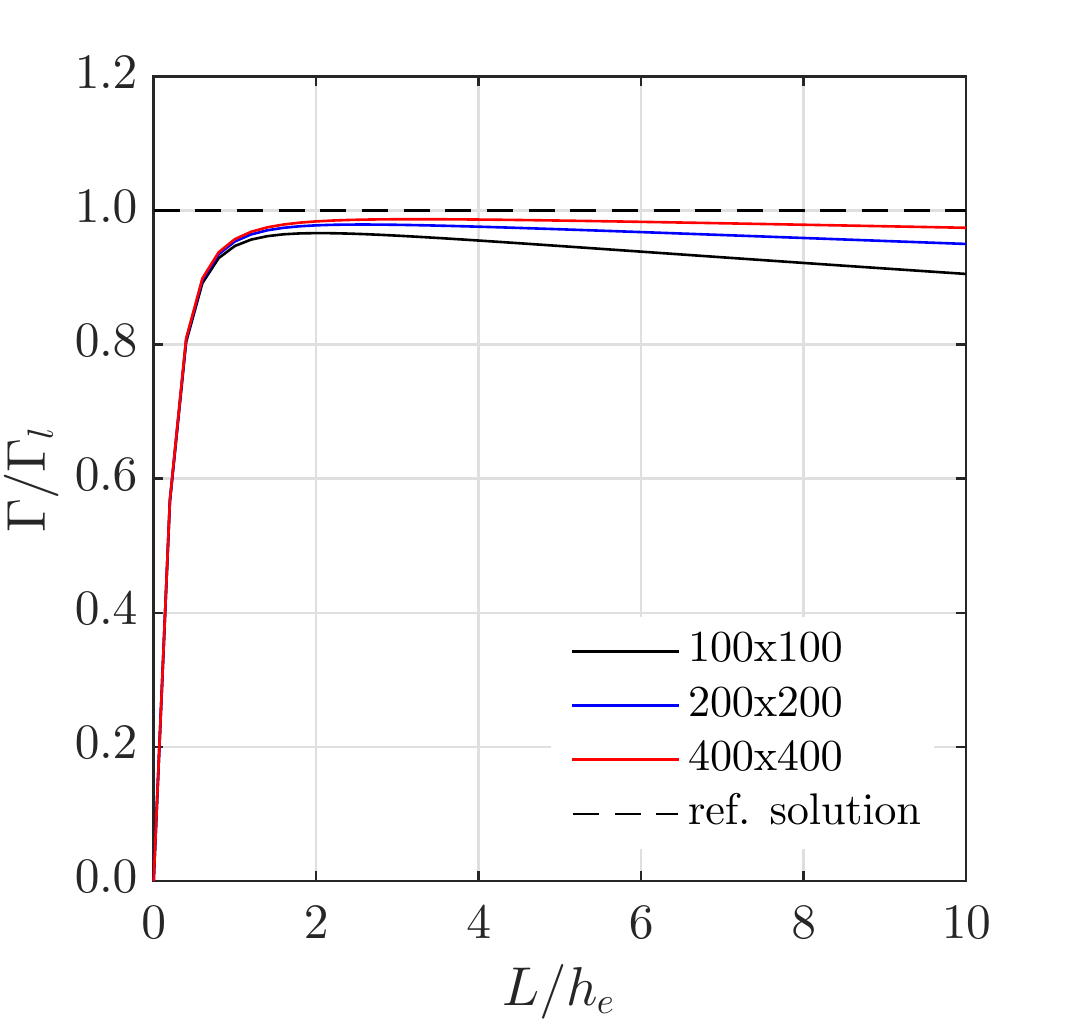}
\put(0,2){(a)}
\end{overpic}
\begin{overpic}[width=0.31\linewidth]{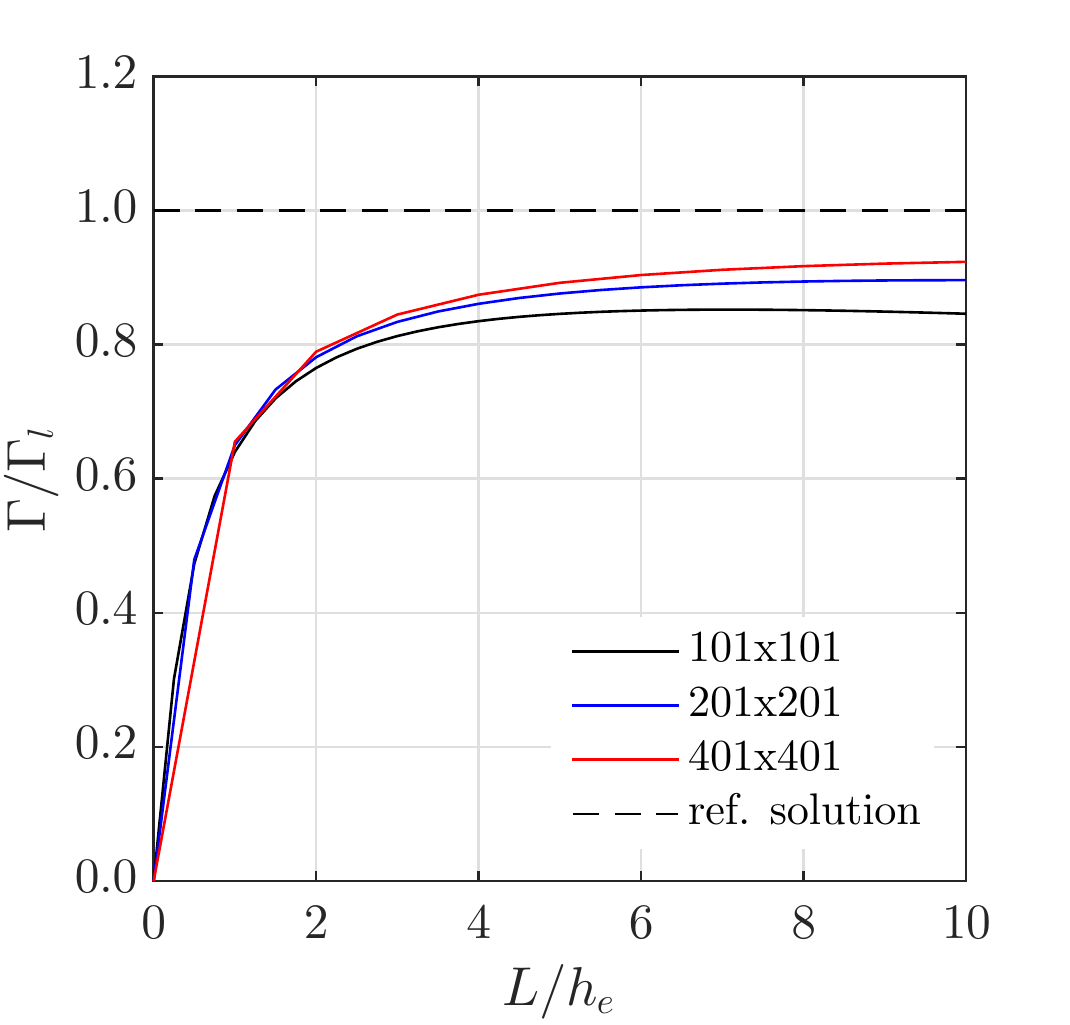}
\put(0,2){(b)}
\end{overpic}
\caption{Regularization results using (a) Dirichlet boundary conditions along the crack surface and (b) prescribing a large driving force along the crack surface.}
\label{fig:regularization_comp}
\end{figure}

In the following numerical example, the model problem from \cite{NME:NME2861} is revisited with the intent of providing some insight into the $\Gamma$-convergence properties of the proposed method. Consider a two-dimensional continuum $\mathcal{B}$ with an embedded crack surface $\Gamma$ from the left side to the center of the domain as depicted in Figure \ref{fig:model_problem}a. The regularization is driven by prescribing the crack driving forces to a value of $\widetilde{D}_p = 10^3$ at the quadrature points along $\Gamma$. A similar approach was adopted by \cite{BORDEN201277} for modeling preexisting cracks in a continuous body. 
The computations are performed on uniform finite element meshes consisting of bilinear quadrilateral elements. The regularized crack functionals
\begin{equation}
\Gamma_l(d) = \int_{\Omega} \gamma_{L}(d,\nabla d) ~d\mathcal{B},
\end{equation}
are then calculated as a post-processing step of the boundary value problem \eqref{eq:gen1}-\eqref{eq:gen2}.

In Figure \ref{fig:model_problem}b-d the computed regularized cracks are shown for different values of the regularization length. 
The results are in excellent agreement with the ones reported in \cite{NME:NME2861}.
The results for the numerical $\Gamma$-convergence investigation are summarized in Figure \ref{fig:regularization_comp}. In the Dirichlet case, the calculated fracture energy approaches the theoretical estimate as the mesh is sufficiently refined.  The results suggest that better results are obtained for smaller ratios $L/h_e$ of the regularization length to the mesh size.  This is counter-intuitive, and the results when the problem is driven by a prescribed driving force, Figure \ref{fig:regularization_comp}b, suggest that they are simply an artifact of the Dirichlet boundary condition.  The driving force results indicate that while the error between the sharp and regularized crack surfaces may eventually asymptote to zero, much more refined meshes will be required.

\section{Finite element implementation}
\label{sec:finite_element_implementation}

We now discuss the numerical discretization of the evolution equations \eqref{eq:strong_form2}.  We begin by discussing existing strategies for enforcing the irreversibility constraint, \eqref{eq:strong_form2}$_3$, and then present the finite-element formulation.  Finally, details of the staggered solution algorithm are provided.  

\subsection{Augmented Lagrangian implementation}
\label{sec:alm}

A now standard approach to enforcing the irreversibility constraint $\dot{d} \ge 0$ is to replace the crack driving function $\widetilde{D}$ in \eqref{eq:strong_form2}$_2$ with a monotonic version $\mathcal{H}$:
%
%
%
%
\begin{equation}
\mathcal{H}(\mathbf{x},t) = \max_{s \in [0,t]} \widetilde{D}(\mathbf{x},s),
\label{eq:history2}
\end{equation}
over the full temporal history $s \in [0,t]$. 
\cite{Miehe20102765} demonstrated that such a strategy leads to a thermodynamically consistent phase-field model for fracture. However, while such an approximation effectively enforces irreversibility, it is  variationally inconsistent.  An alternative approach using an augmented Lagrangian method was proposed in \cite{WHEELER201469}, in which the solution to the evolution equations is obtained from a constrained minimization problem.  

In the following, we briefly review the augmented Lagrangian method described in \cite{WHEELER201469}. The formulation starts with the energy function $\widetilde{\Psi}(\mathbf{u},d,\nabla d)$ from \eqref{total_energy2} which is minimized with respect to the unknown solution variables, which are the displacements $\mathbf{u}(\mathbf{x},t)$ and continuous damage variable $d(\mathbf{x},t)$. Differentiation with respect to $\mathbf{u}$ and $d$ leads to the Euler-Lagrange equations. Given a previous solution $(\mathbf{u}^{n-1},d^{n-1})$ we seek a solution
\begin{equation}
\begin{aligned}
\min \, &\widetilde{\Psi}(\mathbf{u},d, \nabla d) \\
&\text{such that } d \in \left\lbrace d \, | \, 0 \leq d^{n-1} \leq d \leq 1 \right\rbrace
\end{aligned}
\end{equation}
which is approximated as
\begin{equation}
\min \, \widetilde{\Psi}(\mathbf{u},d,\nabla d) + \dfrac{1}{2\gamma} \|  \left\langle \lambda + \gamma \left(d^{n-1}-d\right) \right\rangle_+ \|^2 + \dfrac{1}{2\gamma} \| \left\langle \lambda + \gamma \left(1-d\right) \right\rangle_- \|^2,
\label{eq:nrg}
\end{equation}
where $\lambda \in L^2 \left( \Omega \right)$ are the Lagrange multipliers and $\gamma \in \mathbb{R}_{>0}$ is a penalty kernel. An iterative scheme for solving this constrained optimization problem is described in Section \ref{sec:staggered}.

We note that the first penalty term in \eqref{eq:nrg} enforces the monotonicity constraint \eqref{eq:strong_form2}$_3$ on the damage field, while the second prevents the damage field from exceeding unity. While we have not found the second condition to be necessary with the quasi-quadratic degradation function, we have found it to be essential with the quasi-linear degradation function. Additional comments to this effect are provided in Section~\ref{sec:numerical_examples}. 

\subsection{Galerkin finite element discretization}
To  approximate the solution to the energy functional from \eqref{eq:nrg} using a finite element method, we employ a finite element discretization as follows. The admissible function spaces $\mathcal{U}$ for the displacements, $\mathcal{D}$ for the damage field, and $\Lambda$ for the Lagrange multiplier field are defined as
\begin{align}
\mathcal{U} &= \left\lbrace \mathbf{u}(t) \in \left(H^1(\Omega)\right)^n~|~ \mathbf{u} = \bar{\mathbf{u}}~\text{on}~\partial \Omega_d \right\rbrace, \\
\mathcal{D} &= \left\lbrace d(t) \in H^1(\Omega) \right\rbrace; \quad \text{and} \quad \Lambda = \left\lbrace \lambda(t) \in L^2(\Omega) \right\rbrace,
\label{function_spaces_adm}
\end{align}
respectively, while the associated weighting spaces are given by
\begin{align}
\mathcal{U}_0 &= \left\lbrace \mathbf{v} \in \left(H^1(\Omega)\right)^n~|~ \mathbf{u} = \mathbf{0}~\text{on}~\partial \Omega_d \right\rbrace, \\
\mathcal{D}_0 &= \left\lbrace w \in H^1(\Omega) \right\rbrace; \quad \text{and} \quad \Lambda_0 = \left\lbrace z \in L^2(\Omega) \right\rbrace.
\label{weight_function_spaces_adm}
\end{align}
Next the differentiation of \eqref{eq:nrg} is performed with respect to $\mathbf{u}$ and $d$. After multiplication by the appropriate weighting functions and integrating by parts, we obtain
\begin{equation}
\mathcal{A}_1(\mathbf{u},\mathbf{v}) = \left(\boldsymbol{\sigma}(\mathbf{u}),\boldsymbol{\varepsilon}(\mathbf{v}) \right) + \left(\rho\ddot{\mathbf{u}},\mathbf{v}\right) -\left(\bar{\mathbf{t}},\mathbf{v}\right)_{\partial \Omega_t} = 0 \quad \forall \mathbf{v} \in \mathcal{U}\,,
\label{eq:bilinear1}
\end{equation}
as well as
\begin{equation}
\begin{aligned}
\mathcal{A}_2(d,w) = \left( g'(d)\widetilde{D}(\boldsymbol{\varepsilon}), w\right) &+ \dfrac{3\mathcal{G}_c}{4L} \left( \left(1, w\right) + \left( \dfrac{L^2}{2} \nabla d, \nabla w\right) \right) \\
& \quad - \left( \left\langle \lambda + \gamma \left(d^{n-1}-d\right) \right\rangle_+ + \left\langle \lambda + \gamma \left(1-d\right) \right\rangle_-, w \right) = 0 \quad \forall w \in \mathcal{D}\,.
\end{aligned}
\label{eq:bilinear2}
\end{equation}
Where the bilinear forms $\mathcal{A}_1(\mathbf{u},\mathbf{v})$ and $\mathcal{A}_2(d,w)$ were introduced for the displacement field and damage field problems, and $(\cdot,\cdot)$ denotes the $\mathcal{L}_2$ inner product on $\Omega$. 

Following a standard Galerkin method, the problem is recast in finite-dimensional subspaces $\mathcal{U}^h \subset \mathcal{U},~\mathcal{U}_0^h \subset \mathcal{U}_0,~\mathcal{D}^h \subset \mathcal{D}$,~$\mathcal{D}_0^h \subset \mathcal{D}_0$ and $\Lambda^h \subset \Lambda,~\Lambda_0^h \subset \Lambda_0$. Given any $d^h \in \widetilde{\mathcal{U}}^h$ we find $\mathbf{u}^h \in \mathcal{U}^h$ satisfying
\begin{equation}
\mathcal{A}_1\left( \mathbf{u}^h, \mathbf{v} \right) = 0 \quad \forall \mathbf{v} \in \mathcal{U}_0^h,
\end{equation}
and similarly $d^h \in \widetilde{\mathcal{D}}^h$ satisfying
\begin{equation}
\mathcal{A}_2\left( d^h, w \right) = 0 \quad \forall w \in \widetilde{\mathcal{D}}_0^h.
\end{equation}

In this work, standard finite elements are used along with full integration.  In particular, we employ bilinear quadrilateral elements to approximate the displacement and damage fields.  The Lagrange multiplier is also approximated with bilinear quadrilateral elements.  

We note that in the phase-field literature, it is common to introduce a small regularization parameter $\eta \approx 0$ which provides a lower bound on the tensile stresses as the damage reaches unity, as in \cite{Miehe20102765, NME:NME2861}. This is a form of numerical regularization that provides completely damaged elements with a small degree of stiffness as a means to circumvent a loss of ellipticity in the discrete equations. In this work, however, we have not found such a parameter to be necessary. We note that this is not necessarily the result of our cohesive formulation. Indeed, recent phase-field implementations based on Griffith models have reported similar findings, as in \cite{BORDEN201277}. As a result, none of the calculations presented in Section \ref{sec:numerical_examples} rely on any form of regularization beyond the aforementioned penalty kernel. 

\subsection{Staggered solution scheme} \label{sec:staggered}
We now describe an efficient solution procedure based on a staggered solution scheme for the successive update of the damage and displacement fields. In the first part of the algorithm, an outer loop is considered on the update for the damage evolution equation and the explicitly calculated Lagrange multipliers, while the displacement field is held fixed. A new damage field is obtained once the algorithm converges within a predefined tolerance $\mathcal{R}_d$. The second part of the algorithm then holds this damage field fixed while considering the elastodynamics problem, which is advanced using an explicit algorithm based on the central difference time integration scheme.  As is standard with explicit algorithms, we employ a lumped mass matrix.  

We note that this approach is distinct from that of  \cite{WHEELER201469}, in which staggered iterations were used in an attempt to more closely capture the true equilibrium solution through an iterative procedure. In general, such a strategy is significantly more expensive per time or load step, and a detailed investigation into the optimal trade-off between accuracy and computational time has yet to be performed. The approach presented here, summarized in Algorithm \ref{alg:staggered_solution_scheme}, relies on using sufficiently small time steps to ensure accuracy.  In general, we have not found the need to use time steps that are significantly below the stability limit as established by the usual CFL condition in explicit dynamics.  

\begin{algorithm}[!htbp]
\caption{Staggered solution scheme in the time interval $[t^{n-1},t^n]$. }
\begin{algorithmic}
\State Choose $\lambda_0$ = $\lambda^{n-1}$, $\gamma \in \mathbb{R}_{d>0}$, and let $k=0$ 
\Repeat 
\State {Solve the nonlinear damage equation using a Newton-type scheme and initial guess $\lambda_0$
\begin{equation*}
\mathcal{A}_2\left( d^h, w \right) = 0 \quad \forall w \in \widetilde{\mathcal{U}}_0^h
\end{equation*}}
\State {Update the Lagrange multipliers
\begin{equation*}
\lambda_{k+1} = \left\langle \lambda_k + \gamma \left(d^{n-1}-d\right) \right\rangle_+ + \left\langle \lambda_k + \gamma \left(1-d\right) \right\rangle_-
\end{equation*}}
\State $k \leftarrow k+1$
\Until{$\left\lvert \left\lvert d^k-d^{k-1}\right\rvert \right\rvert _2 \leq \mathcal{R}_d$}
\State Solve the elastodynamics problem
\begin{equation*}
\mathcal{A}_1\left( \mathbf{u}^h, \mathbf{v} \right) = 0 \quad \forall \mathbf{v} \in \mathcal{U}_0^h
\end{equation*}
\State Increment $t^n \leftarrow t^{n+1}$
\end{algorithmic}
\label{alg:staggered_solution_scheme}
\end{algorithm}

\section{Numerical results and validation}
\label{sec:numerical_examples}

We now present the results from a series of representative numerical experiments designed to demonstrate the ability of the proposed approach to capture pertinent aspects of cohesive fracture processes. The goal of these examples is threefold: (i) to numerically verify the convergence properties of the proposed formulation with respect to the regularization length as well as to provide insight into its interpretation in a multidimensional setting; (ii) to examine various options at imposing irreversibility, as discussed in Section \ref{sec:alm}; and (iii) to demonstrate the performance of the regularized cohesive model for  benchmark problems in quasi-static and dynamic fracture. 


In our calculations, we rely on meshes that are sufficiently refined to capture the variation in the damage field around crack surfaces.  Unless specified otherwise,  we use element sizes of $h_e = L/10$ as a rule of thumb.  We also rely on the quasi-quadratic degradation function \eqref{eq:gq} with $p=1$, except where indicated. For all two-dimensional numerical experiments that are considered in this Section, plane strain conditions are assumed to hold.
Finally, all calculations employing the augmented Lagrangian method use a tolerance of $\mathcal{R}_d=10^{-4}$  in the outer loop. In our experience, some problem-dependent fine-tuning of the penalty kernel $\gamma$ is required to obtain good convergence behavior in the augmented Lagrangian approach.  In practice, we have employed penalty kernels in the range of $\gamma = 10^{2} - 10^{5}$, which is consistent with those reported in \cite{WHEELER201469}.  In all cases, the monotonicity penalization is employed, whereas the threshold penalization is only employed in conjunction with the quasi-linear degradation function.




\subsection{A one-dimensional bar under tension}
\label{subsec:1d_num}

\begin{figure}[!bp]
\centering \small
\begin{overpic}[width=.44\linewidth]{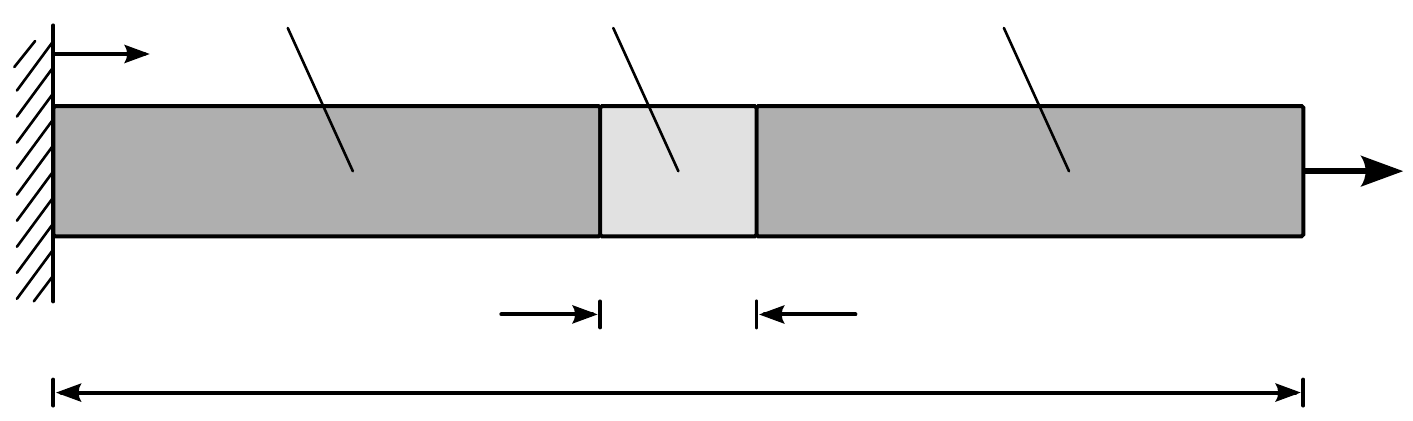}
\put(44,6.25){$\ell/10$}
\put(47,-2.5){$\ell$}
\put(16,29){$A$}
\put(35,29){$A/2$}
\put(67,29){$A$}
\put(5,28){$x$}
\put(101,16.5){($F,\bar{u}$)}
\end{overpic}
\caption{Bar with reduced cross sectional area in the middle, subjected to a tensile load $F$.}
\label{1d_example}
\end{figure}

Consider a one-dimensional bar with a reduced cross sectional area in  the middle, subjected to a tensile load $F$ applied to the right end, as shown in Figure \ref{1d_example}. The material parameters are $E=10$~MPa, $\mathcal{G}_c=0.1$~N/mm and $\sigma_c=2$~MPa. The bar has a length of $\ell = 1~$mm and a nominal cross-sectional area of $A=1~$mm$^2$. 

Numerically demonstrating convergence towards a cohesive zone formulation for a vanishing regularization length is a non-trivial task. To ensure the equilibrium solution is captured accurately at every load step, we depart slightly from the algorithm described in Section~\ref{sec:staggered} and implement a monolithic solution scheme for this problem. A path-following constraint was introduced to control the loading process, as in \cite{NME:NME2447}. In a one-dimensional setting, a constant mesh size of $h_e = \ell/2000$ was employed, providing sufficient resolution to resolve the damage profile for all of the regularization lengths under consideration. 

\begin{figure}[!tbp]
\centering \small
\begin{overpic}[width=.44\linewidth]{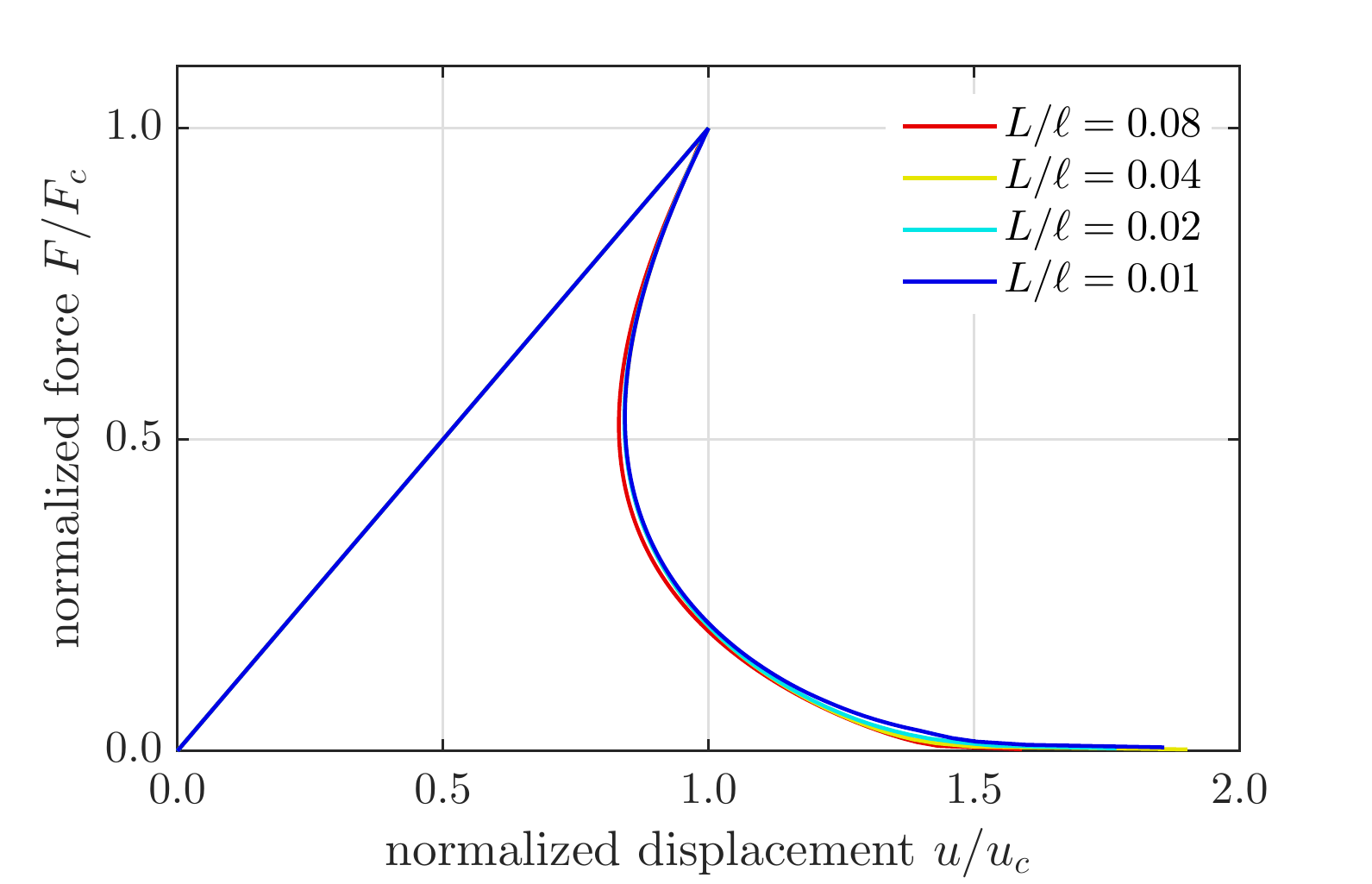}
\put(0,2){(a)}
\end{overpic}
\begin{overpic}[width=.44\linewidth]{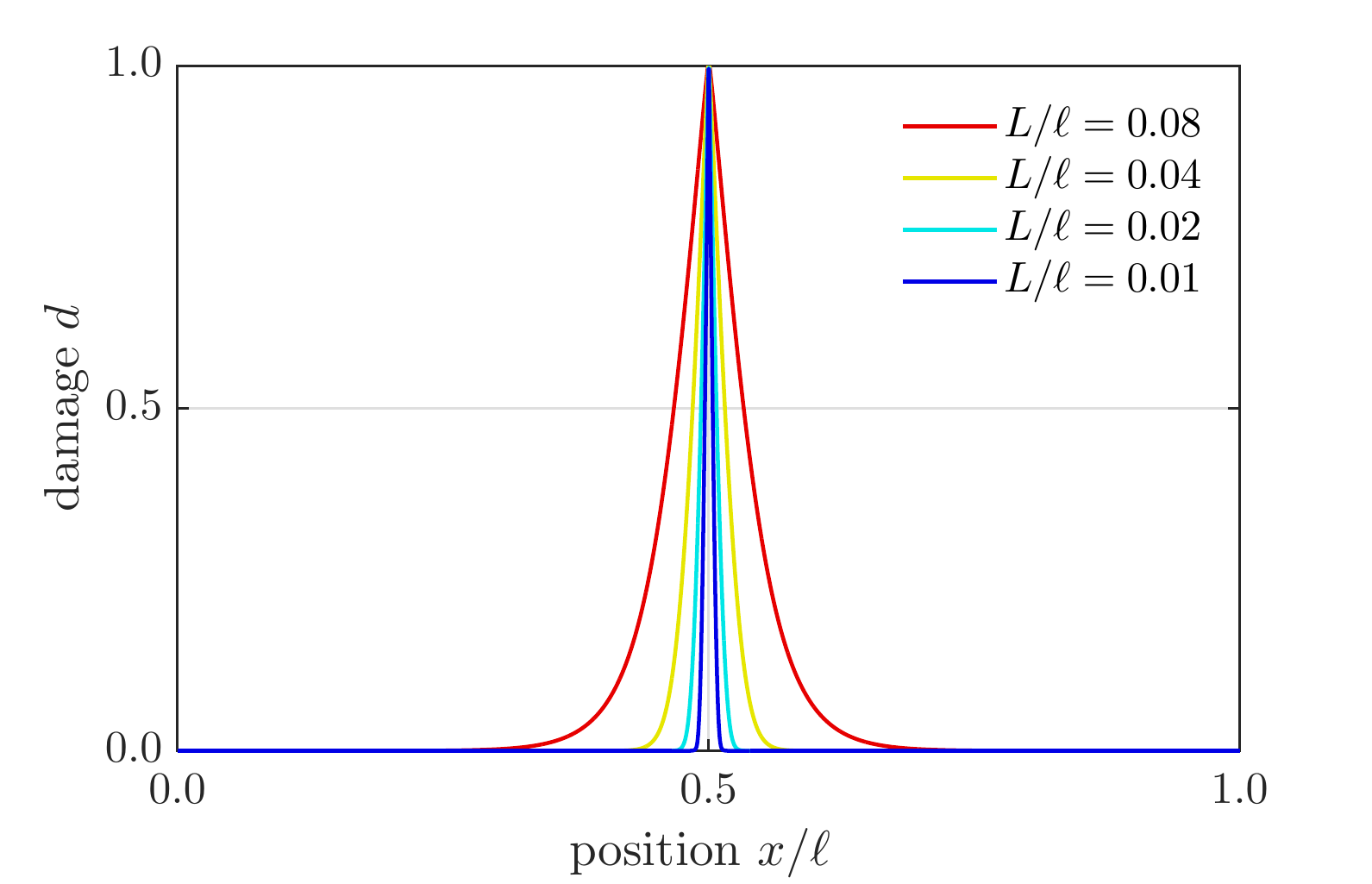}
\put(0,2){(b)}
\end{overpic}
\caption{Effect of the regularization length on (a) the normalized force-displacement curve, and (b) the damage field using a constant shape parameter $p=1$.}
\label{1d_example_res1}
\end{figure}

The influence of the regularization length on the force-displacement curve and damage profile is shown in Figure \ref{1d_example_res1}. Here the reaction force $F$, normalized by a peak force $F_c$, is plotted as a function of the displacement $u$ at the right end of the bar, normalized by its value $u_c$ at the onset of damage. Prior to softening, the force-displacement curves superimpose, as the constitutive response is purely elastic. Once the critical stress in the bar is exceeded, damage is initiated, and strain softening begins. Despite significant differences in the corresponding damage distributions, the regularization length clearly has a negligible effect on the structural response.  These results confirm similar ones made in \cite{LORENTZ201120}, in which it was demonstrated that the one-dimensional response of the proposed phase-field/gradient damage model converges toward that of a cohesive zone model in the limit of vanishing regularization length. 

\begin{figure}[!tbp]
\centering \small
\begin{overpic}[width=.44\linewidth]{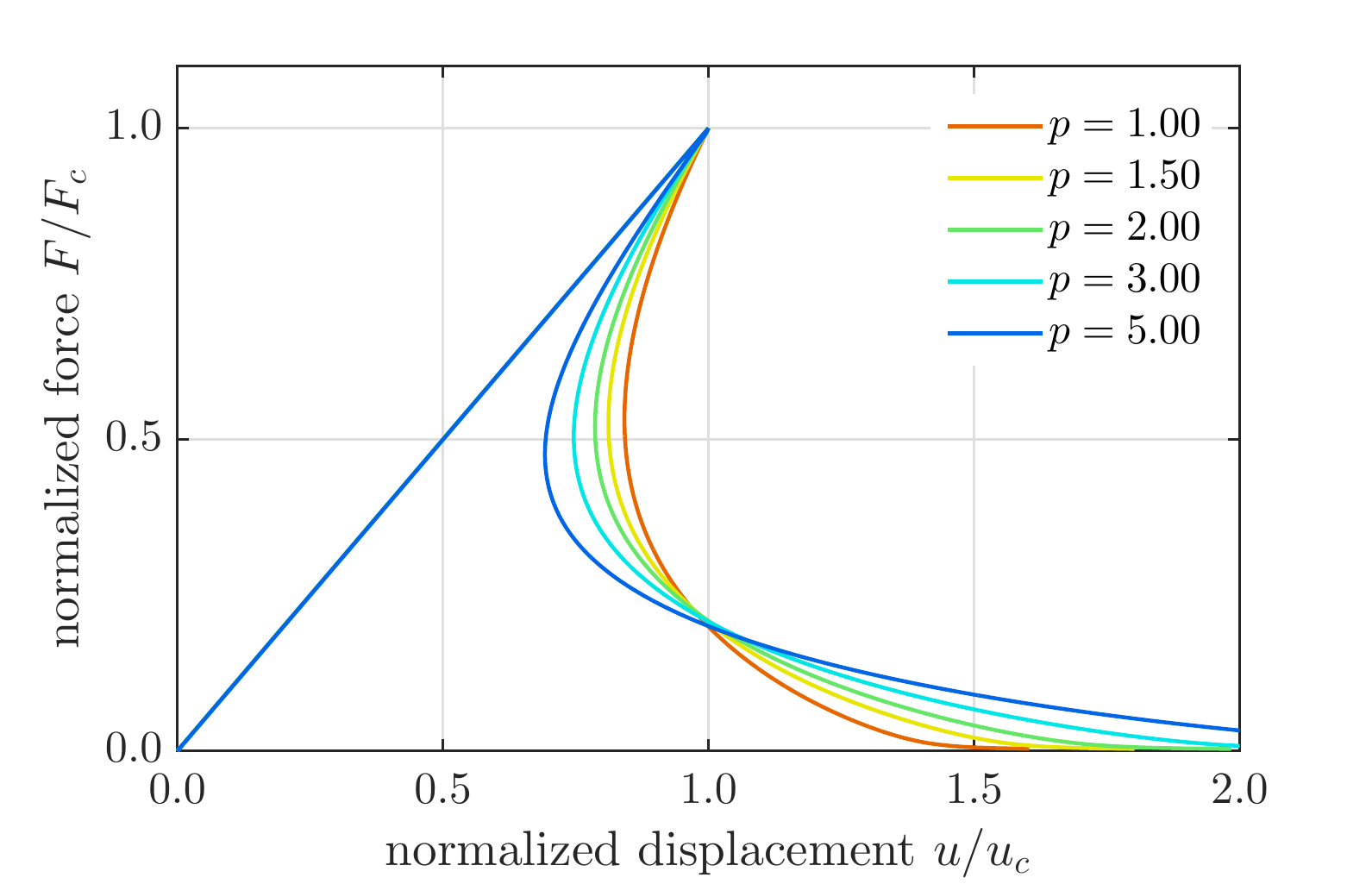}
\put(51,30){\vector(-2,-1){12}}
\put(53,30){$p$}
\put(0,2){(a)}
\end{overpic}
\begin{overpic}[width=.44\linewidth]{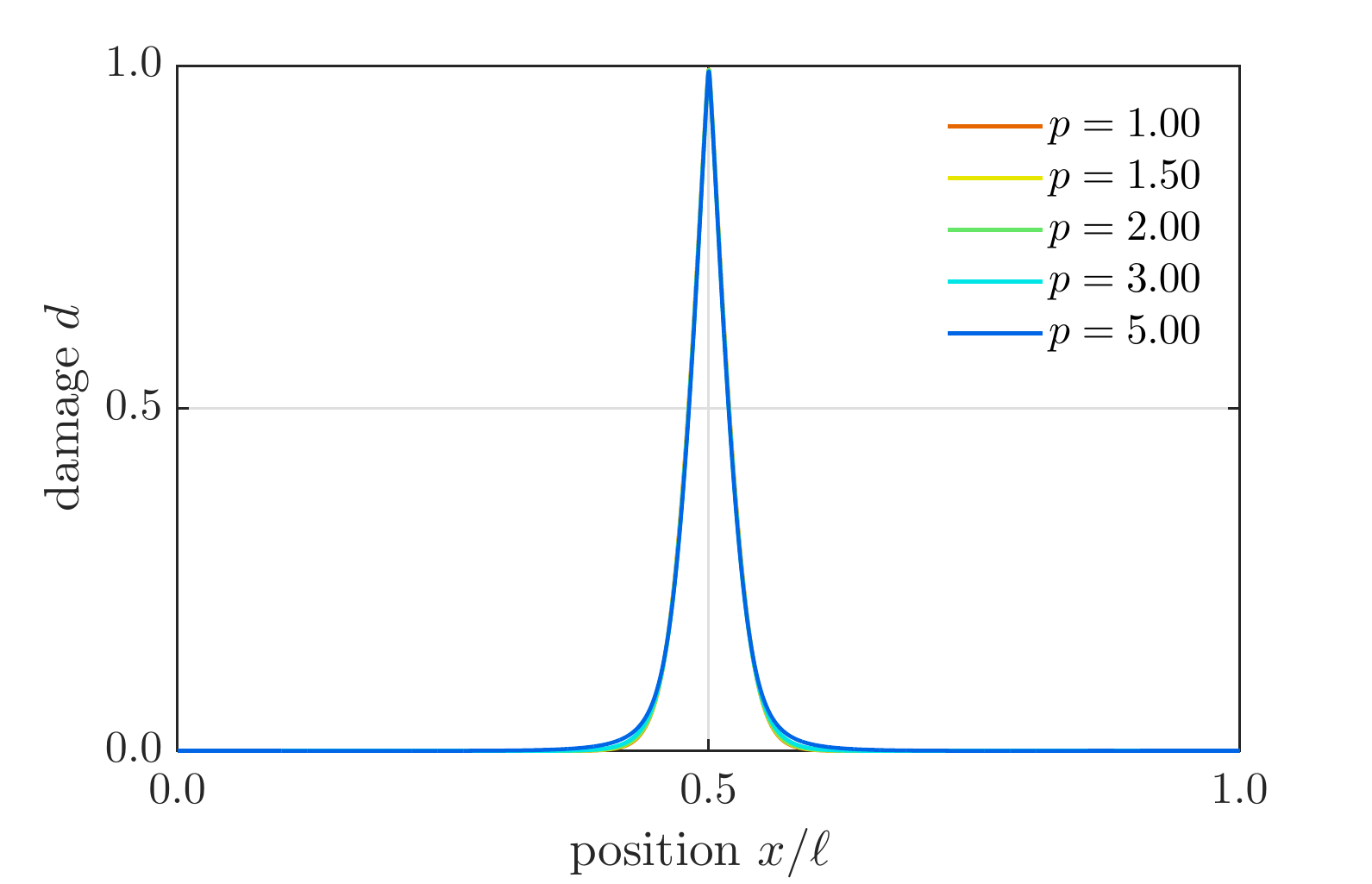}
\put(0,2){(b)}
\end{overpic}
\caption{Effect of the cohesive shape parameter $p$ on (a) the force-displacement curve, and (b) the damage field when the regularization length is held fixed at $L/\ell=0.05$.}
\label{1d_example_res2}
\end{figure}

\begin{figure}[!tbp]
\centering \small
\begin{overpic}[width=.44\linewidth]{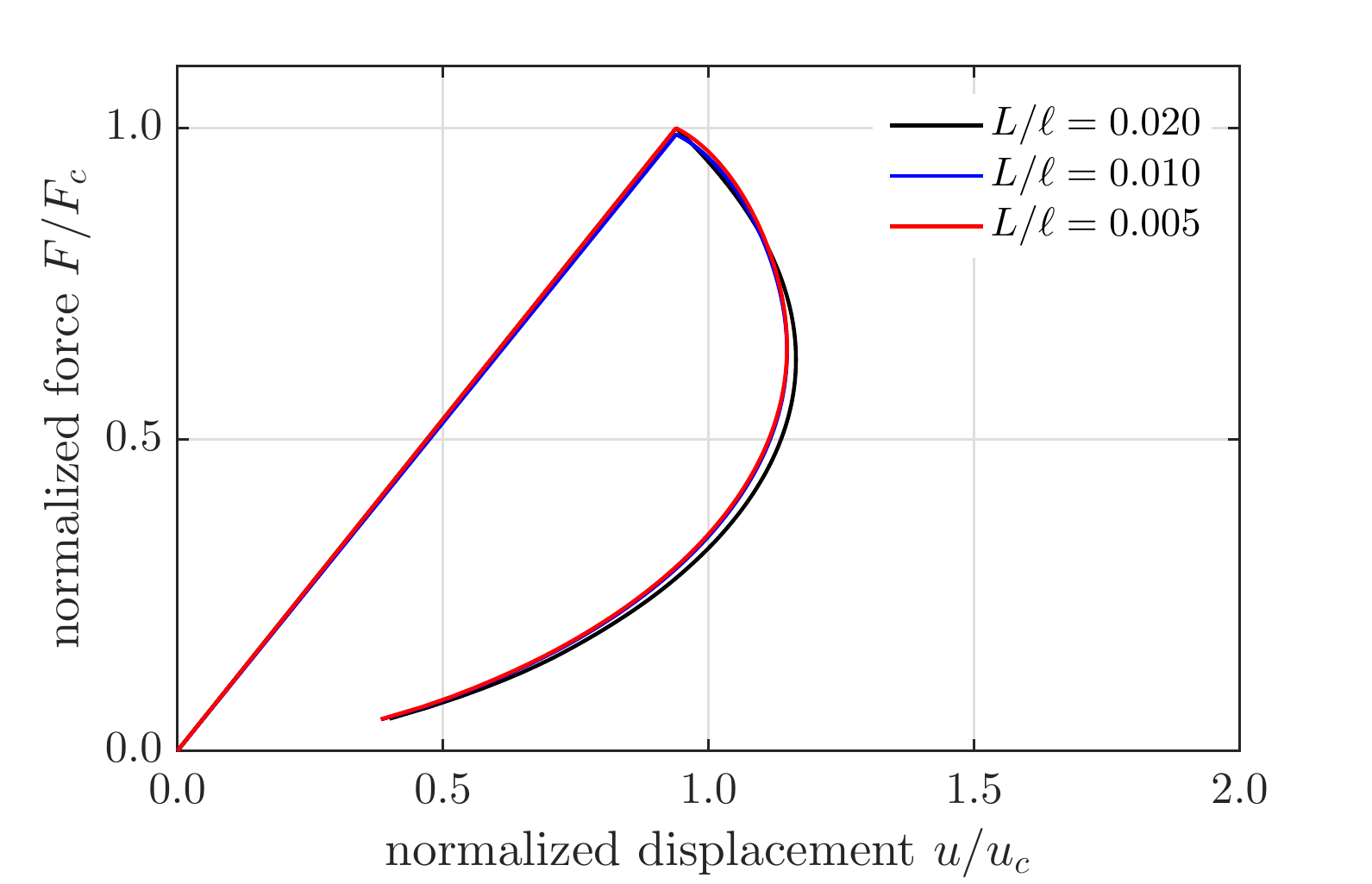}
\put(0,2){(a)}
\end{overpic}
\begin{overpic}[width=.44\linewidth]{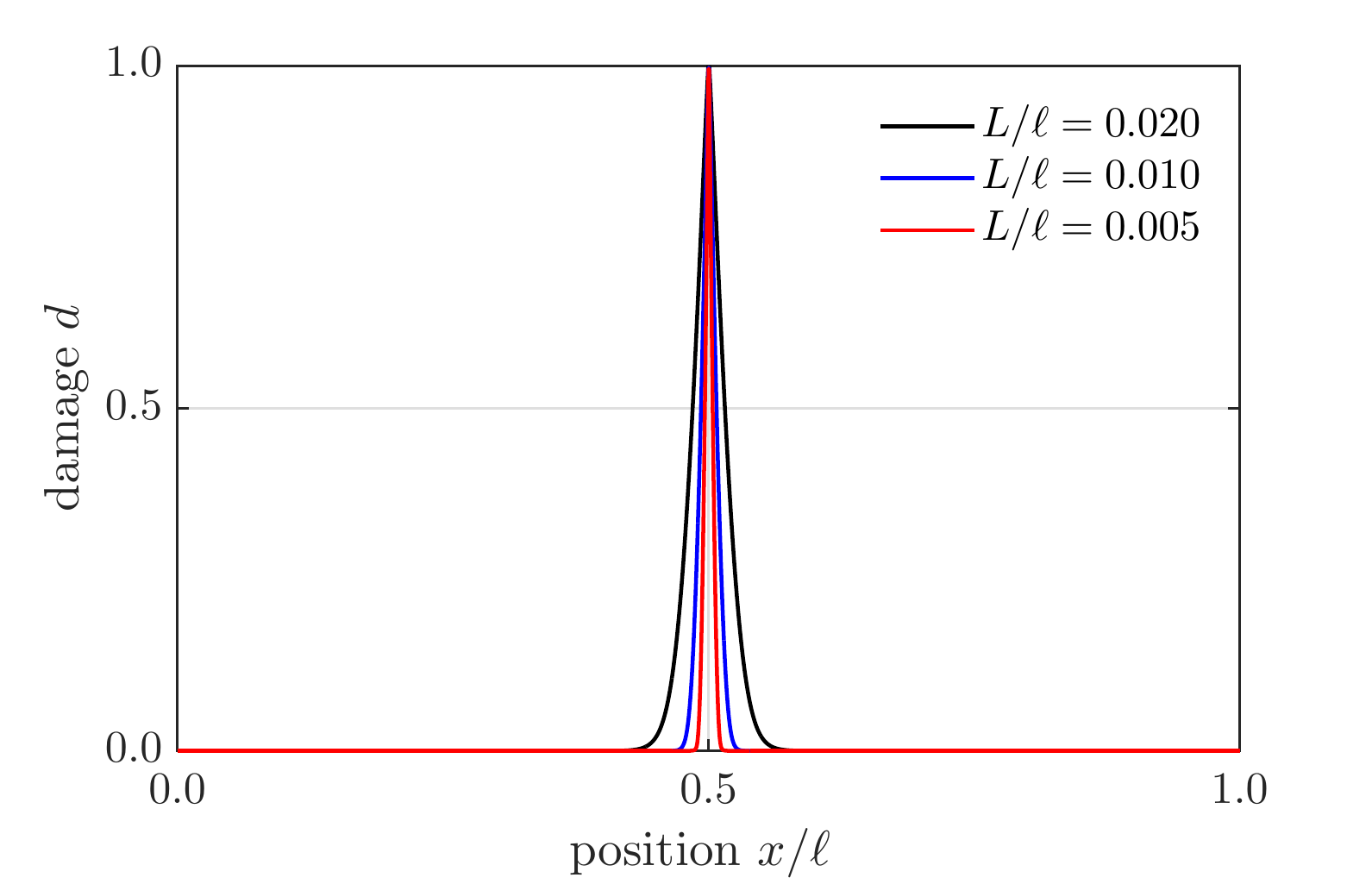}
\put(0,2){(b)}
\end{overpic}
\caption{Effect of the regularization length on (a) the normalized force-displacement curve, and (b) the damage field using quasi-linear degradation function \eqref{eq:gl}.}
\label{1d_example_gl}
\end{figure}

Next we study the impact of the softening shape parameter $p$ on the constitutive response. For a fixed regularization length, an increasing value for this parameter leads to a more rapid decay of the stress in the early stages of the strain-softening process, as demonstrated in Figure \ref{1d_example_res2}. This behavior is also implied by the form of the quasi-quadratic degradation function \eqref{eq:gq}. In addition, we note that the shape parameter has minimal influence on the ultimate damage distribution.

We now study the same one-dimensional problem, but using the  quasi-linear degradation function \eqref{eq:gl}. In this case, the upper bound on the regularization length is computed to be $L < 0.375~$mm. We therefore consider regularization length scales of $L=0.020~$mm, $L=0.010~$mm and $L=0.005$~mm. We recall that \eqref{eq:gl} does not possess any shape parameters as it is solely designed to approximate a bilinear stress-stress response upon strain-softening.

\begin{figure}[!tbp]
\centering \small
\begin{overpic}[width=.44\linewidth]{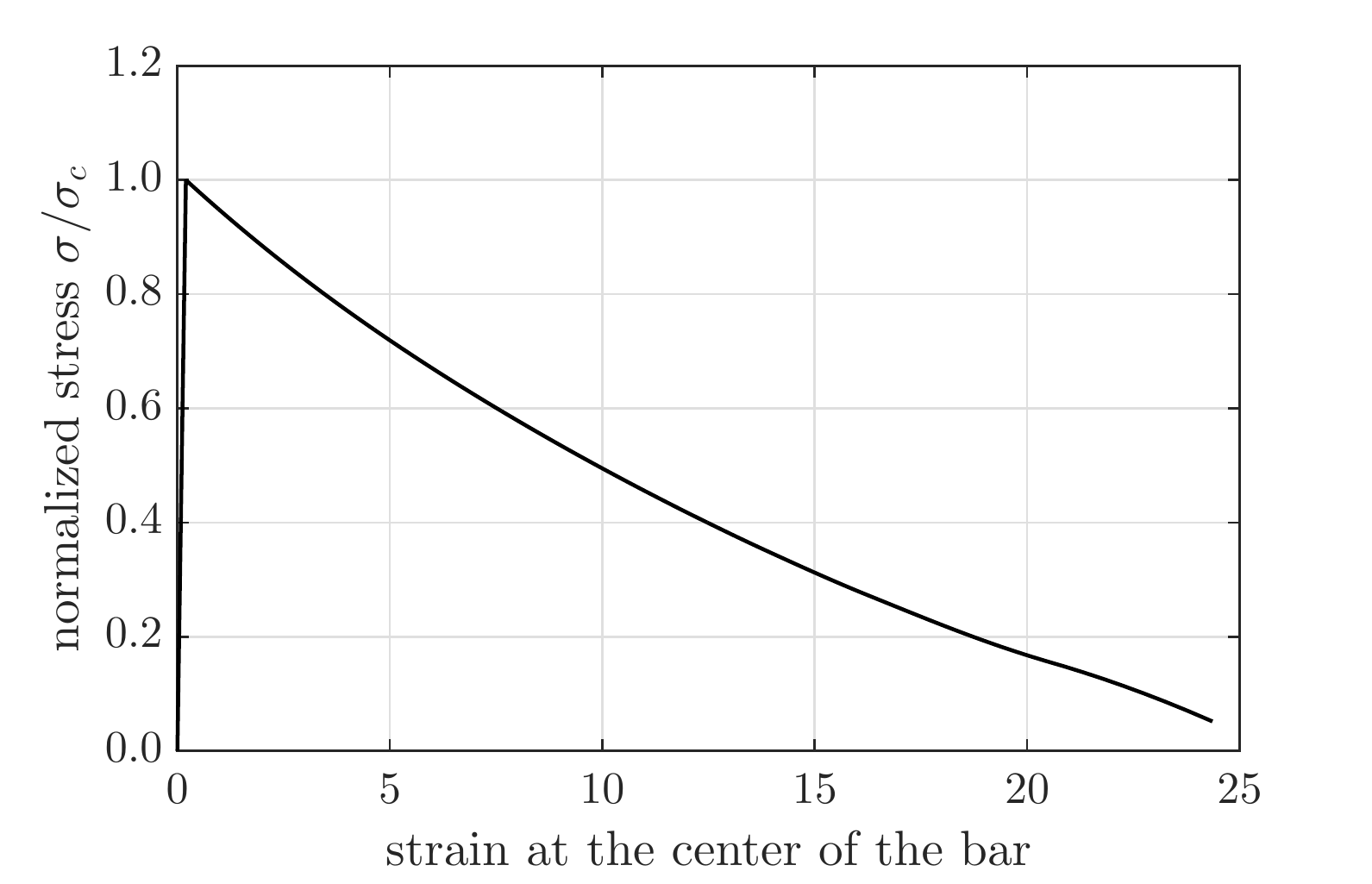}
\end{overpic}
\caption{A quasi-linear stress-strain response at the center of the bar for the cohesive model when employing the quasi-linear degradation function \eqref{eq:gl} and $L/\ell = 0.01$.}
\label{fig:stress-strain}
\end{figure}

The non-dimensional force-displacement curves for the quasi-linear degradation function are given in Figure \ref{1d_example_gl}. Similar to the results obtained with the quasi-quadratic degradation function, we see virtually no sensitivity to the regularization length scale. However, the constitutive response markedly differs from the one obtained through the quasi-quadratic degradation function. The force-displacement response is a lot softer, while the stresses rapidly drop to zero as $d\rightarrow 1$. The findings are consistent with the arguments outlined in Section \ref{sec:ql-deriv}. In addition, Figure \ref{fig:stress-strain} demonstrates that the stress-strain response at the center of the bar effectively approximates a linear decay in the post-peak regime.

\subsection{Single edge notched tests}
\label{subsec:single_edge_tests}

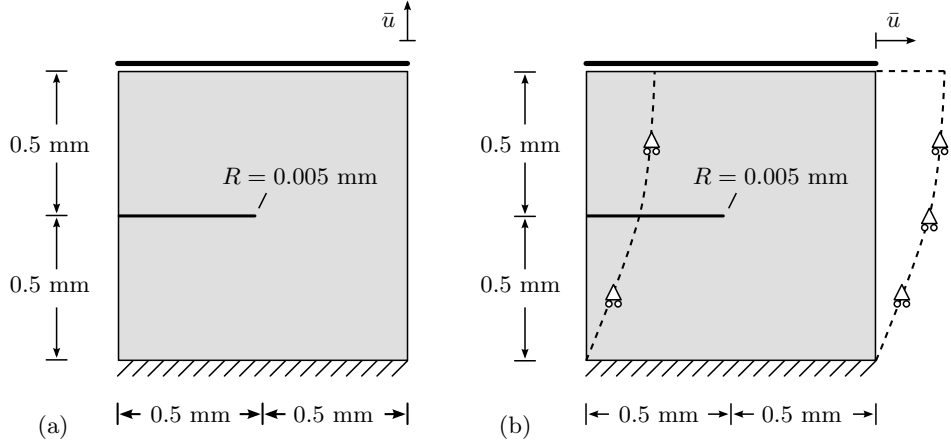
\begin{figure}[!tbp]
\centering \small
\begin{overpic}[width=0.7\linewidth]{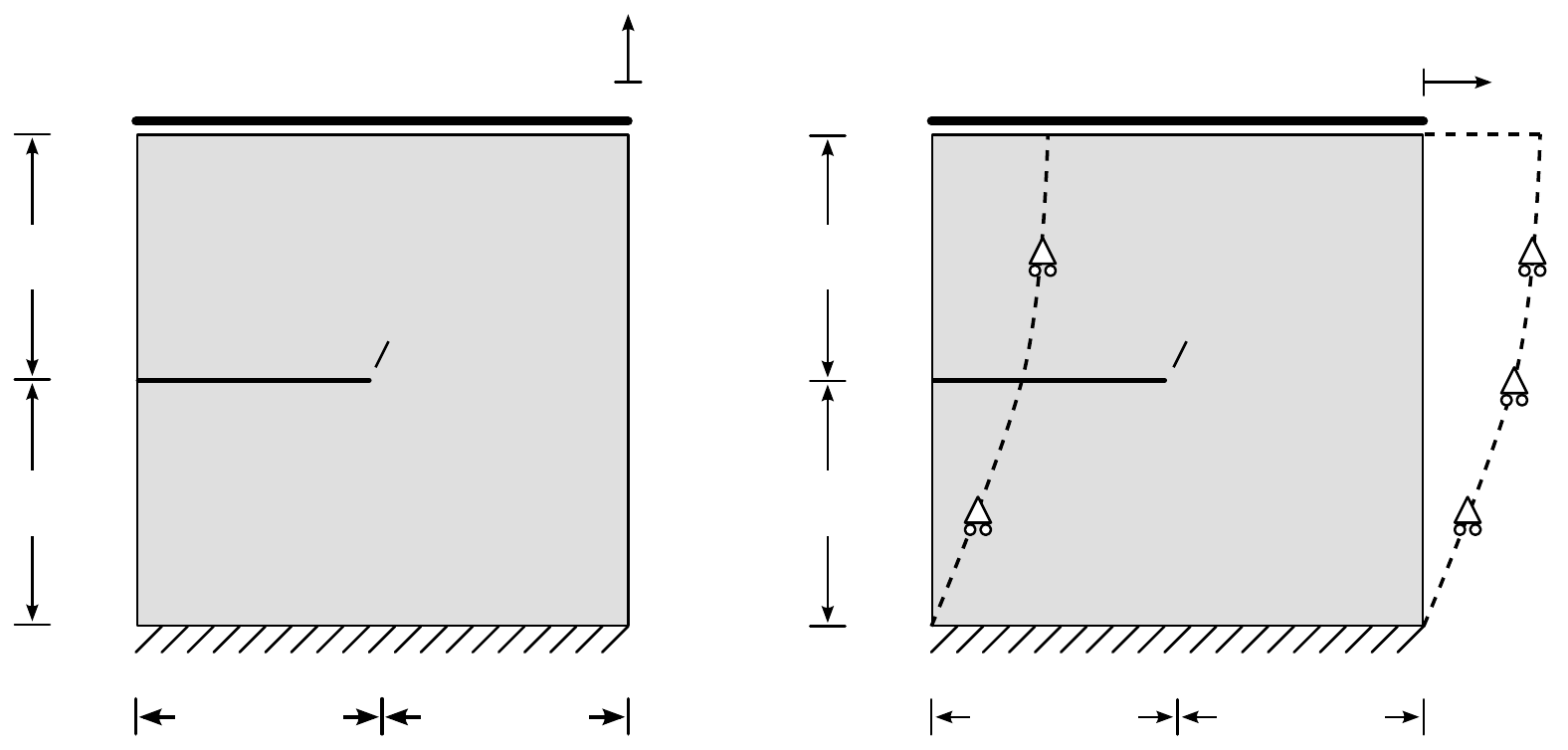}
\put(0,0){(a)}
\put(50,0){(b)}
\put(-3,30.5){0.5~mm}
\put(-3,15.25){0.5~mm}
\put(48,30.5){0.5~mm}
\put(48,15.25){0.5~mm}
\put(12.25,1.25){0.5~mm}
\put(27.75,1.25){0.5~mm}
\put(63,1.25){0.5~mm}
\put(78.5,1.25){0.5~mm}
\put(20,27){$R=0.005~$mm}
\put(71,27){$R=0.005~$mm}
\put(92,44){$\bar{u}$}
\put(37.25,44){$\bar{u}$}
\end{overpic}
\caption{Geometry and boundary conditions for the single-edge notched tests (a) in tension, and (b) in shear.}
\label{mode_12}
\end{figure}

We now investigate two benchmark problems, which have become canonical in the phase-field for fracture literature. Consider a square plate with an initially horizontal edge crack extending to the middle of the specimen, as shown in Figure \ref{mode_12}a. Following \cite{NME:NME2861}, the material parameters are chosen to be $E=210~$GPa, $\nu=0.3$, and $\mathcal{G}_c=2.7~$N/mm. In the following, the cohesive strength is set to $\sigma_c = 2.5~$GPa. The corresponding estimate for the length of the process zone is $\ell_{\text{FPZ}}\approx E\mathcal{G}_c/\sigma_c^2 = 0.091~$mm. The computations are performed in a displacement driven context where the displacement increment is adjusted upon approaching the peak load to ensure accuracy. To accurately capture the evolution of the damage field, the mesh is refined along the anticipated crack path. In addition, the initial notch was modeled with a small radius $R = 0.005~$mm to obtain a more realistic representation of the stresses in the vicinity of the initial crack tip. 

\begin{figure}[!tbp]
\centering \small
\begin{overpic}[width=0.75\linewidth]{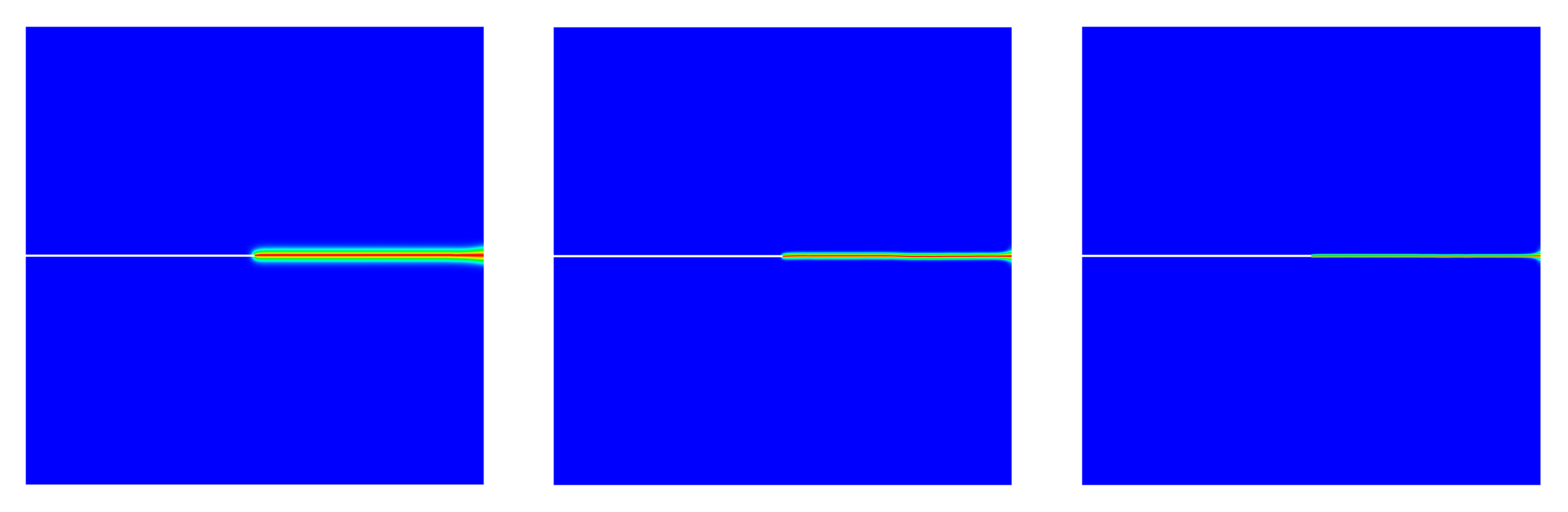}
\put(-2,1.5){(a)}
\put(31.6,1.5){(b)}
\put(65.6,1.5){(c)}
\end{overpic}
\hspace{1em}
\raisebox{2.1em}{
\begin{overpic}[width=0.015\linewidth]{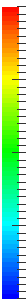}
\put(-9,-11){$d=0$}
\put(-9,103.5){$d=1$}
\end{overpic}}
\caption{Single edge notched test under tension for (a) $L=0.020$~mm, (b) $L=0.010$~mm and (c) $L=0.005$~mm.}
\label{fig:mode_1_results}
\end{figure}

First we apply a vertical displacement to the complete top edge, as demonstrated in Figure \ref{mode_12}a. The fracture patterns for different values of the regularization length are shown in Figure \ref{fig:mode_1_results}. The blue and red contour levels indicate an intact and fully damaged material state, respectively. In contrast to results obtained for this problem using phase-field approximations of a Griffith model of fracture, we do not observe any evolution of the damage field away from the regularized fracture surface.

The same specimen is now subjected to a shear load, as depicted in Figure \ref{mode_12}b. We note that the crack trajectories presented here are in excellent agreement with the ones reported in \cite{Miehe20102765} and display little sensitivity to the regularization length scale. For a more comprehensive investigation into the effect of a vanishing regularization length, we refer to the trapezoid problem investigated in the next subsection.

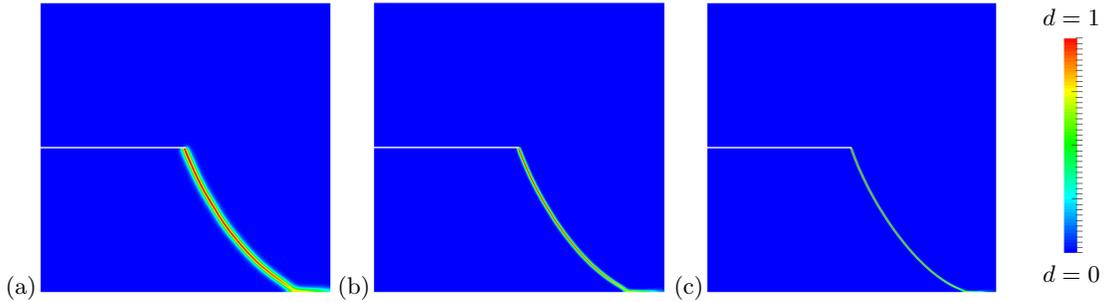
\begin{figure}[!tbp]
\centering \small
\begin{overpic}[width=0.75\linewidth]{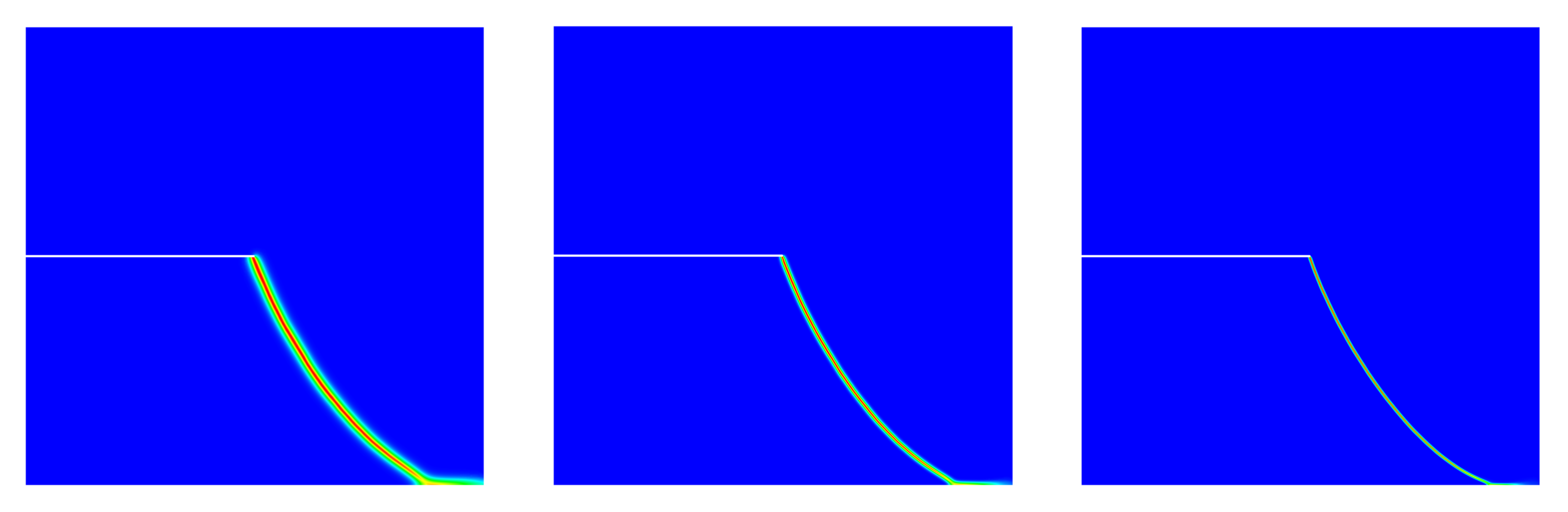}
\put(-2,1.5){(a)}
\put(31.6,1.5){(b)}
\put(65.6,1.5){(c)}
\end{overpic}
\hspace{1em}
\raisebox{2.1em}{
\begin{overpic}[width=0.015\linewidth]{legend.png}
\put(-9,-11){$d=0$}
\put(-9,103.5){$d=1$}
\end{overpic}}
\caption{Single edge notched test under shear for (a) $L=0.020$~mm, (b) $L=0.010$~mm and (c) $L=0.005$~mm using shape parameter $p=1$.}
\label{fig:mode_2_results}
\end{figure}

\subsection{Trapezoid problem}

We now examine a problem first proposed by \cite{Lorentz20111927} in which a particular geometric setup is designed to facilitate stable crack propagation  without exhibiting any snap-back behavior. In addition to this, the dimensions of the virtual specimen, several meters in size, are significantly larger then the characteristic size of the fracture process zone in plain concrete. This separation of scales is less distinct in the single-edge notched specimens from Section \ref{subsec:single_edge_tests}. For these reasons, the trapezoid problem is viewed as being particularly well-suited for an investigation into the effects of a vanishing regularization length.

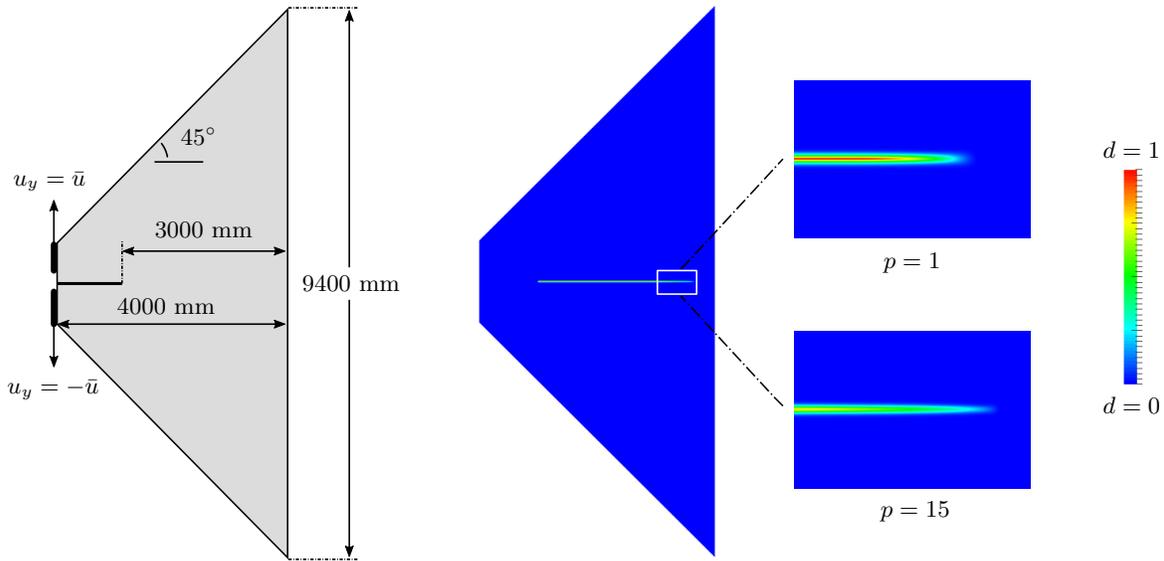
\begin{figure}[!bp]
\centering \small
\begin{overpic}[width=0.28\linewidth]{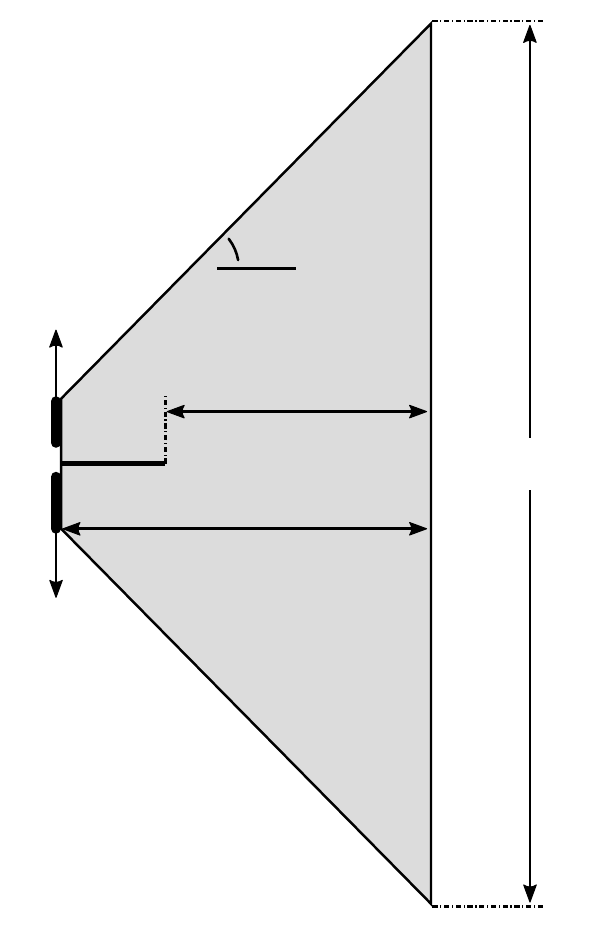}
\put(-1,67){$u_y = \bar{u}$}
\put(-2,31){$u_y = -\bar{u}$}
\put(17,44.75){4000~mm}
\put(23.5,58){3000~mm}
\put(49,48.75){9400~mm}
\put(28,74){$45^{\circ}$}
\end{overpic}
\hspace{2em}
\raisebox{0.2em}{
\begin{overpic}[width=0.45\linewidth]{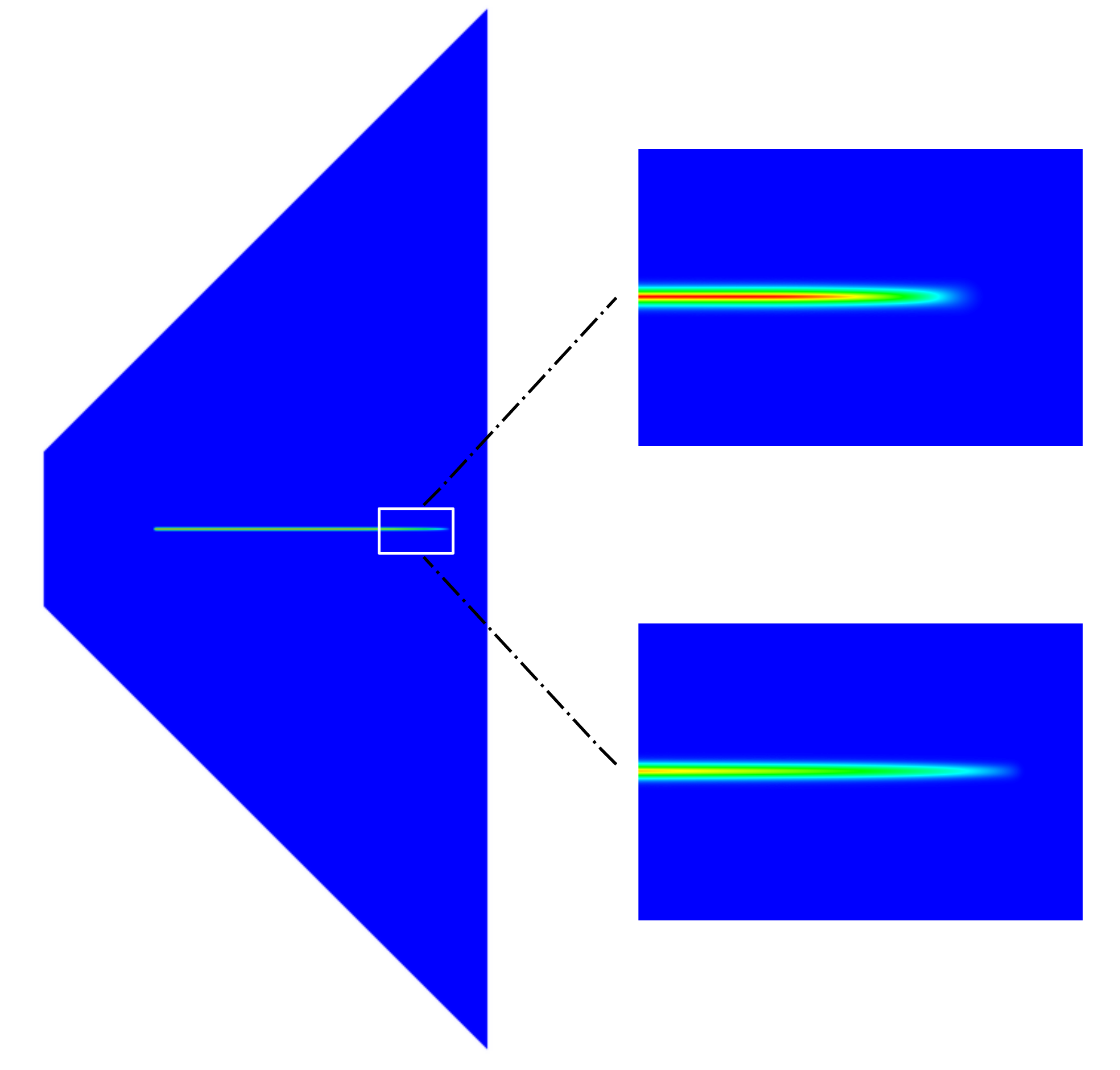}
\put(71.5,8.5){$p=15$}
\put(72,50.5){$p=1$}
\end{overpic}}
\hspace{2em}
\raisebox{7.5em}{
\begin{overpic}[width=0.015\linewidth]{legend.png}
\put(-9,-11){$d=0$}
\put(-9,103.5){$d=1$}
\end{overpic}}
\caption{(a) Geometry and boundary conditions for a trapezoidal-shaped fracture specimen, taken from \cite{Lorentz20111927}. (b) Fracture pattern and the impact of the shape parameter on the effective process zone.}
\label{fig:trapezoid}
\end{figure}

The material parameters are chosen to be representative of concrete, namely $E = 30~$GPa, $\nu=0.2$, $\mathcal{G}_c=0.1~$N/mm and $\sigma_c=3~$MPa. The corresponding size of the fracture process zone is $\ell_{\text{FPZ}} \approx 333~$mm. The analysis was performed for two different values of the cohesive shape parameter, namely $p=1$ and $p=15$. The finite element mesh is refined along the expected crack path.

\begin{figure}[!tbp]
\centering \small
\begin{overpic}[width=0.44\linewidth]{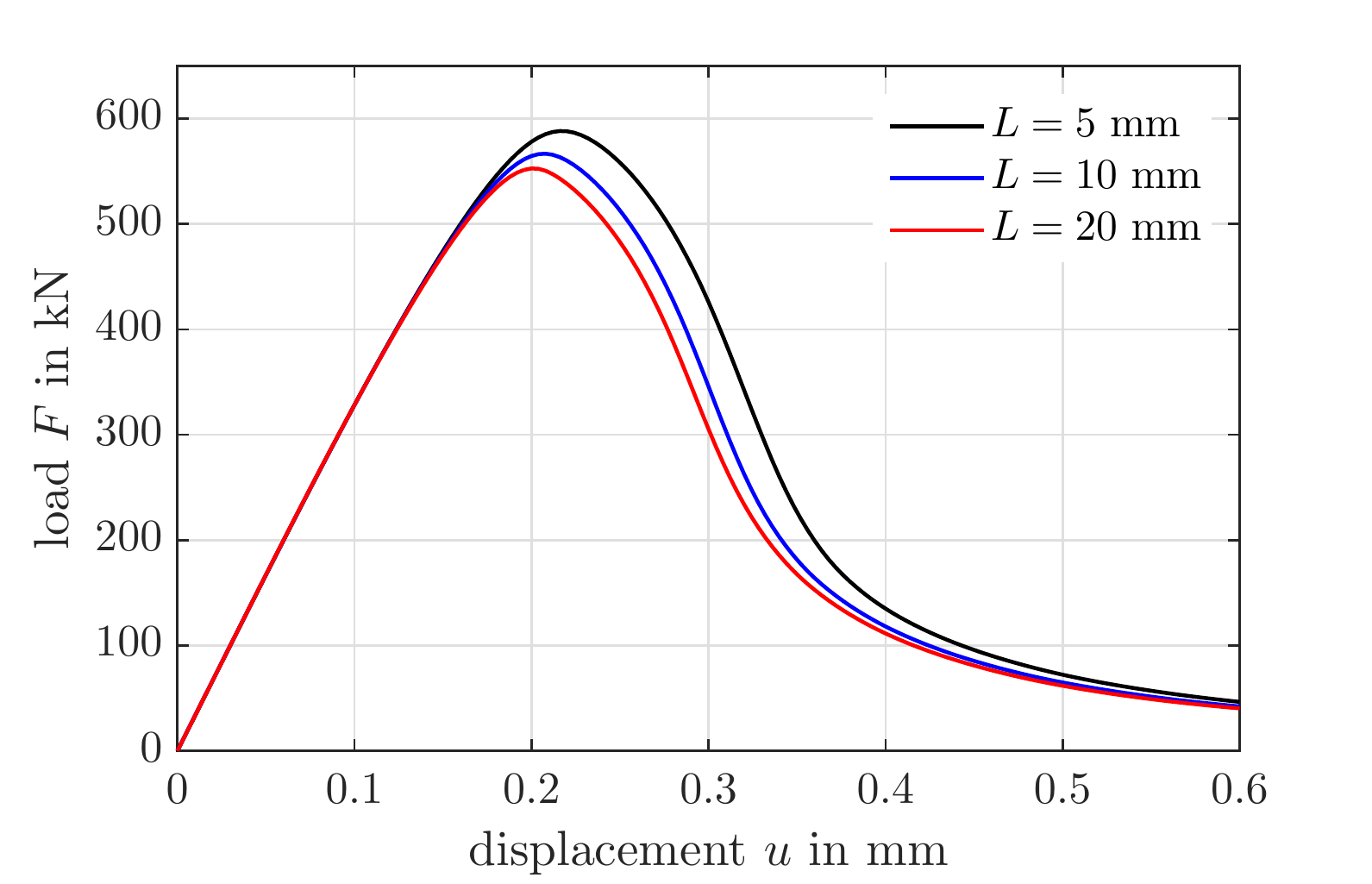}
\put(0,2){(a)}
\end{overpic}
\begin{overpic}[width=0.44\linewidth]{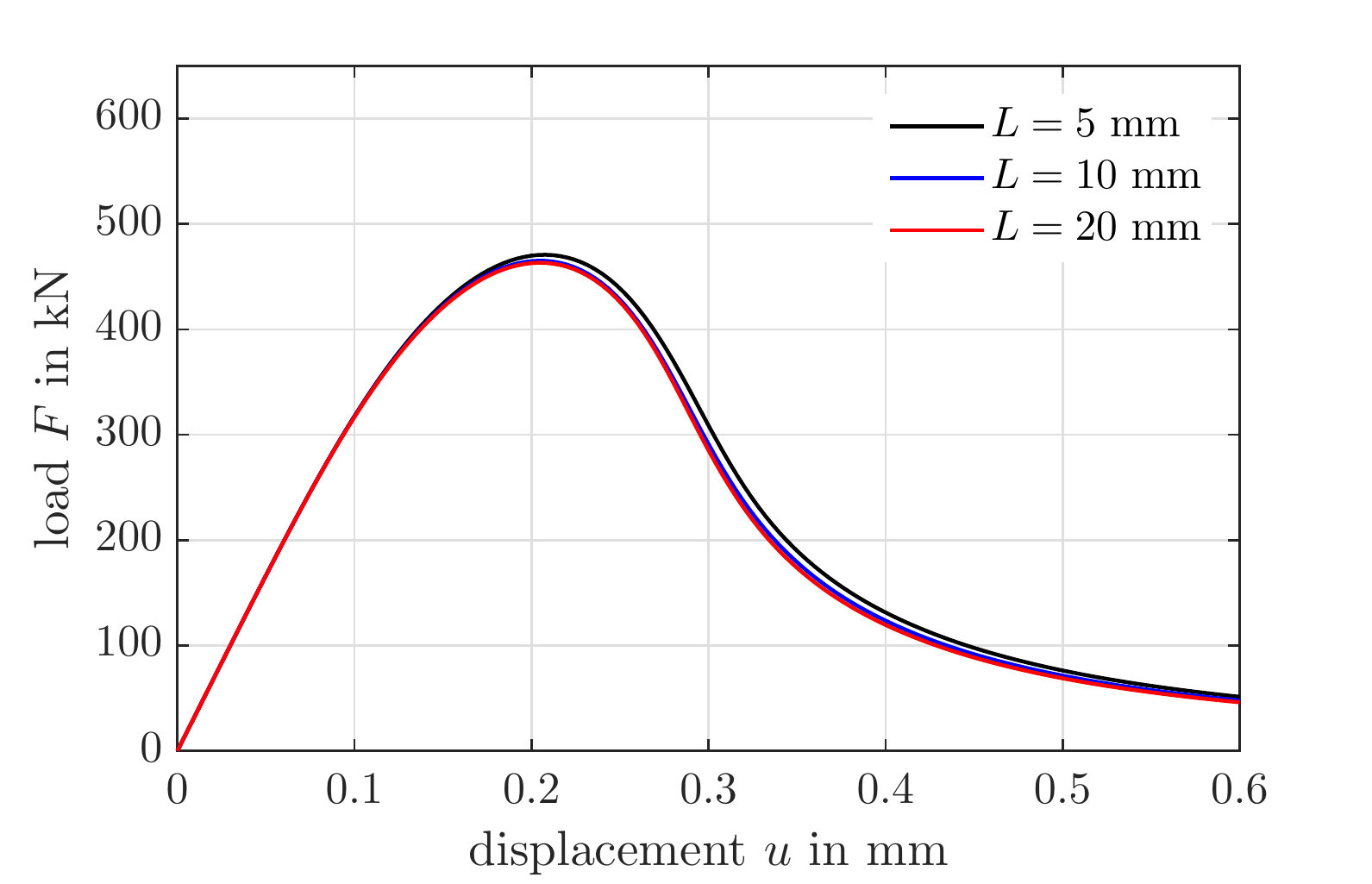}
\put(0,2){(b)}
\end{overpic}
\caption{Load-displacement curves for different values of the regularization length $L$ of the trapezoid problem. The shape parameter under consideration equals (a) $p=1$, and (b) $p=15$.}
\label{fig:trapezoid_fd}
\end{figure} 

The force-displacement curves for this problem are shown in Figure \ref{fig:trapezoid_fd}, and are in excellent agreement with those reported in \cite{Lorentz20111927} and \cite{Lorentz2012}. The results illustrate the significance of the shape parameter $p$ and its impact on the critical load the structure is able to sustain. An increasing value for $p$ can also considerably elongate the length of the fracture process zone, as indicated in Figure \ref{fig:trapezoid}b. In this case, the stresses in the diffusive process zone are effectively smeared over a larger distance, which actually facilitates a convergence study with respect to the regularization length. For a value of $p=15$, the differences in the force-displacement curves for the various regularization lengths are negligible. As the fracture process zone decreases in size, strain components in the plane parallel to the crack become more significant.  As a result, there is greater separation in the force displacement curves for $p=1$, as indicated by \ref{fig:trapezoid_fd}. Given these conditions, we can effectively demonstrate the insensitively of the regularization length to the constitutive response. We refer the reader to \cite{Lorentz2012} for a more comprehensive analysis of this problem.


\subsection{Three-point bending test}

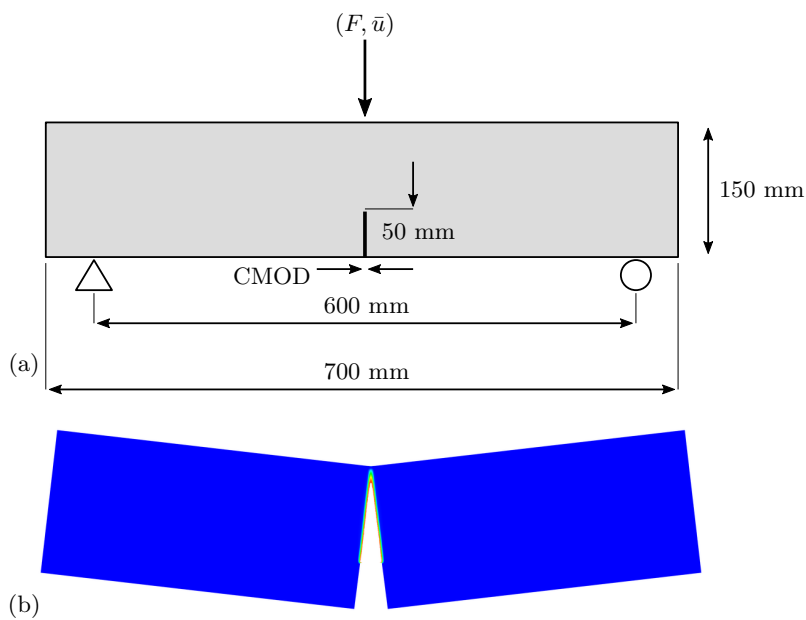
\begin{figure}[!bp]
\centering \small
\begin{overpic}[width=0.55\linewidth]{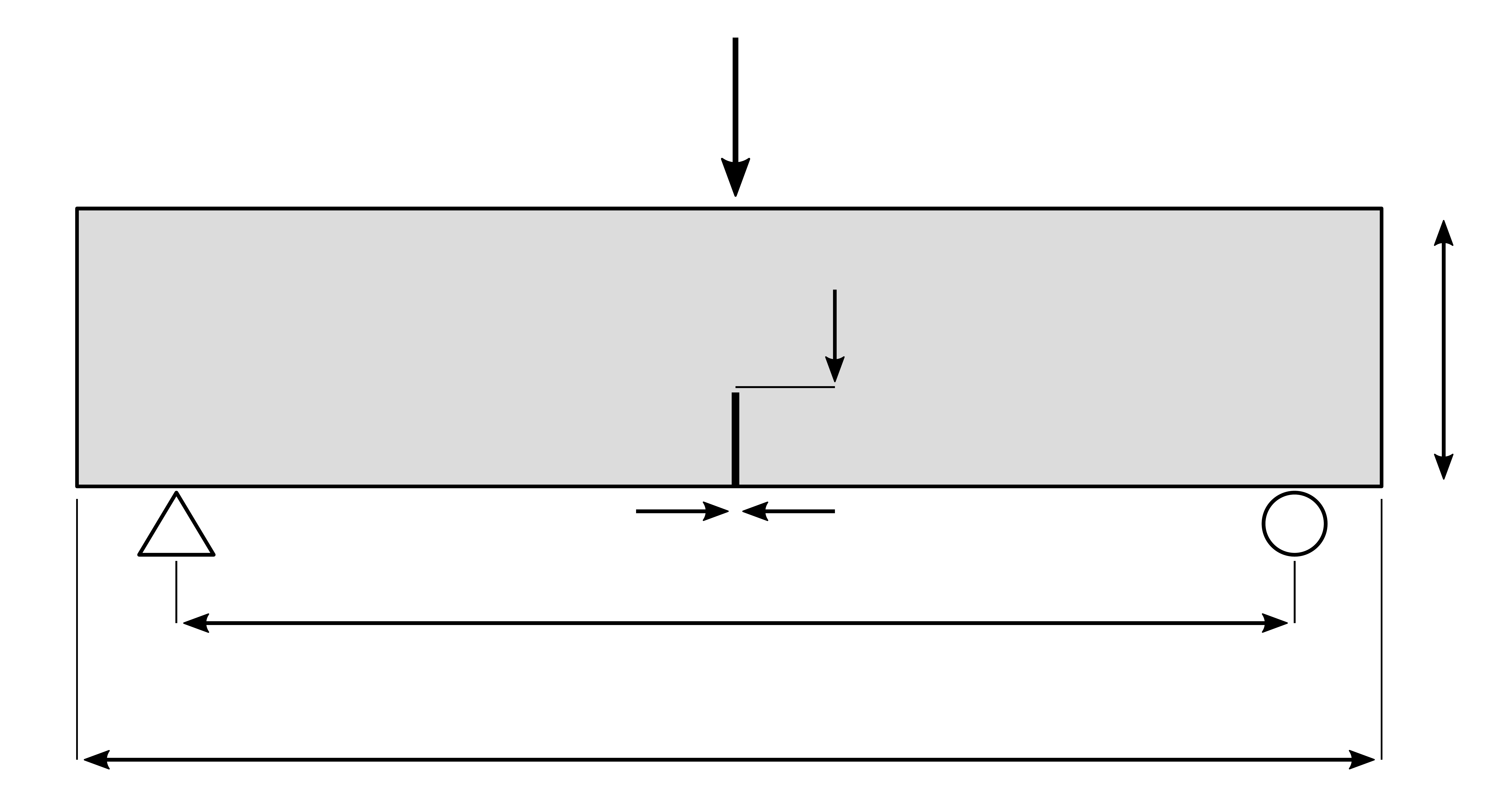}
\put(43.5,14){600~mm}
\put(31,18){CMOD}
\put(43.5,5){700~mm}
\put(98,30){150~mm}
\put(51.5,24.25){50~mm}
\put(45,53){($F,\bar{u}$)}
\put(0,6){(a)}
\end{overpic}
\begin{overpic}[width=0.51\linewidth]{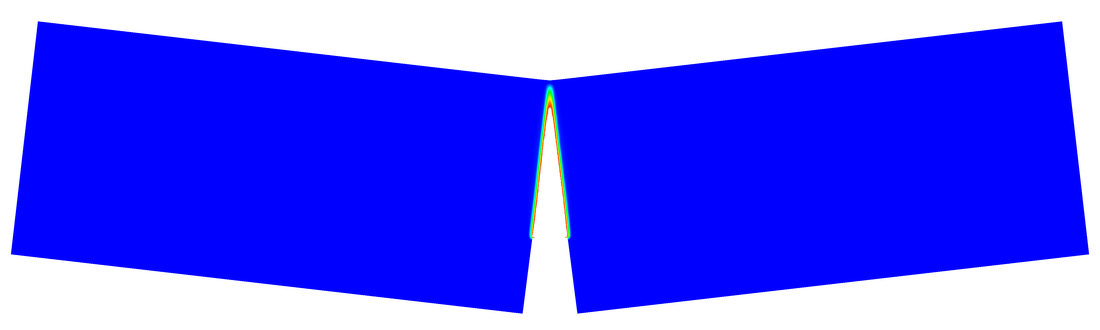}
\put(-4,1){(b)}
\end{overpic}
\caption{(a) Dimensions and boundary conditions for the three-point bending test, from \cite{ROESLER2007300}. (b) Associated damage pattern for the three-point bending test in the deformed configuration with degradation function \eqref{eq:gq} and $p=1$. The displacements have been scaled by a factor of 50 and areas of the model where $d\geq0.95$ have been removed in order to show a representation of the fractured geometry. }
\label{fig:3pb}
\end{figure}

We now consider a classical benchmark problem in cohesive fracture mechanics, the three-point bending (TPB) test. Consider the problem indicated in Figure \ref{fig:3pb}a.  The single notched concrete specimen is pushed downwards with imposed displacements $\bar{u}$.  Experimental results of interest for this problem have been  reported in \cite{ROESLER2007300}.  Previous numerical results based on cohesive zone approaches have been reported for this problem by \cite{ROESLER2007300} and \cite{duarte}, among others.  

The material parameters are chosen as $E=32~$GPa, $\nu=0.25$, $\sigma_c = 4.15~$MPa, and $\mathcal{G}_c=0.16~$N/mm. The regularization length is set to $L=5~$mm. The length of the fracture process zone is $\ell_{\text{FPZ}}\approx 149~$mm, which is fairly large relative to the overall dimensions of the beam. The mesh is refined in the center of the beam where the crack is expected to propagate.

\begin{figure}[!tbp]
\centering \small
\begin{overpic}[width=0.44\linewidth]{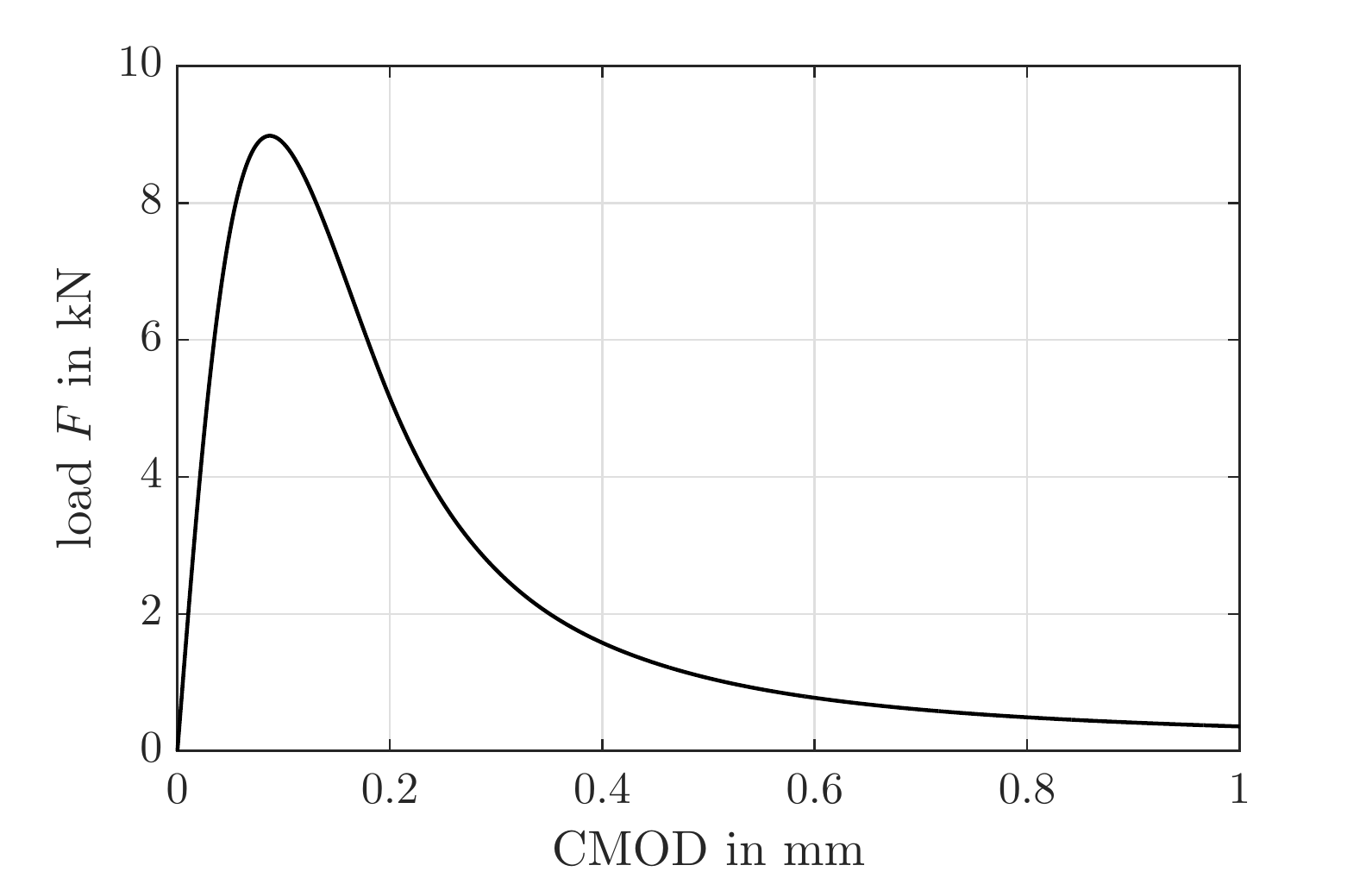}
\put(0,2){(a)}
\end{overpic}
\begin{overpic}[width=0.44\linewidth]{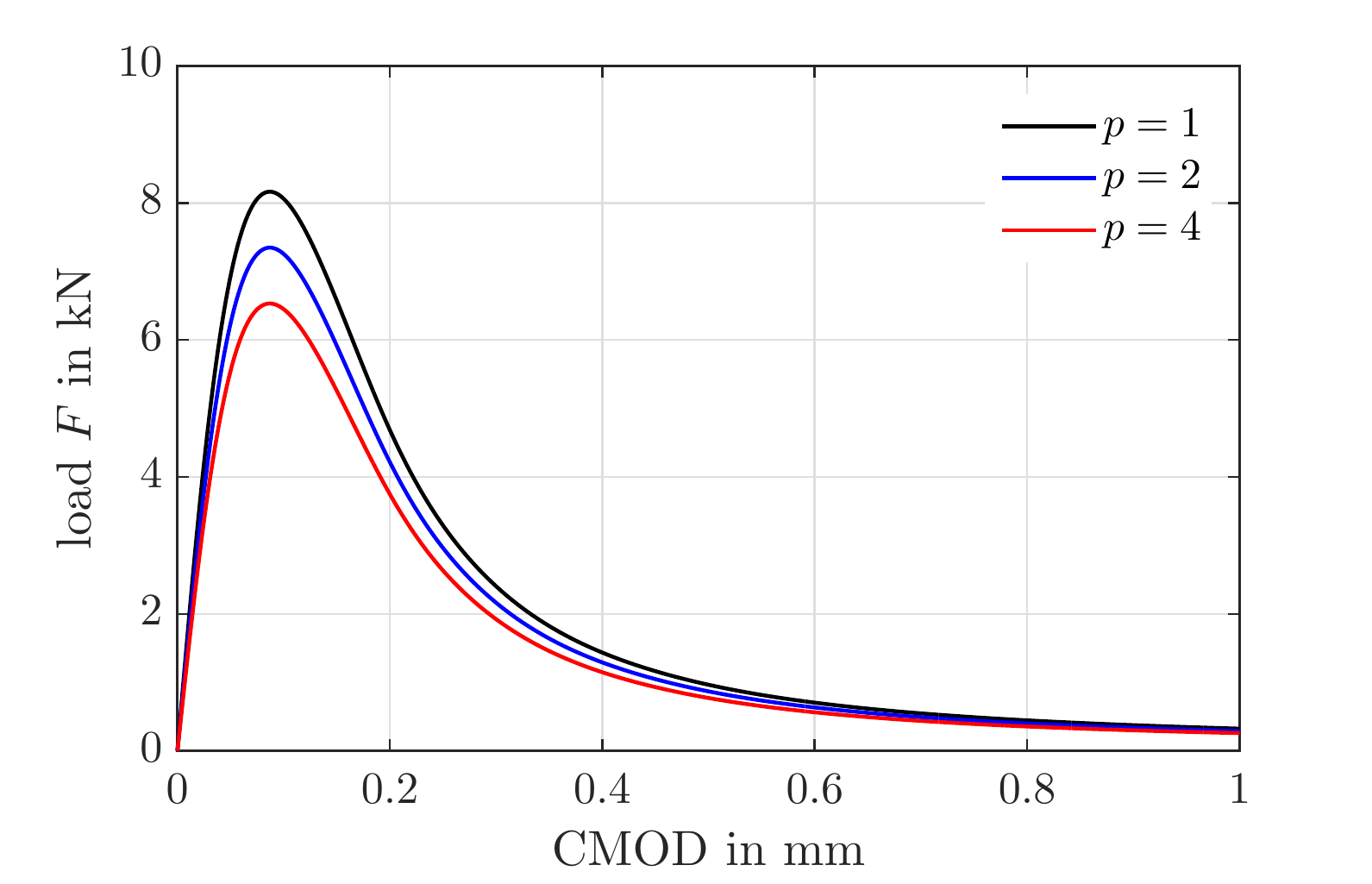}
\put(0,2){(b)}
\end{overpic}
\caption{Three-point bending test: reaction force $F$
versus crack mouth opening displacement (CMOD) using (a) the quasi-linear degradation function (b) the quasi-quadratic degradation function.}
\label{fig:3pb_fd}
\end{figure}

We examine the influence of the quasi-linear and quasi-quadratic degradation functions on the constitutive response. It is emphasized that the former function requires additional penalization to prevent the approximation to the damage field from exceeding unity, as discussed in Section \ref{sec:alm}. 

Figure \ref{fig:3pb_fd} shows the computed force-displacement curves for both degradation functions under consideration.
We emphasize that all calculations employ the same values for the critical energy and cohesive strength.  Nevertheless, the results shown in Figure \ref{fig:3pb_fd} indicate that the peak load sustained by the structure is sensitive to the particular degradation function. The quasi-linear degradation function gives rise to the largest peak load.  As the shape parameter $p$ in the quasi-quadratic function is increased, the peak load decreases. We note that among the results shown here, those obtained using the quasi-quadratic degradation function and $p=2$ compare most favorably to the experimental load-displacement results reported by \cite{ROESLER2007300}.  

\subsection{Dynamic crack branching}

\begin{figure}[!bp]
\centering \small
\begin{overpic}[width=0.52\linewidth]{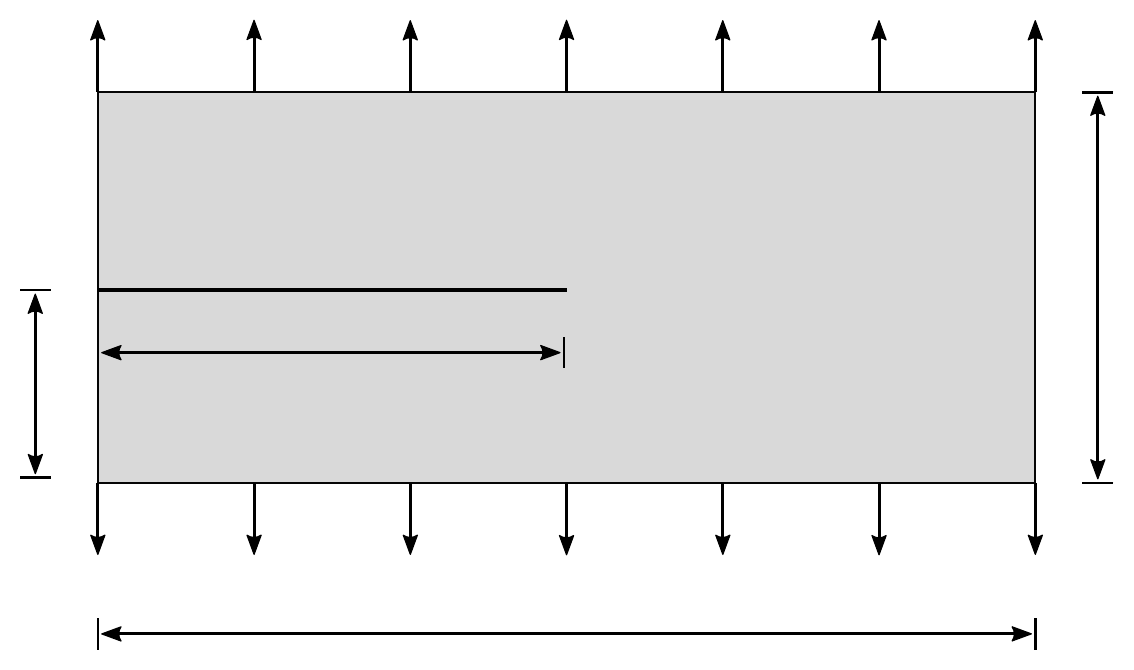}
\put(98,33){40~mm}
\put(-9.5,24.5){20~mm}
\put(44,5){100~mm}
\put(24,23){50~mm}
\put(40,59){$\sigma=1$~MPa}
\end{overpic}
\caption{Geometry and boundary conditions for a dynamic crack branching problem.}
\label{fig:branching}
\end{figure}
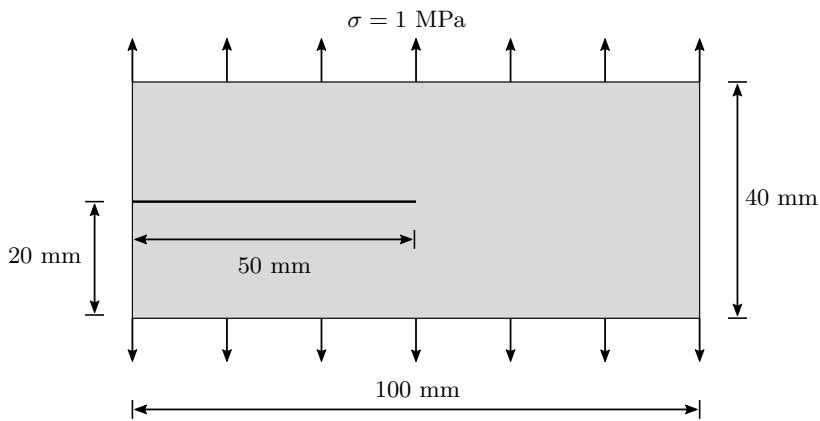

We now examine a standard benchmark problem in dynamic fracture that has been extensively investigated in previous works, including \cite{NME:NME941} and \cite{BORDEN201277}, to name a few. Consider a notched rectangular plate of dimensions $100\times40$~mm dynamically loaded in tension, as shown in Figure \ref{fig:branching}. 
The domain contains an initial crack that is horizontal and spans from the left edge of the plate to its center. A tensile load of $\sigma=1$~MPa is applied to the top and bottom surfaces and held constant throughout the course of the simulation.

The material parameters are set to $E=32$~GPa, $\nu=0.2$, $\mathcal{G}_c=0.003$~N/mm, $\sigma_c=3.08~$MPa, and $\rho=2450$~kg/m$^3$. The corresponding dilatational, shear and Rayleigh wave speeds are $v_d=3810$~m/s, $v_s=2333$~m/s and $v_R=2125~$m/s, respectively. The characteristic length of the fracture process zone is approximately  $\ell_{\text{FPZ}} \approx 10$mm. The regularization length is set to $L=1.25~$mm. The computation is performed on a uniform mesh with spacing $h_e = L/10$.

\begin{figure}[!tbp]
\centering \small
\begin{overpic}[width=0.95\linewidth]{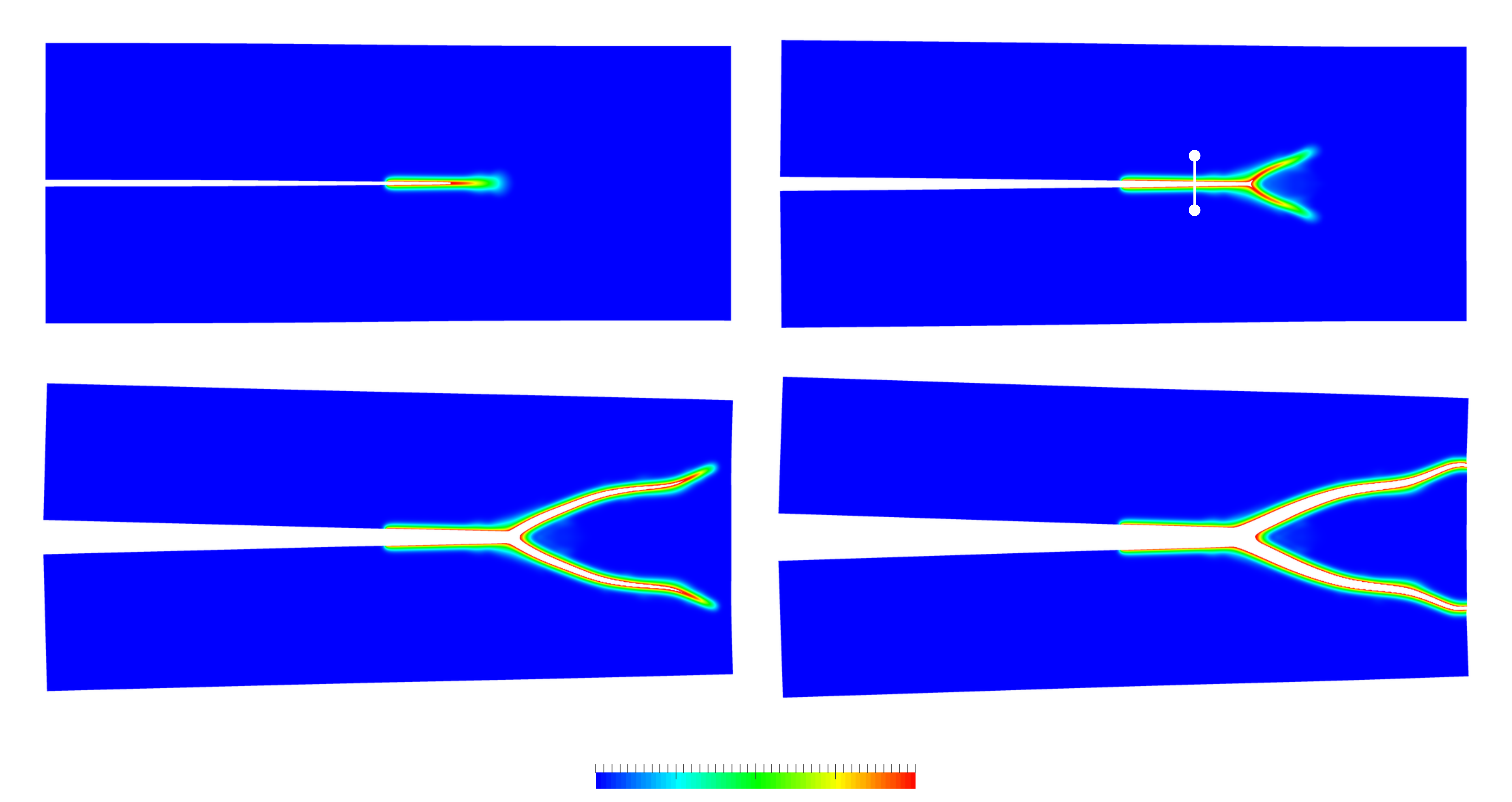}
\put(0.5,30){(a)}
\put(0.5,6) {(c)}
\put(49,30) {(b)}
\put(49,6)  {(d)}
\put(33.5  ,0.75){$d=0$}
\put(62,0.75){$d=1$}
\put(78.25,36.5){\color{white} $A$}
\put(78.25,44){\color{white} $A^*$}
\end{overpic}
\caption{Evolution of the damage field with time for the crack branching problem: (a) $t=28~\mu$s; (b) $t=40~\mu$s; (c) $t=63~\mu$s and (d) $t=75~\mu$s. The displacements have been scaled by a factor of 50. In addition, areas of the model where $d\geq0.95$ have been removed in order to show a representation of the fractured geometry.}
\label{fig:branching_results}
\end{figure}

\begin{figure}[!tbp]
\centering
\begin{overpic}[width=0.44\linewidth]{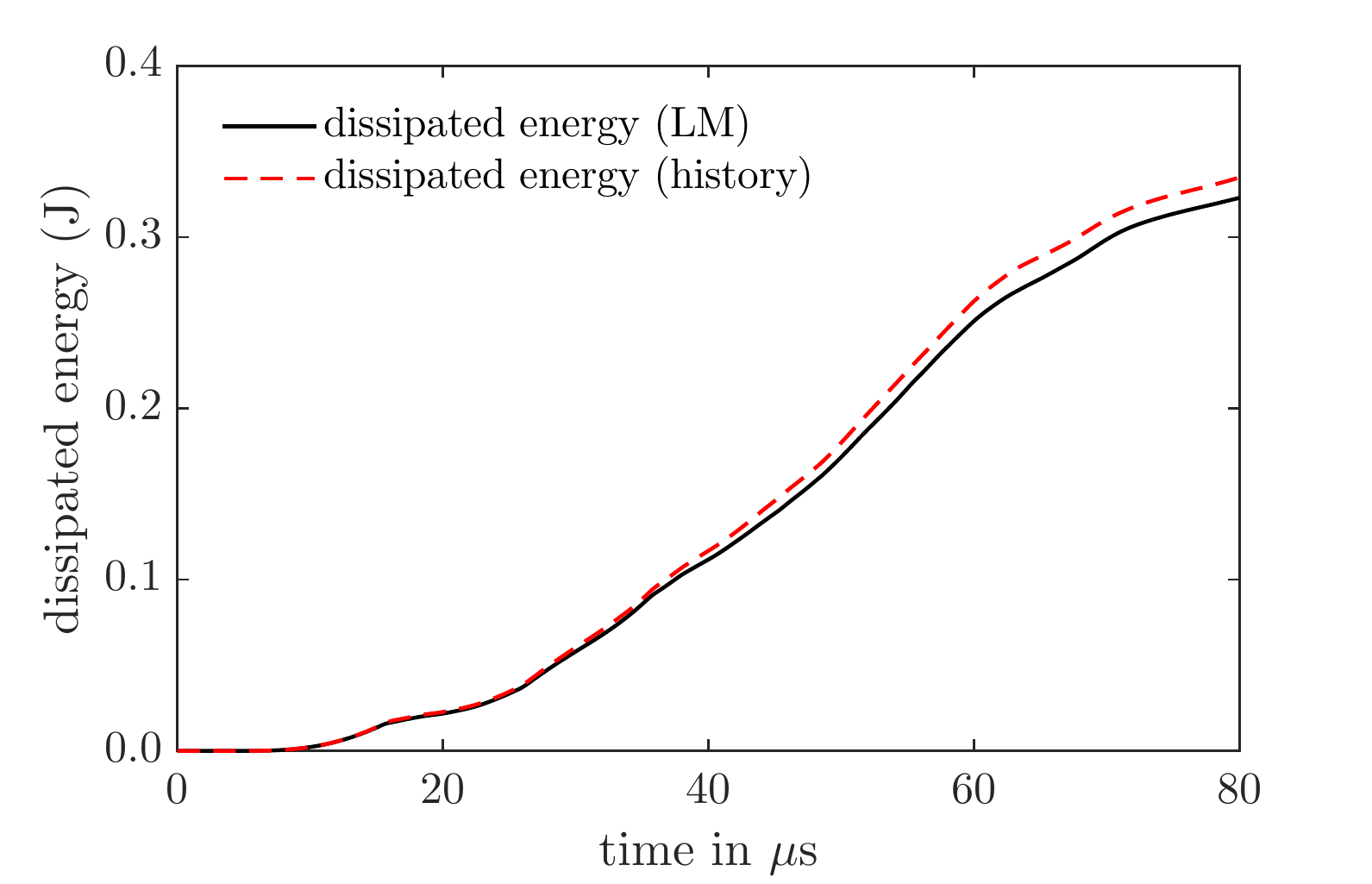}
\put(0,2){(a)}
\end{overpic}
\begin{overpic}[width=0.44\linewidth]{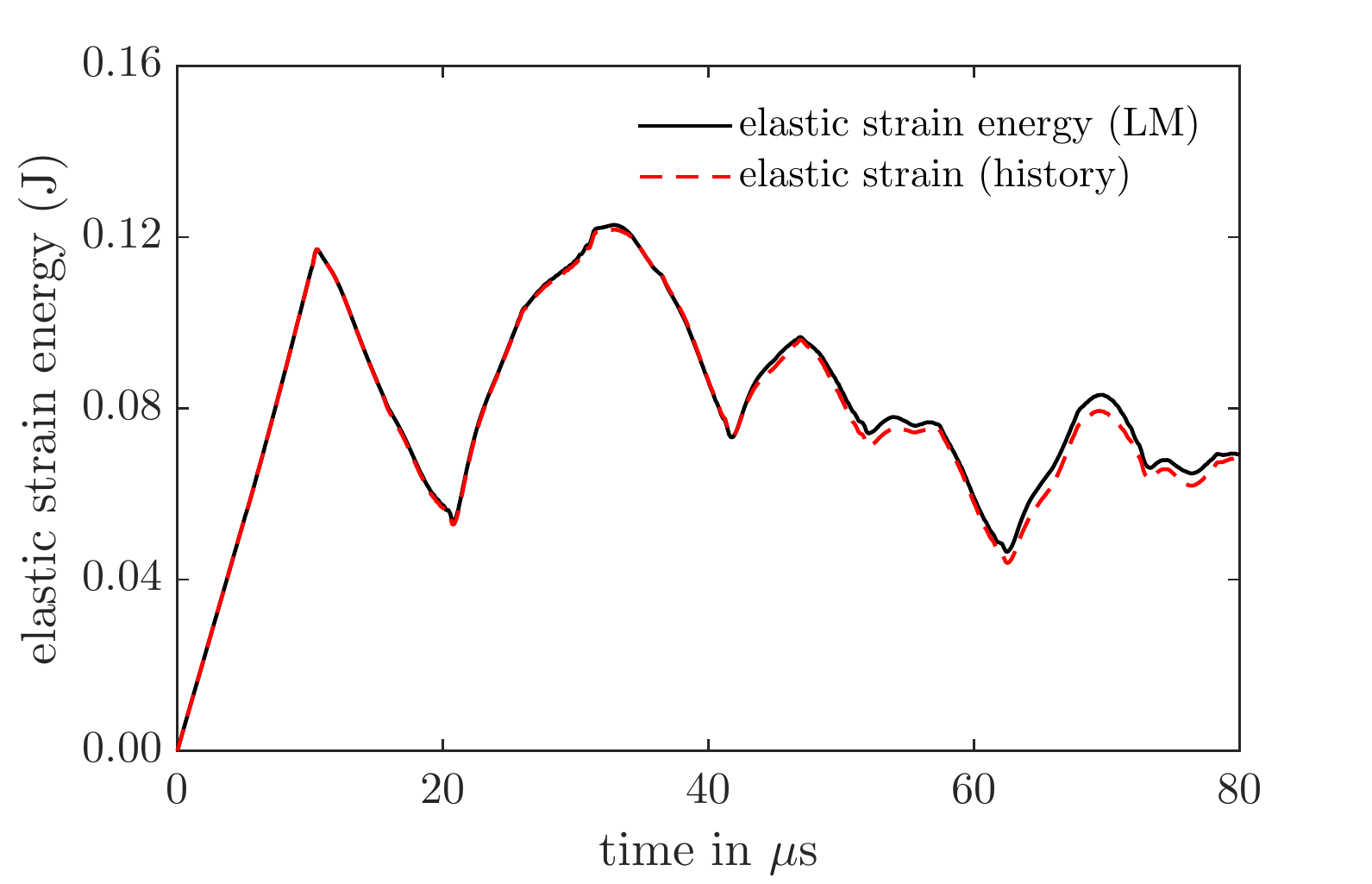}
\put(0,2){(b)}
\end{overpic}
\caption{Evolution of (a) the dissipated energy and (b) the elastic strain energy through time for the Lagrange multiplier (LM) and history-based formulations for the crack branching problem.}
\label{fig:branching_energy}
\end{figure}

\begin{figure}[!tbp]
\centering
\begin{overpic}[width=0.44\linewidth]{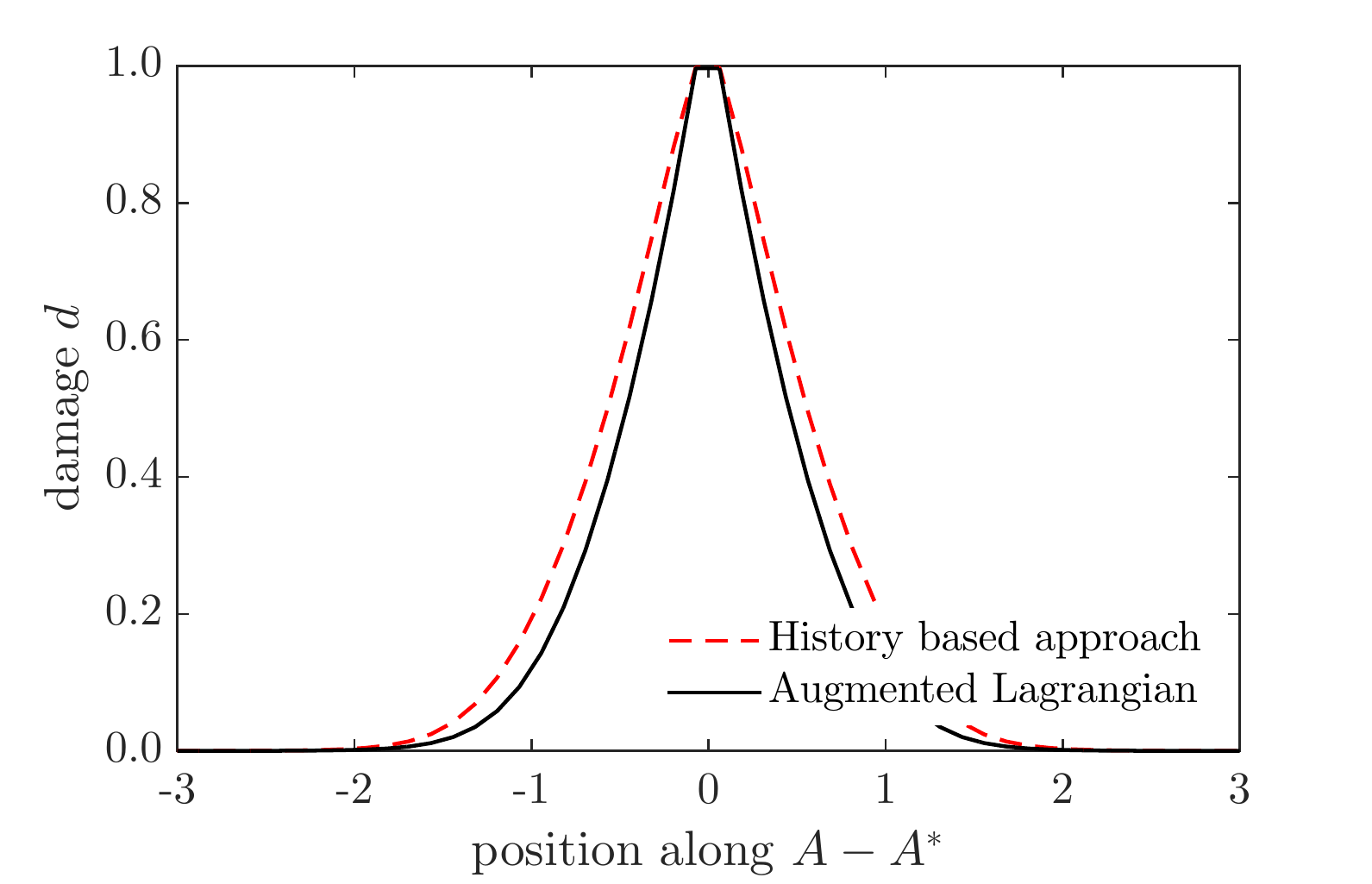}
\end{overpic}
\caption{Comparison of the damage distributions across line segment $A-A^*$ for the two different strategies for enforcing the irreversibility constraint.}
\label{fig:damage_comparison}
\end{figure}

The damage field at selected time steps is shown in Figure \ref{fig:branching_results}. The proposed model for cohesive fracture is clearly capable of capturing the bifurcation of a rapidly evolving cohesive crack. Although cracks are not tracked algorithmically in this approach, crack-tip velocities can be estimated by post-processing a fixed damage iso-contour.  Our results indicate that the crack tip velocity remains below 60\% of the Rayleigh wave speed at all times, agreeing with previous numerical investigations of this problem as well as the experimental observations of \cite{Ravi-Chandar1984}.

This particular problem also lends itself to an investigation of the impact of the various strategies for enforcing the irreversibility constraint.   In Figure \ref{fig:branching_energy}, the evolution of the elastic strain and dissipated energies is displayed for both the history and Lagrange multiplier methods of imposing a monotonically increasing damage field. While the history formulation is computationally more efficient (requiring a single solve of the nonlinear damage equation per time step), it results in a greater degree of energy dissipation. In the calculations, this manifests itself mainly through a widening of the regularized crack surface as compared to the results obtained with the Lagrange multiplier implementation. This is clearly demonstrated by looking at a cross-section A-A$^*$ across the regularized fracture surface, shown in Figure \ref{fig:damage_comparison}.

\subsection{The Kalthoff-Winkler experiments}

\begin{figure}[!bp]
\centering \small
\begin{overpic}[width=0.35\linewidth]{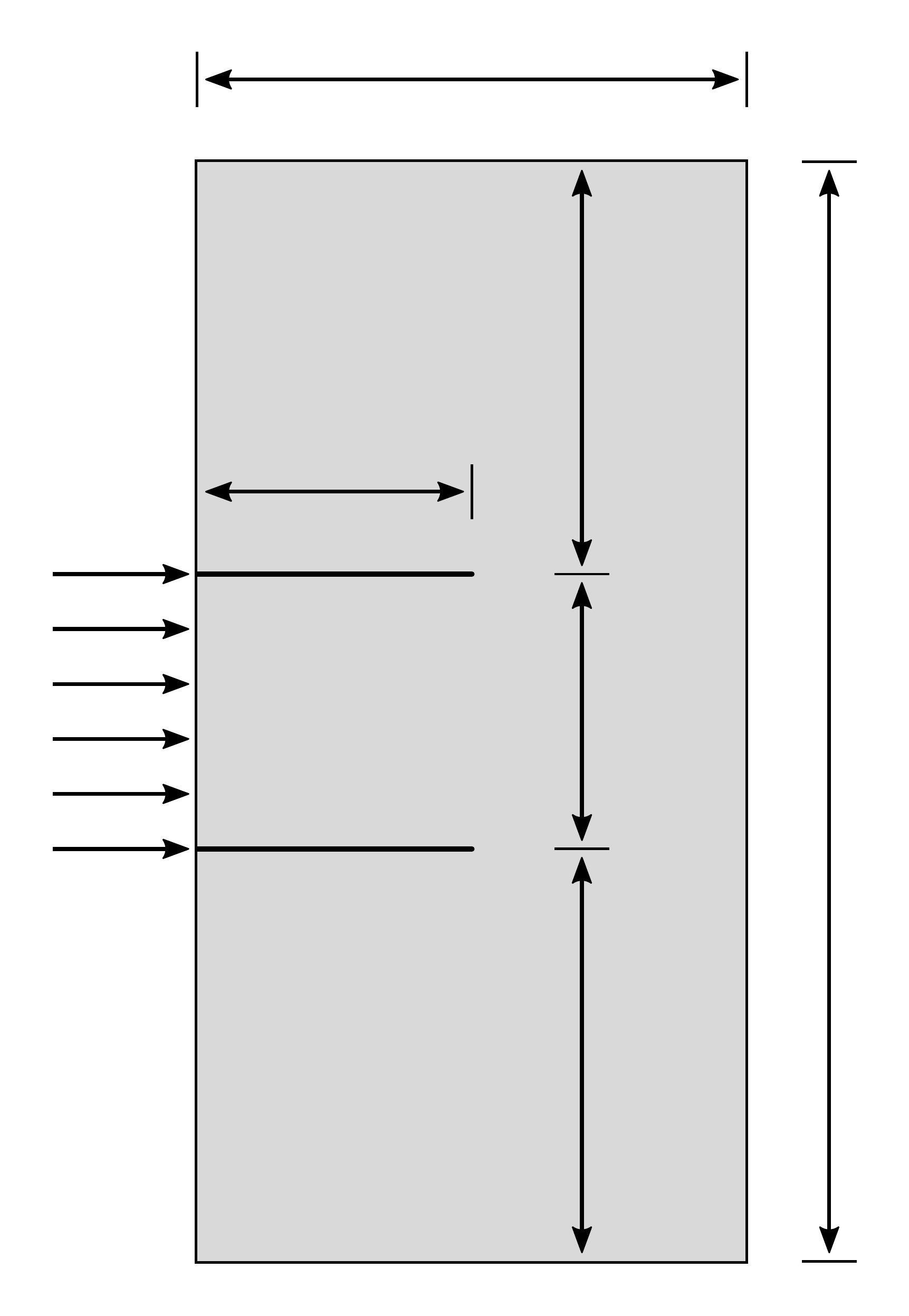}
\put(-5,44.5){$v(t)$}
\put(45.25,18){$75\,$mm}
\put(45.25,45){$50\,$mm}
\put(45.25,72){$75\,$mm}
\put(20,64.5){$50\,$mm}
\put(64,45){$100$~mm}
\put(29,96){$100\,$mm}
\end{overpic}
\caption{The Kalthoff problem: geometry and boundary conditions.}
\label{fig:kalthoff}
\end{figure}

We now focus on a second benchmark problem in dynamic fracture, namely models of the experiments by \cite{kalthoff1987}.  Consider a plate with two edge notches that is impacted by a projectile, as shown in Figure \ref{fig:kalthoff}.  The notches extend halfway through the plate width. Due to symmetry, only the upper half of the specimen is explicitly modeled in the simulations that follow.


\begin{figure}[!tbp]
\centering \small
\begin{overpic}[width=0.9\linewidth]{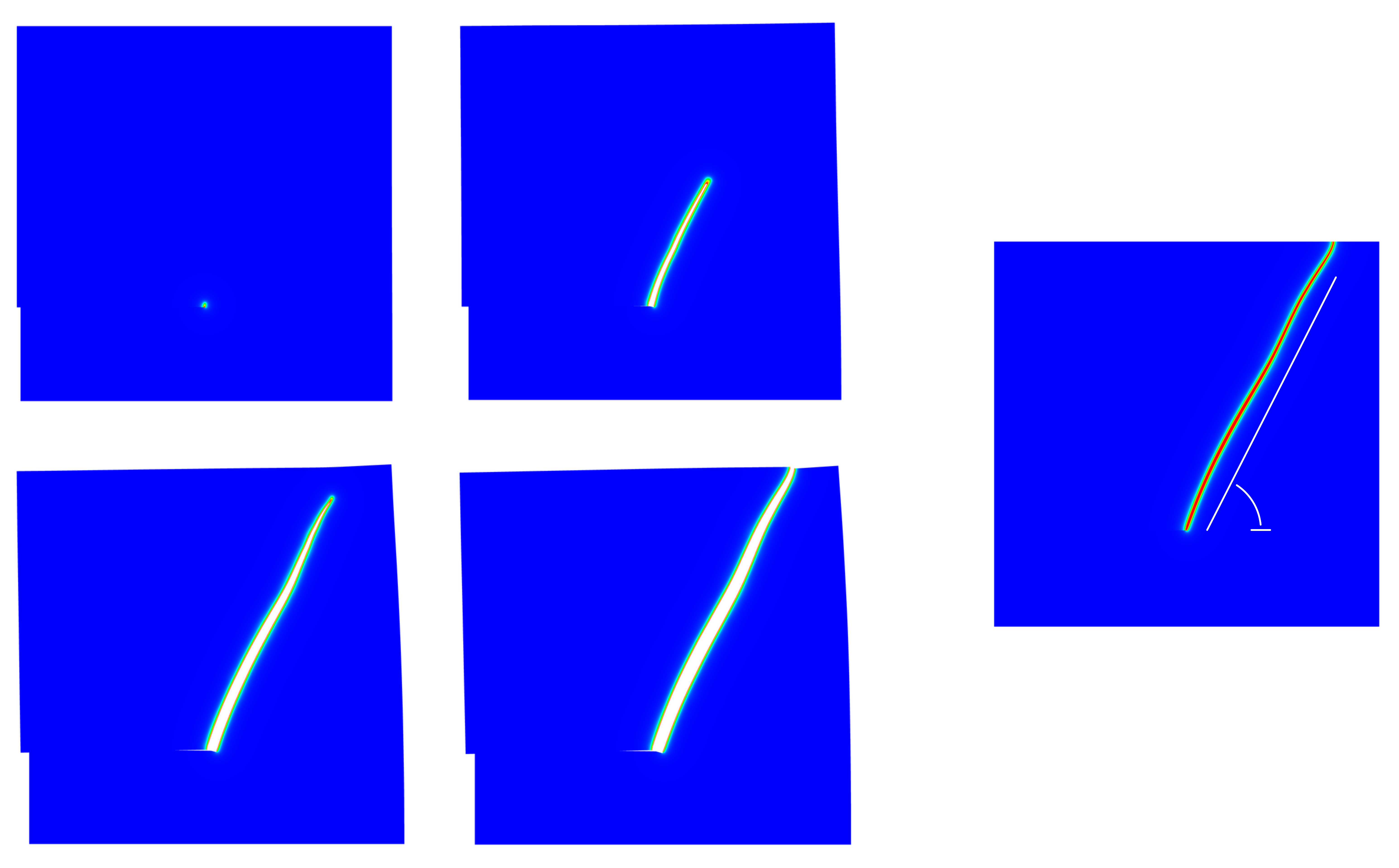}
\put(-1.5,33.25){(a)}
\put(-1.5,1.5) {(c)}
\put(30,33.25){(b)}
\put(30,1.5) {(d)}
\put(68,17.25){(e)}
\put(90,26.5){{\color{white}$63^{\circ}$}}
\end{overpic}
\hspace{1em}
\raisebox{10.25em}{
\begin{overpic}[width=0.015\linewidth]{legend.png}
\put(-9,-11){$d=0$}
\put(-9,103.5){$d=1$}+
\end{overpic}}
\caption{Left: damage fields for the Kalthoff problem at various times: (a) $t=20~\mu$s; (b) $t=45~\mu$s; (c) $t=70~\mu$s and (d) $t=85~\mu$s. The displacements are magnified by a factor of 3. In addition, areas of the model where $d\geq0.95$ have been removed in order to show a representation of the fractured geometry. (e) The resulting crack propagation occurs at an angle of approximately 63$^{\circ}$.}
\label{fig:kalthoff_results}
\end{figure}

The experiments of \cite{kalthoff1987} exhibited different fracture/damage behaviors of a maraging steel material as a function of the impact velocity. For relatively low impact velocities, brittle fracture was observed with propagation angles of approximately 70$^{\circ}$ from the original crack plane. For higher impact velocities, failure was observed to occur due to  shear localization originating from shear band formation ahead of the notch. In this work,  we study the brittle failure mode. In particular, we are concerned with the load case where the impact velocity of the projectile is $33\,$m/s. Assuming that the projectile has the same impedence as the specimen, we apply only half of its velocity.  Following \cite{NME:NME941}, we therefore consider a velocity boundary condition applied to the notched area with magnitude $v_0=16.5\,$m/s, linearly ramped up from zero over a time interval of $t_R=1~\mu$s.
 
\begin{figure}[!tbp]
\centering
\begin{overpic}[width=0.45\linewidth]{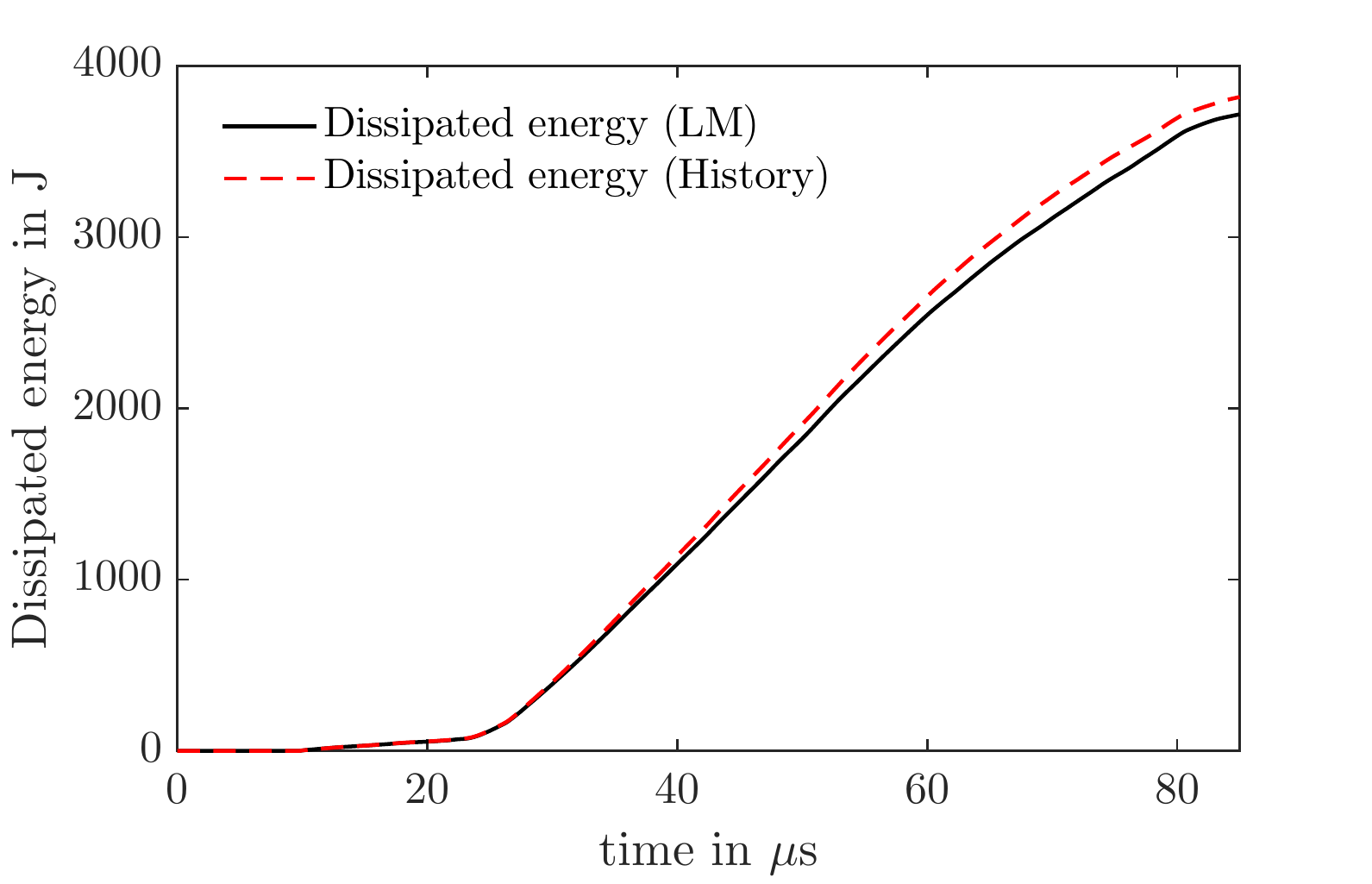}
\put(0,2){(a)}
\end{overpic}
\begin{overpic}[width=0.45\linewidth]{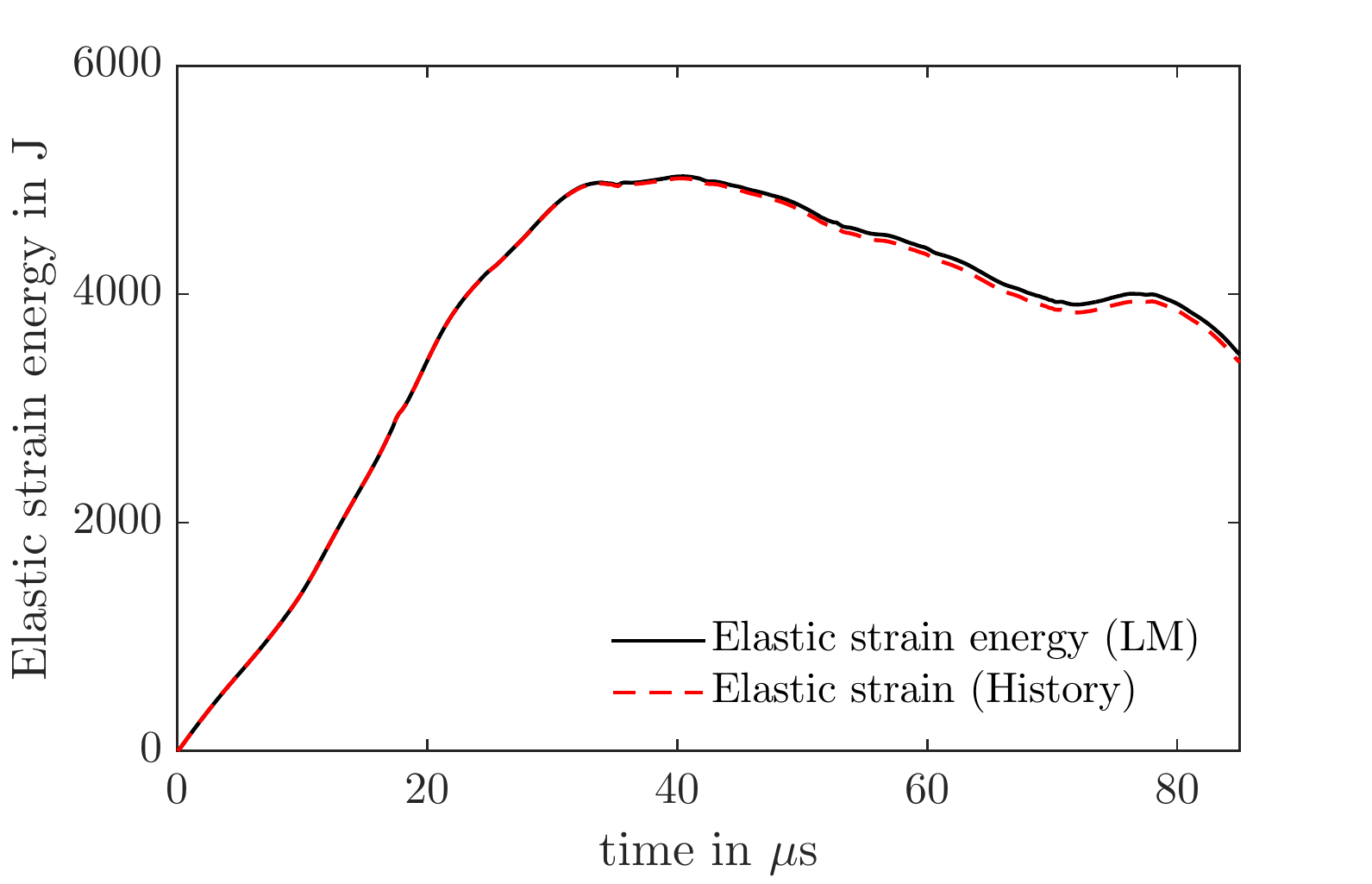}
\put(0,2){(b)}
\end{overpic}
\caption{Evolution of (a) the dissipated energy and (b) the elastic strain energy with time for the Lagrange multiplier and history-based formulations, for the Kalthoff problem.}
\label{fig:kalthoff_energy}
\end{figure}

The maraging steel used in the original experiments is modeled using the following material properties: $E=190$~GPa, $\nu=0.3$, $\mathcal{G}_c = 22.2~$N/mm and $\rho=8000\,$kg/m$^3$. This leads to dilatational, shear and Rayleigh wave speeds of $v_d=5654\,$m/s, $v_s=3022~$m/s and $v_R=2803~$m/s, respectively.  The cohesive strength has a value of $\sigma_c = 1.733$~GPa. To avoid the use of a fairly small regularization length, a frequently adopted strategy in the phase-field community is to lower the effective strength of the material to facilitate a computational analysis, as in \cite{BORDEN201277}. In that work, the maximum uniaxial tensile strength was set to 1.07~GPa. We found that the use of an artificially small cohesive strength gives rise to secondary cracking at the bottom of the specimen, very similar to what was reported in \cite{borden2012isogeometric} and \cite{moreau2015explicit}. 
Here, we report simulation results using the ``true'' cohesive strength of $\sigma_c = 1.733$~GPa, which gives rise to a fracture process zone with characteristic size $\ell_{\text{FPZ}} \approx 1.4~$mm.  We conduct simulations using a regularization length of $L=0.7$~mm, using a uniform mesh with mesh spacing $h = L/10$.

The damage field at selected times during the simulation is shown in Figure~\ref{fig:kalthoff_results}a-d. Our results indicate a crack propagation with an average orientation of approximately $63^{\circ}$, as shown in Figure \ref{fig:kalthoff_results}e. The result compares favorably with experimental observations of an angle of roughly $70^{\circ}$, as well as the $65^{\circ}$ result from the phase-field simulation of \cite{BORDEN201277}. Figure \ref{fig:kalthoff_energy} shows the evolution of the dissipation and elastic energy functionals through time for the various irreversibility implementations. The history implementation once again is seen to dissipate more energy than the Lagrange multiplier approach advocated in this work.

\subsection{Fragmentation of a thick cylinder}

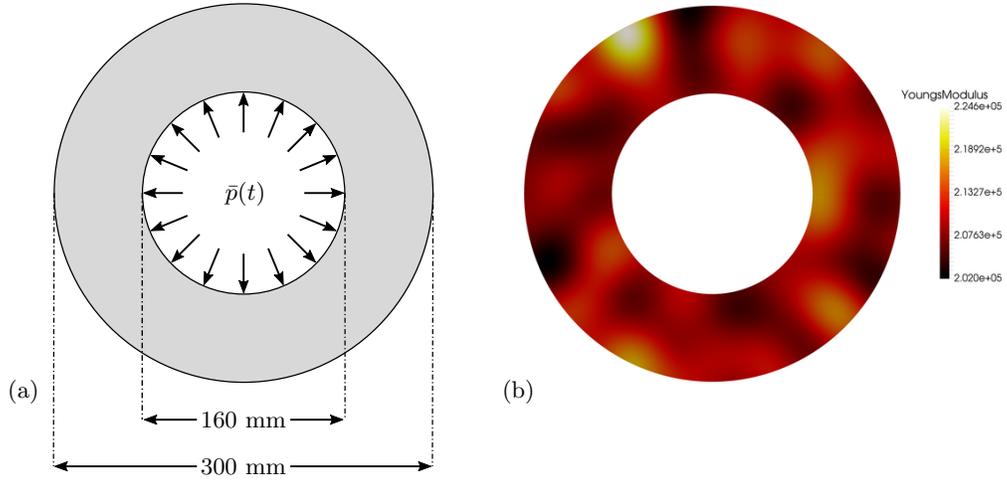
\begin{figure}[!tp]
\centering \small
\begin{overpic}[width=0.75\linewidth]{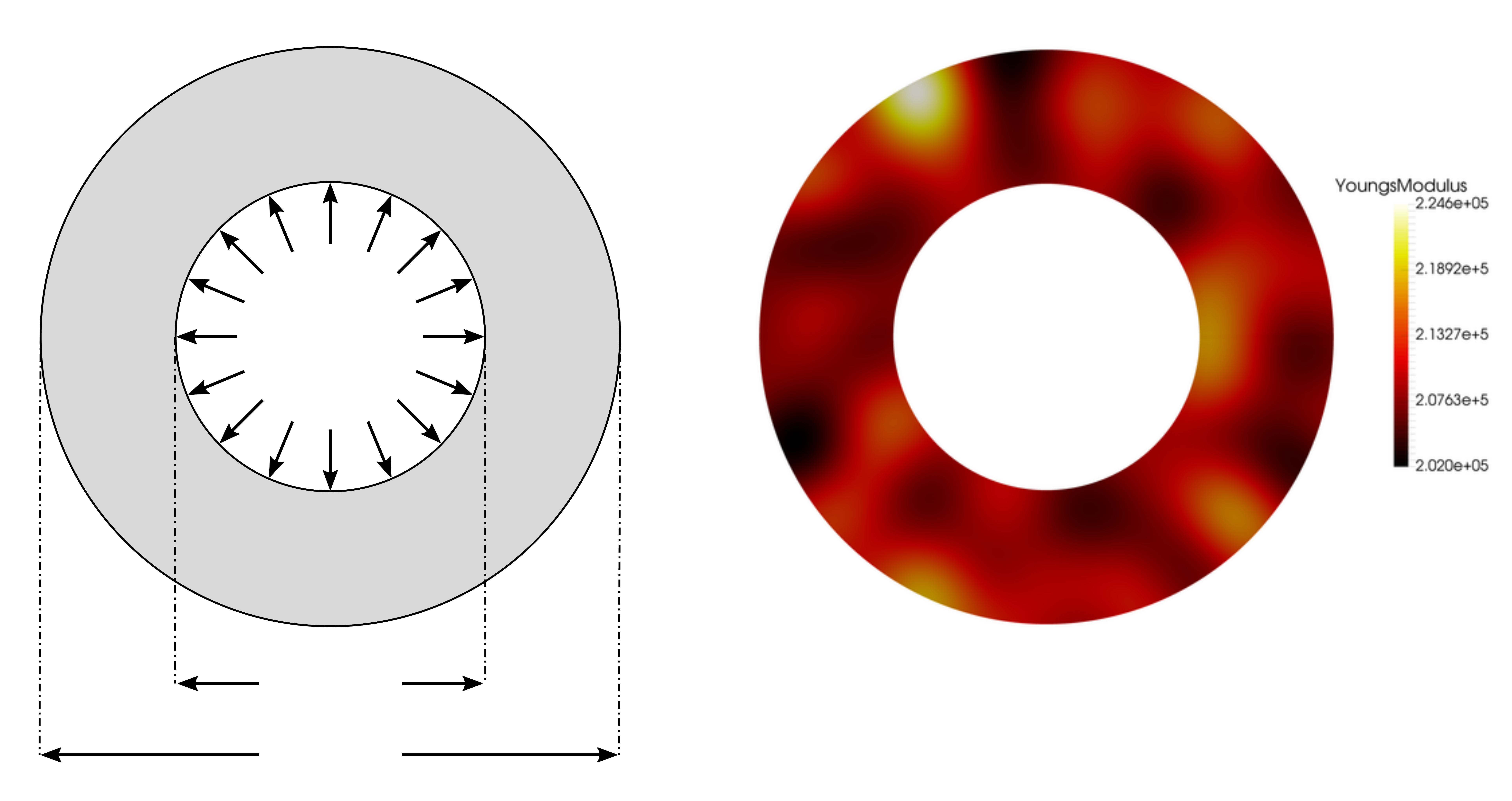}
\put(-2,10){(a)}
\put(48,10){(b)}
\put(20,30){$\bar{p}(t)$}
\put(17.5,7){160~mm}
\put(17.5,2.25){300~mm}
\end{overpic}
\caption{(a) Geometry and boundary conditions for the fragmentation of a thick cylinder. (b) one sample of a spatially random Young's modulus field.}
\label{fig:song_set-up}
\end{figure}

In order to demonstrate the capabilities of the proposed method for simulating fragmentation problems, we consider a thick cylinder subjected to an impulsive internal pressure, as shown in Figure~\ref{fig:song_set-up}a.  The cylinder has inner and outer radii of 80 and 150~mm, respectively.  The inner surface is subjected to a spatially constant pressure that evolves in time as $p(t)= p_0 \exp \left( −t/t_0 \right) $, after an initial linear ramp to $p_0 =400~$MPa over a rise time of $t_0 =100~\mu$s. This problem has been studied previously by \cite{song2009cracking}, using a cracking-node algorithm, and by \cite{doi:10.1002/nme.5819} using a particular cohesive zone model, to mention a few.

The material parameters are $E=210~$GPa, $\rho=7850~$kg/m$^3$, and $\nu=0.30$. This leads to dilatational, shear and Rayleigh wave speeds of $v_d=6001\,$m/s, $v_s=3208~$m/s and $v_R=2971~$m/s, respectively.  Given the critical strain of $0.5~\%$ provided by \cite{song2009cracking}, the cohesive strength was set to $\sigma_c = 1~$GPa. The fracture energy was set to $\mathcal{G}_c = 20~$N/mm, while the regularization length scale was set to $L=2~$mm. The size of the process zone associated with this set of material parameters is  $\ell_{\text{FPZ}} \approx 4.2~$mm.

\begin{figure}[!bbp]
\centering \small
\begin{overpic}[width=0.95\linewidth]{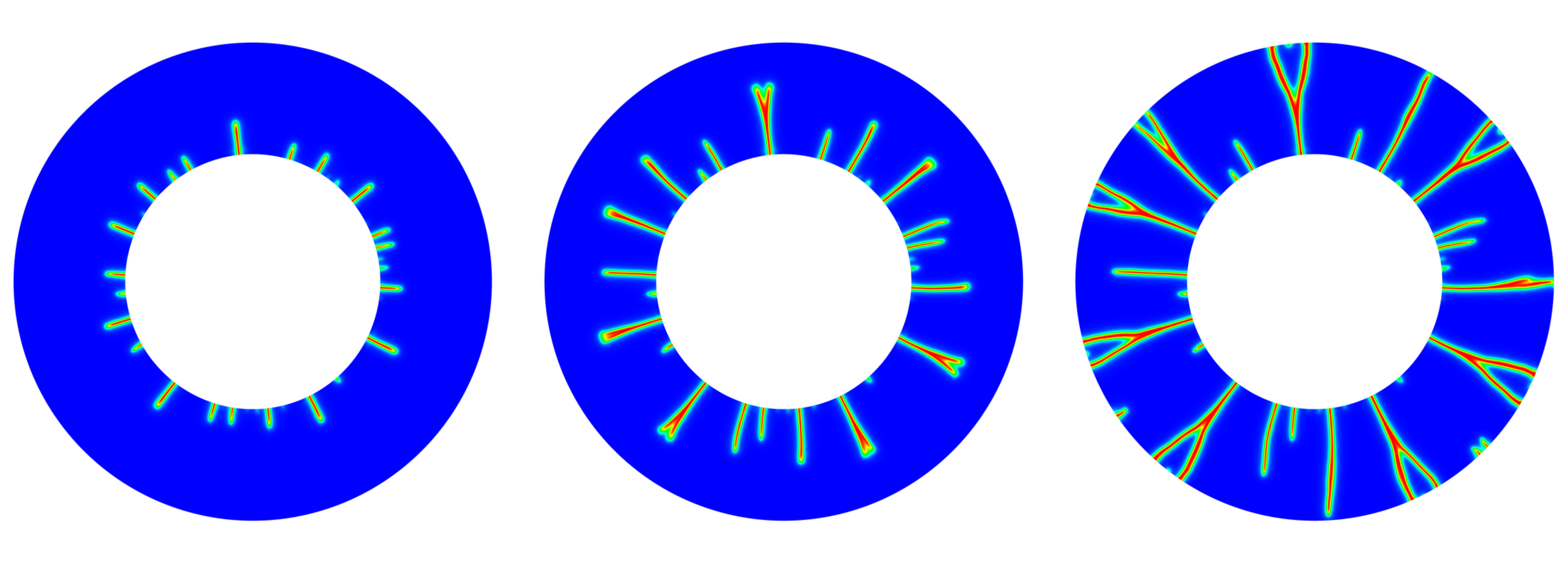}
\put(0,2){(a)}
\put(34,2){(b)}
\put(68,2){(c)}
\end{overpic}
\caption{Damage field for the thick-walled cylinder fragmentation problem on mesh 3 at times: (a) $t=55~\mu$s; (b) $t=65~\mu$s; and (c) $t=80~\mu$s.}
\label{fig:song_results}
\end{figure}

\begin{figure}[!bbp]
\centering \small
\begin{overpic}[width=0.95\linewidth]{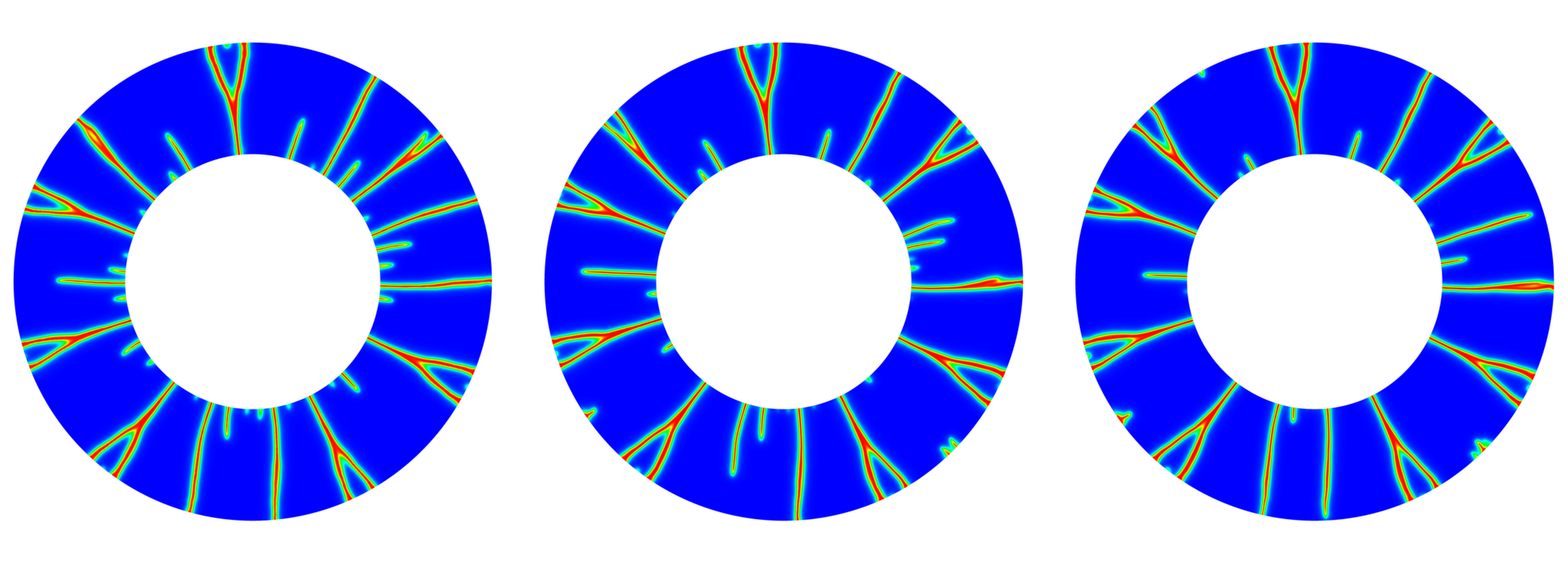}
\put(0,2){(a)}
\put(34,2){(b)}
\put(68,2){(c)}
\end{overpic}
\caption{Final crack patterns in the reference configuration for (a) mesh 1 (b) mesh 2 and (c) mesh 3.}
\label{fig:song_results2}
\end{figure}

\begin{figure}[!tbp]
\centering \small
\begin{overpic}[width=0.45\linewidth]{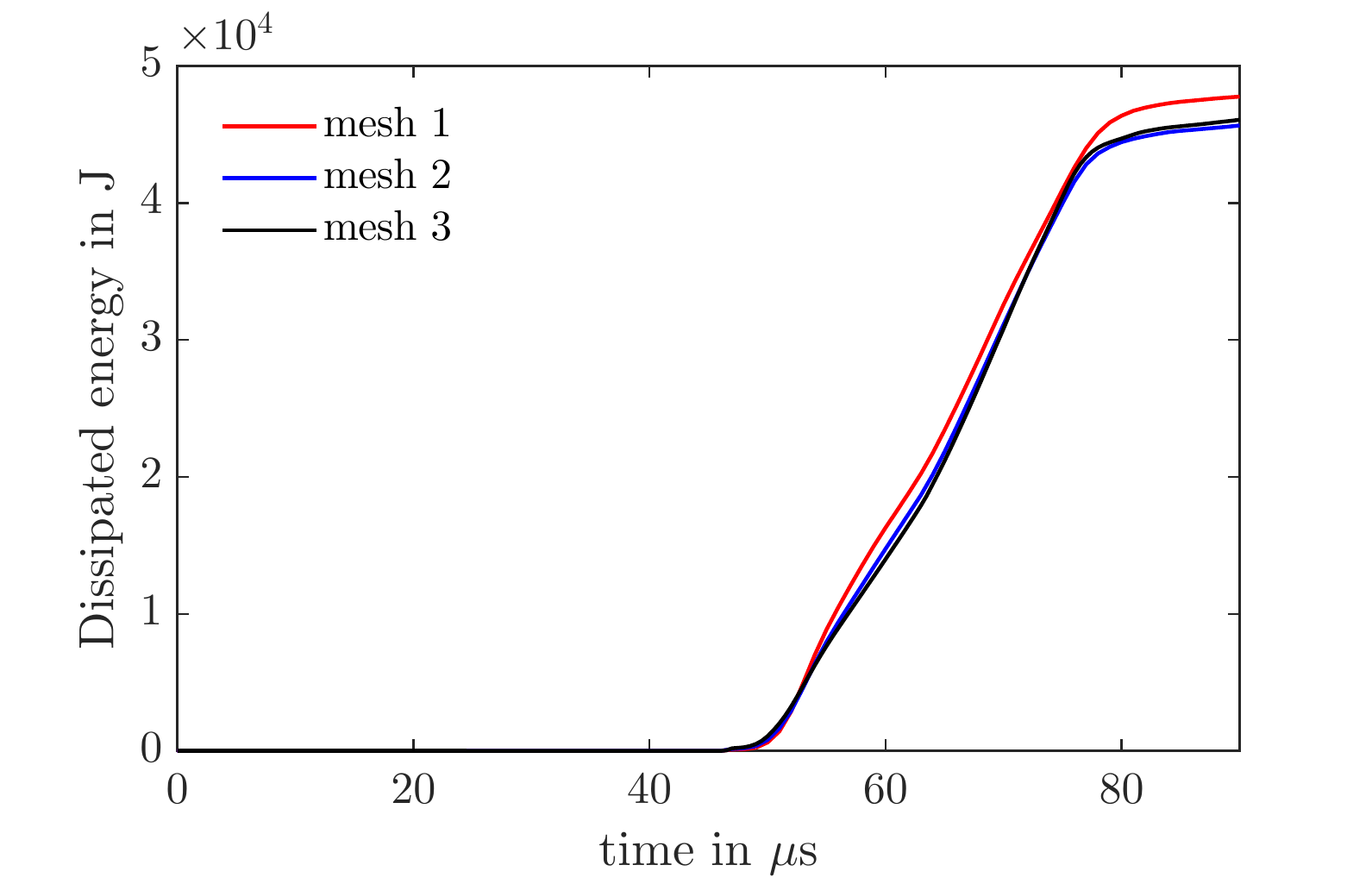}
\put(0,2){(a)}
\end{overpic}
\begin{overpic}[width=0.45\linewidth]{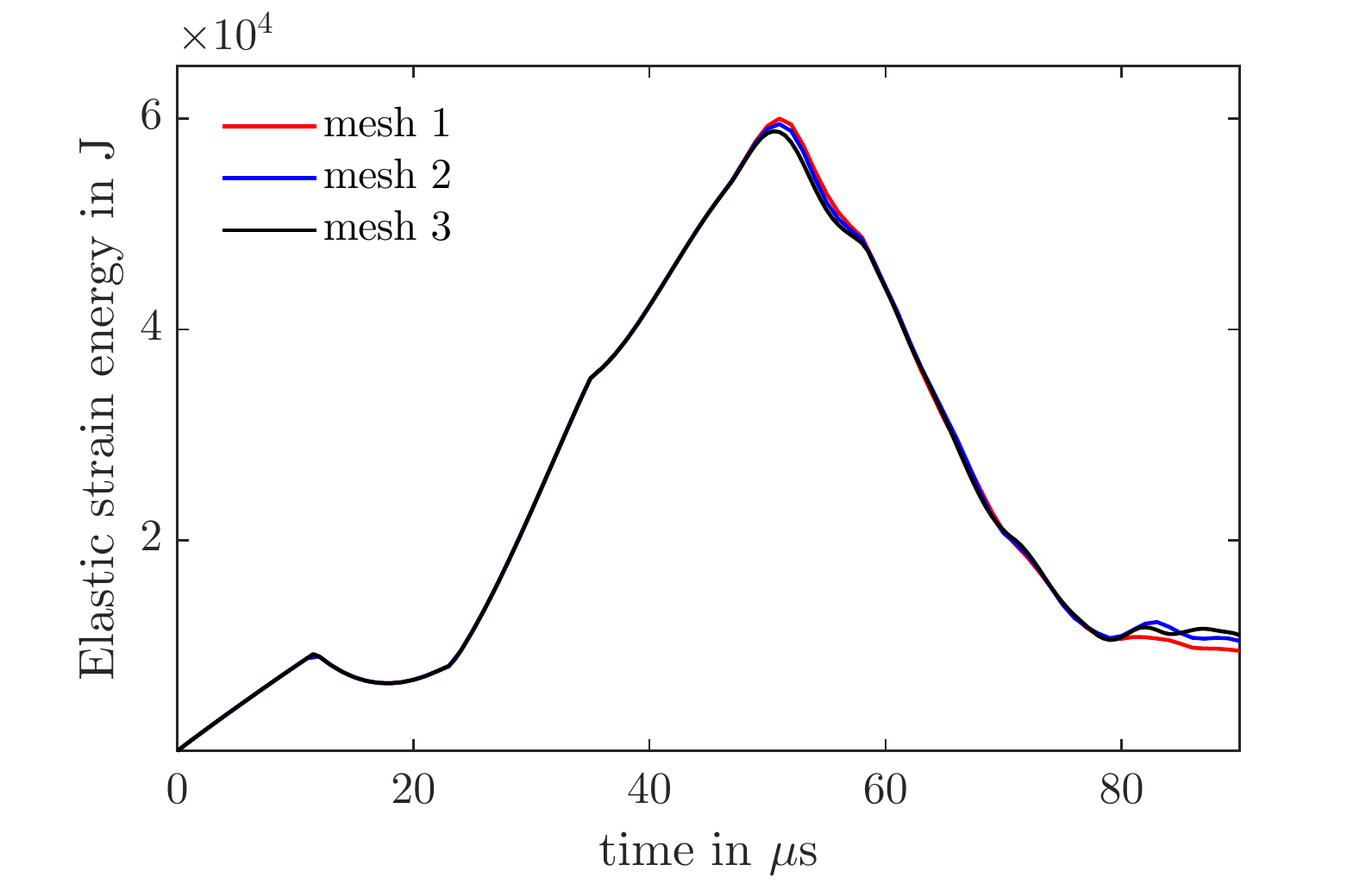}
\put(0,2){(b)}
\end{overpic}
\caption{Evolution of the (a) dissipated energy and (b) elastic strain energy with time for the thick cylinder fragmentation problem for a sequence of meshes of increasing refinement.}
\label{fig:song_analysis}
\end{figure}

To introduce a stochastic element to the simulations, we construct a spatially-varying Young's modulus field according to 
\begin{equation}
E(\mathbf{x},\theta) = \bar{E}(\mathbf{x}) + \sum_{i=1}^n \sqrt{\lambda} \phi_i (\mathbf{x}) \xi_i(\theta),
\end{equation}
where $\bar{E}$ is the mean value of the random field; $\xi(\theta)$ the uncorrelated Gaussian random variable and $\lambda_i, \psi_i(\mathbf{x})$ the set of eigenvalues and basis functions associated with the Fredholm equation of the second kind, respectively; and $n$ is the number of basis functions. We construct the spatially varying Young's modulus field shown in Figure~\ref{fig:song_set-up}b using $n=50$ basis functions.  The resulting field possesses a mean of $\bar{E} = 210~$GPa, with a correlation length of $\ell_{\text{cor}} = 20~$mm, and a standard deviation of 5\%. Additional details concerning the construction of such fields are provided in \cite{SHANG201365}. 

We now hold the regularization length and spatially varying Young's modulus field fixed, and examine the fracture patterns predicted by our simulations over a sequence of increasingly refined meshes. Table \ref{table:meshes} provides the number of elements for each mesh and the corresponding ratio of regularization length to element size.  

\begin{table}[!htbp]
\centering
\caption{The sequence of meshes considered in the convergence study of the thick cylinder fragmentation problem.}
\label{table:meshes}
\begin{tabular}{l|lll}
                   & Mesh 1 & Mesh 2 & Mesh 3 \\ \hline
Number of elements & 218400 & 400000 & 728000 \\
$L/h_e$ ratio      & 4      & 6      & 8     
\end{tabular}
\end{table}

Figure \ref{fig:song_results} shows the fragmentation process at different times for the most refined mesh, mesh 3. The fragmentation process begins around $t=50~\mu$s with a large number of small cracks along the inner surface. However, with increasing time, some of the initial cracks arrest, and the remainder continue to propagate to the outer surface of the cylinder. The fragmentation process completes around $t=80~\mu$s, when the multiple fragments move outward from the center of the domain with no further crack initiation and virtually no growth of the arrested cracks. 

The final damage fields for each mesh are shown in Figure \ref{fig:song_results2}. As can be seen, the fragmentation patterns are very similar and, in fact, the number of large fragments does not appear to change upon mesh refinement. This in contrast to the results presented in \cite{song2009cracking}.

Finally, Figures \ref{fig:song_analysis}a-b compares the dissipated and elastic strain energies for the uniformly refined sequence of meshes from Table \ref{table:meshes}. The dissipated energy shown in Figure \ref{fig:song_analysis}a is over-predicted on the coarsest mesh 1, and appears to converge quickly as the mesh is refined.

\section{Conclusions}
\label{sec:conclusions}
In this work, we have extended a phase-field/gradient damage model, based on the work of \cite{Lorentz20111927}, to the dynamic case.  Such a formulation is particularly well-suited to operate within the context of cohesive fracture, i.e.\ when a Griffith description of fracture is insufficient. We demonstrated that the combination of an alternative phase-field approximation, supported by a suitable stiffness degradation function, effectively leads to a regularized description of cohesive fracture. Such a formulation is distinctly characterized by a linear elastic regime prior to the onset of damage and controlled strain-softening thereafter. The governing equations are derived according to macro- and microforce balance theories, naturally addressing the irreversibility of crack growth in time by introducing suitable constraints for the kinetics of the underlying microstructural changes.

Our work departs from that of \cite{Lorentz20111927} in a number of key respects.  In particular, we have developed a new degradation function that recovers a linear stress-strain response in the localization zone.  We have also employed a tension-compression decomposition of the strain, similar to what is employed in phase-field models of fracture.  And we have explored options for enforcing the irreversibility of the damage field.  Finally, we have demonstrated the ability of the proposed method to reproduce the results of both idealized benchmark and experimental problems in quasi-static and dynamic fracture.

The current manuscript gives rise to several areas for future investigation. First and foremost, we highlight that no calibration of the degradation and/or corresponding strain-softening parameters was conducted in this work. It is anticipated that such an approach, possibly supported by data-driven optimization methods, could further reduce any discrepancies between simulations and experimental observations. Secondly, we mention the extension of some of the basic elements of a cohesive fracture based phase-field description to an anisotropic setting. Such an extension is expected to be non-trivial, with particular regard to the calculation of the effective properties along given material directions.

\section*{Acknowledgements}
This work was performed under a research grant from Sandia National Laboratories, to Duke University.  That support is gratefully acknowledged.  The work was also partially supported by the Laboratory Directed Research and Development program at Sandia National Laboratories.
 

\bibliographystyle{elsarticle-harv}	
 
\end{document}